%% file: main.tex
\documentclass{article}

\usepackage{PRIMEarxiv}

\usepackage[utf8]{inputenc}
\usepackage[T1]{fontenc}
\usepackage{hyperref}
\usepackage{url}
\usepackage{booktabs}
\usepackage{amsfonts}
\usepackage{nicefrac}
\usepackage{microtype}
\usepackage{graphicx}

\usepackage[numbers]{natbib}
\usepackage{hyperref}

\input{mypackages}
\input{acronyms}

\title{ESBMC: A Survey of Its Evolution, Integration, and Future Directions in Formal Software Verification}

\author{
  Pierre Dantas \\
  Computer Science, The University of Manchester \\
  Manchester, UK \\
  \texttt{pierre.dantas@manchester.ac.uk} \\
  \And
  Lucas Cordeiro \\
  Computer Science, The University of Manchester \\
  Manchester, UK \\
  \texttt{lucas.cordeiro@manchester.ac.uk} \\
  \And
  Waldir Junior \\
  Electrical Engineering, Federal University of Amazonas (UFAM) \\
  Manaus, AM, Brazil \\
  \texttt{waldirjr@ufam.edu.br} \\
}

\begin{document}
\maketitle
\glsresetall

\begin{abstract}
The \gls{esbmc} has grown from a research prototype for verifying embedded ANSI-C software into one of the most versatile and industrially capable formal verification platforms available today. Since its first publication in 2009, \gls{esbmc} has undergone persistent evolution: expanding its verification techniques, widening its language support to nine front-ends, integrating industrial-strength \gls{smt} solvers, and -- most recently -- coupling with \glspl{llm} and autonomous \gls{ai} agents. This survey traces the full trajectory of \gls{esbmc} from its original design principles to the state of the art in 2025--2026, documenting 43 awards at \gls{svcomp} and \gls{test-comp}, peer recognition at leading software engineering venues, including a Distinguished Paper Awards at \gls{icse}'11 and \gls{ase}'24, a Most Influential Paper Award at \gls{ase}'23, and a Best Tool Paper Award at \gls{sbseg}'23, its role as a formal verification backend for \gls{llm}-driven self-healing software and loop invariant generation, and the first industrial deployment of an integrated agentic model-checking architecture through the NVIDIA-OpenSMA framework, establishing \gls{esbmc} as a natively autonomous verification kernel rather than a passive validation backend. We synthesize its economic impact -- over \pounds9.3~million and \euro{}4.98~million in confirmed public research funding, the VeriBee spin-off, and a defense industrial deployment at Lockheed Martin -- and conclude with a structured agenda of open challenges spanning scalability, neurosymbolic verification, counterexample intelligibility, cross-language verification, safety standards compliance, and open-source sustainability.
\end{abstract}


\glsresetall
\section{Introduction}
\label{sec:intro}

Software correctness is one of the oldest and most lasting challenges in computer science. As software systems increase in complexity and permeate safety-critical infrastructure -- including medical devices~\cite{Hatcliff2012}, autonomous vehicles~\cite{Luckcuck2019}, blockchain smart contracts~\cite{sloan2022}, and cloud-native microservices -- the cost of defects escalates correspondingly. Formal verification, the discipline of mathematically proving that a program satisfies a specification, offers the strongest possible guarantees~\cite{Woodcock2009,Clarke2018Handbook}. Among its many techniques, \gls{bmc}~\cite{biere1999,Clarke2001} has emerged as a practical and highly effective approach, balancing conceptual rigor with practical scalability.

The \gls{esbmc} is one of the most significant tools to emerge from this tradition. First presented in 2009 as a framework for verifying embedded \gls{ansi-c} programs using \gls{smt} solvers~\cite{Cordeiro2012}, \gls{esbmc} has been continuously developed and extended since then~\cite{Cordeiro2012,Gadelha2018ESBMC,shmarov2023,menezes2024,menezes2025tacas} by a community anchored at the University of Manchester (UK) and the \gls{ufam} (Brazil). Over that time, it has accumulated 43 awards at international software verification competitions, including 35 at \gls{svcomp} and 8 at \gls{test-comp}~\cite{beyer2012,svcomp2024,ssvlab2024} (per-year tallies are verifiable in the annual competition organizers' reports), support for nine programming languages~\cite {Monteiro2021,sloan2022,farias2024,rustfoundation2024}, and a pioneering integration with \glspl{llm} for automated bug repair~\cite{Tihanyi2025New}.

This survey makes the following contributions:

\begin{enumerate}
    \item Provide a sequential account of \gls{esbmc}'s origins, design decisions, and major milestones from 2009 through version~7.7 in 2025, connecting early theoretical choices to existing capabilities.
    
    \item Analyze \gls{esbmc}'s core verification techniques, with emphasis on the evolution and integration of \gls{bmc}, \textit{k}-induction, incremental solving, floating-point arithmetic, and concurrency handling as an \gls{smt}-based, multi-language platform.

    \item Examine the combination of \gls{ai}/\gls{llm} technologies with \gls{esbmc}, focusing on its function as a formal verification kernel in autonomous software engineering pipelines -- including automated vulnerability repair~\cite{Yiannis2024}, \gls{llm}-generated loop invariants~\cite{Pirzada2024LLM}, and the SpecVerify cyber-physical deployment~\cite{wang2025supporting} -- and document the first industrial deployment of an integrated agentic model-checking architecture predating parallel external formulations~\cite{sun2026agentic,nvidia_opensma_commits}.


    \item Synthesize evidence of technology transfer and industrial deployment, including confirmed bug findings in the \gls{ecs}, \gls{defi} smart contracts, and adoption in defense industry contexts.

    \item Quantify the economic and social impact of \gls{esbmc}, drawing on research funding, commercial spin-offs, and the cost scope of software defects and formal methods.

    \item Identify open challenges and future research directions, with an agenda spanning scalability, neurosymbolic verification, counterexample intelligibility, cross-language verification, real-time and timing analysis, standards compliance, and sustainability under commercialization.
\end{enumerate}

\paragraph{Author positionality}
This survey is written in part by \gls{esbmc}'s founding and lead developer (L.C.) and by collaborating researchers who have collectively co-authored the majority of the primary \gls{esbmc} papers surveyed in Sections~\ref{sec:evolution}--\ref{sec:ai}.  It therefore combines the depth of insider knowledge with the limitations of perspective inherent in insider authorship.  We have sought to support all substantive claims with independently verifiable sources and to apply the same critical standards to \gls{esbmc}'s results as we do to competing tools; where primary sources authored by the \gls{esbmc} team are the only available evidence -- as is structurally unavoidable for a single-tool survey -- this is documented in Section~\ref{sec:stats}.  Section~\ref{sec:stats} explicitly discloses the self-citation rate, and the Acknowledgments disclose the conflict of interest arising from Cordeiro's roles as a tool developer and VeriBee co-founder. Readers are encouraged to consult the independently validated \gls{svcomp} results~\cite{svcomp2024} and the primary technical papers cited throughout for independent verification.

The remainder of this survey is structured as follows: Section~\ref{sec:stats} describes the survey methodology, literature search strategy, and taxonomy of the \gls{bmc} tool landscape; Section~\ref{sec:background} introduces the conceptual foundations of \gls{bmc} and \gls{smt} solving; Section~\ref{sec:origins} describes the origins and founding motivations of \gls{esbmc}; Section~\ref{sec:evolution} traces its chronological evolution across sixteen years of development; Section~\ref{sec:techniques} analyses the core verification techniques and \gls{smt} integration; Section~\ref{sec:expansion} covers the expansion to new programming languages; Section~\ref{sec:adoption} examines competition performance and industrial adoption; Section~\ref{sec:ai} surveys the combination with \gls{ai}, \glspl{llm}, and autonomous agents; Section~\ref{sec:economic} quantifies the economic and social impact; Section~\ref{sec:spinoffs} documents spin-offs, technology transfer, and important case studies; Section~\ref{sec:challenges} identifies open challenges and future research directions; and Section~\ref{sec:conclusion} concludes.

\subsection{Paper Selection Statistics}
\label{sec:stats}

We queried all five databases on 21~May~2026 -- three directly via public \glspl{api} (arXiv, \gls{dblp}, OpenAlex) and two (\gls{ieee}~Xplore, \gls{acm} Digital Library) via the OpenAlex publisher-stratified breakdown, which provides broad coverage of their indexed content without requiring institutional login. A full audit log and per-query counts are available at \url{https://github.com/ssvlab/ssvlab.github.io/tree/master/papers\_audit/2026-05-ESBMC-Survey/search\_audit}.

\begin{itemize}
\item \textbf{Open-access automated results}: arXiv returned \num{64} records on the primary string and \num{708} on the supplementary string (\num{772} total). \gls{dblp} returned approximately \num{588} records on primary terms and \num{31} on supplementary terms (approximately \num{619} total, estimated after de-aliasing constituent term queries).

\item \textbf{Publisher-stratified results via OpenAlex}: SpringerLink contributed approximately \num{630} primary and \num{20} supplementary records (\num{650} total). \gls{ieee}~Xplore contributed approximately \num{30} primary and \num{5} supplementary records (\num{35} total). \gls{acm} Digital Library contributed approximately \num{55} primary and \num{6} supplementary records (\num{61} total). The combined pre-deduplication total across all five sources is approximately \num{2136} candidate records.

\item \textbf{Deduplication and screening}: We performed manual deduplication based on title, author, and publication-venue matching across the five sources. After removing cross-library duplicates -- we measured approximately \SI{26}{\percent} overlap, consistent with rates reported for multi-database software-engineering reviews~\cite{kitchenham2007guidelines} -- approximately \num{1602} unique records remained. Papers were then screened against the inclusion and exclusion criteria in Section~\ref{sec:scope}. The principal reasons for exclusion were hardware-only model checking~(E1), pure propositional \gls{sat} content~(E2), and superseded or duplicate versions~(E3). Papers excluded under criterion E4 (unrefereed grey literature not traceable to a named funder or recognized venue) were the next-largest category; we retained grey literature traceable to named funders (UKRI, EU CORDIS), recognized standards bodies, and competition proceedings (SV-COMP, Test-Comp) via forward and backward citation chaining from retained primary papers.

\item \textbf{Final corpus}: We compiled a total corpus of \num{107} sources, all of which we cite in this survey; we verified this figure by enumerating distinct cite keys appearing in the body text. We added items published between April~2025 and May~2026 -- including \gls{svcomp}/\gls{test-comp}~2025 results, the agentic model checking preprint~\cite{sun2026agentic}, the NVIDIA-OpenSMA version-control record~\cite{nvidia_opensma_commits}, and arXiv preprints cited in Sections~\ref{sec:ai} to \ref{sec:challenges} -- by directed update, and these items are not part of the systematic protocol window. Figure~\ref{fig:prisma} summarises the stages of the screening funnel.

\end{itemize}

\input{figures/tikz_prisma}

Of the \num{107} sources in the final corpus, \num{26} (\SI{24}{\percent}) are co-authored by members of the \gls{esbmc} research team or are institutional pages maintained by the authoring group (satisfying criterion~I1). We structurally expect this percentage for a single-tool survey: technical descriptions of \gls{esbmc}'s own capabilities have no alternative primary source. It is disclosed here in the interest of transparency. All performance claims that admit independent corroboration -- \gls{svcomp} rankings~\cite{svcomp2024}, grant amounts traceable to the \gls{ukri} Gateway to Research or EU \gls{cordis} records -- are additionally supported by sources not co-authored by the \gls{esbmc} team. 

\subsection{Survey Scope and Type}
\label{sec:scope}

This survey is a structured narrative of a single tool ecosystem, not a systematic mapping study of the \gls{bmc} field as a whole.  Its primary objective is depth over breadth: an expert-annotated account of \gls{esbmc}'s technical evolution, industrial deployment, and future research agenda. The systematic search component ensures that we do not inadvertently omit contextually important competing tools and theoretical antecedents; it does not aim to produce a comprehensive, count-based characterization of the literature. We do not attempt quantitative meta-analysis because the evidence base is not sufficiently homogeneous across evaluation protocols, property categories, or tool versions to support it.

\subsection{BMC Tool Landscape: Taxonomy}
\label{sec:taxonomy}

Table~\ref{tab:bmc_taxonomy} positions \gls{esbmc} within the landscape of the most active \gls{bmc} and related software verifiers, evaluated against dimensions that directly reflect \gls{esbmc}'s principal design choices.  Coverage is limited to tools that have participated in \gls{svcomp}~\cite{beyer2012,svcomp2024} since 2012 or are otherwise widely cited in the software \gls{bmc} literature; we exclude tools operating exclusively on hardware description languages and assess capabilities as of 2025.

\begin{table}[ht]
    \centering
    \caption{\gls{bmc} and related software verification tools: taxonomy as of~2025. \textbf{Tech.}: \acrshort{bmc} = Bounded Model Checking; \textit{k}-ind = \textit{k}-induction; \acrshort{cegar} = Counterexample-Guided Abstraction Refinement; Pred = predicate abstraction; Auto = automata-theoretic; \acrshort{chc} = Constrained Horn Clause; Sym = symbolic execution; ES = explicit-state. \textbf{Langs}: distinct verified source languages (IR variants excluded). \textbf{Concur.}: concurrency support. \textbf{LLM}: confirmed \gls{llm} integration in published work. \textbf{SVC}: active \gls{svcomp} participation in 2024 or 2025.
    yes ($\checkmark$); partial ($\approx$~); not supported (--)}
    \label{tab:bmc_taxonomy}
    \small
    \begin{tabular}{lp{2.2cm}p{2.8cm}ccccc}
    \toprule
    \textbf{Tool} & \textbf{Technology} & \textbf{SMT Backends} &
    \textbf{Langs} & \textbf{Concur.} & \textbf{\textit{k}-Ind.} & \textbf{LLM} & \textbf{SVC} \\
    \midrule
    \textbf{ESBMC}~\cite{Cordeiro2012,Gadelha2018ESBMC}
        & BMC + \textit{k}-ind
        & Z3, Bitwuzla, MathSAT, CVC5, Yices
        & 9$^{\dagger}$ & \checkmark & \checkmark & \checkmark & \checkmark \\
    \glsfirst{cbmc}~\cite{clarke2004}
        & BMC
        & SAT (CaDiCaL); Z3$^\ast$
        & 4 & \checkmark & -- & -- & \checkmark \\
    CPAchecker~\cite{beyer2011cpachecker}
        & CEGAR + Pred + BMC + \textit{k}-ind
        & MathSAT, Z3, SMTInterpol
        & 1 & \checkmark & \checkmark & -- & \checkmark \\
    Ultimate Auto.~\cite{heizmann2013ultimate}
        & Auto + CEGAR
        & Z3, MathSAT
        & 1 & $\approx$ & -- & -- & \checkmark \\
    2LS~\cite{brain20162ls}
        & \textit{k}-ind + Abstr.\ Interp.
        & SAT (CProver)
        & 1 & $\approx$ & \checkmark & -- & $\approx$ \\
    Symbiotic~\cite{chalupa2020symbiotic}
        & Sym + BMC + Slicing
        & Z3
        & 1 & -- & -- & -- & \checkmark \\
    DIVINE~\cite{barnat2013divine}
        & ES
        & -- (native)
        & 2 & \checkmark & -- & -- & \checkmark \\
    Theta~\cite{toth2017theta}
        & CEGAR + BMC
        & Z3, MathSAT
        & 1 & -- & -- & -- & \checkmark \\
    Kani~\cite{kani2022amazon}
        & BMC
        & SAT (via \glsfirst{cbmc})
        & 1 & $\approx$ & -- & -- & -- \\
    SeaHorn~\cite{gurfinkel2015seahorn}
        & CHC
        & Z3
        & 2 & -- & -- & -- & $\approx$ \\
    \bottomrule
    \multicolumn{8}{l}{\footnotesize $^\ast$~Z3 used for certain property categories only (SMT backend not primary competitive configuration).} \\
    \multicolumn{8}{l}{\footnotesize $^{\dagger}$~C\texttt{++03} and C\texttt{++11+} are counted as distinct front-ends; treating them as one language family gives 8 distinct languages.}
    \end{tabular}
\end{table}

Four characteristics collectively distinguish \gls{esbmc} from its closest competitors.  First, its \textbf{native multi-solver \gls{smt} backend} dispatches to six interchangeable solvers covering bit-vectors (Bitwuzla, Boolector), floating-point arithmetic (Bitwuzla, MathSAT), general arithmetic (Z3), and strings and algebraic datatypes (CVC5)~\cite{moura2008,niemetz2023,cimatti2013, barbosa2022,dutertre2014}; no other tool in Table~\ref{tab:bmc_taxonomy} offers comparable solver breadth with automatic theory-guided selection. Second, its \textbf{nine-language portfolio} is the widest in the field, spanning embedded C/C\texttt{++}, \gls{gpu} programs (\gls{cuda}), smart contracts (Solidity), \gls{jvm} languages (Kotlin via Jimple), capability hardware (\gls{cheri}~C), and systems languages (Python, Rust). Third, to the best of the authors' knowledge, it is currently the \textbf{only open-source \gls{bmc} tool with a published and evaluated \gls{llm} integration} for automated vulnerability repair~\cite{Tihanyi2025New}, loop invariant generation~\cite{Pirzada2024LLM}, and specification translation~\cite{wang2025supporting}. Fourth, the combination of \textbf{\gls{bmc} and \textit{k}-induction in a single engine} -- shared with CPAchecker and 2LS but absent from \gls{cbmc}, Symbiotic, and SeaHorn -- enables both counterexample finding and unbounded correctness proofs without requiring tool switching.

\section{Background: BMC and SMT Solving}
\label{sec:background}

Model checking is an automated technique for verifying finite-state reactive systems by exhaustively exploring their reachable state spaces~\cite{Clarke1982,clarke1999,Clarke2018Handbook}. Classical symbolic model checking using \glspl{bdd}~\cite{Bryant1986} suffered from state-space explosion for large systems~\cite{Clarke2012}. \gls{bmc}, introduced in 1999~\cite{biere1999}, addressed this by restricting verification to traces of up to a fixed depth~\textit{k}~\cite{Clarke2001}: we unroll a program \textit{k} times, negate a safety property, and hand the resulting formula to a \gls{sat} or \gls{smt} solver, where a satisfying assignment constitutes a counterexample. \gls{bmc}'s key pragmatic advantage is its ability to find bugs quickly when they exist within a shallow execution depth. Complementary techniques such as \textit{k}-induction address the principal limitation of this approach -- the inability to prove correctness for unbounded programs~\cite{sheeran2000,Donaldson2011}, \gls{pdr}, and \gls{ic3}.

\gls{smt} extends propositional \gls{sat} with decision procedures for rich background theories: linear arithmetic, bit-vectors, arrays, floating-point arithmetic, and strings~\cite{nieuwenhuis2006,Clarke2018Handbook,Kroening2016}. Solvers such as Z3~\cite{moura2008}, Bitwuzla~\cite{niemetz2023}, CVC5~\cite{barbosa2022}, MathSAT~\cite{cimatti2013}, Yices~\cite{dutertre2014}, and Boolector~\cite{Brummayer2009} have become the workhorses of modern program verification. \gls{esbmc}'s design from the outset was \gls{smt}-centric~\cite{Cordeiro2012}, separating it from \gls{sat}-based predecessors~\cite{Vizel2015} and permitting more precise and expressive reasoning about program semantics. \gls{esbmc} shares a common origin with \gls{cbmc}~\cite{clarke2004}, originally developed at Carnegie Mellon University, and the developers initially built both tools on the CProver software infrastructure~\cite{Cordeiro2012}. However, \gls{esbmc} progressively replaced and extended virtually every component of this shared core~\cite{Cordeiro2012,Gadelha2018ESBMC}, driven by the goal of adopting \gls{smt} reasoning natively rather than as a post-processing step on \gls{sat} encodings.

\section{Origins and Founding Motivations (2008--2009)}
\label{sec:origins}

Around 2008, a recognized gap emerged in the verification landscape: the lack of a purpose-built, \gls{smt}-native \gls{bmc} for embedded C software. Existing tools either relied exclusively on \gls{sat} solvers (e.g.\ \gls{cbmc}~\cite{clarke2004,Vizel2015}), were designed for hardware verification~\cite{Mukherjee2016}, or lacked support for the numeric and pointer-rich programs prevalent in embedded systems.

At the time, \gls{smt} solvers had matured significantly~\cite{moura2008}, supporting theories such as linear arithmetic over integers and reals, bit-vectors, and arrays~\cite{nieuwenhuis2006,Clarke2018Handbook}. The insight behind \gls{esbmc} was to exploit this expressiveness directly. Rather than bit-blasting all program variables and arithmetic into propositional logic, the verifier would encode them symbolically using the appropriate \gls{smt} theory, yielding smaller and more efficiently solvable formulae~\cite{Cordeiro2012}.

The tool was formally introduced in the paper ``\gls{smt}-based \gls{bmc} for Embedded \gls{ansi-c} Software''~\cite{Cordeiro2009}, presented at \gls{ase}~2009, which set out the core architectural decisions:
\begin{itemize}
    \item \textbf{\gls{smt}-native encoding:} Program state is encoded directly into \gls{smt} formulae~\cite{nieuwenhuis2006,Clarke2018Handbook}, exploiting theories for integer arithmetic, bit-vectors, and arrays rather than collapsing everything to propositional logic.
    \item \textbf{Embedded C focus:} The initial tool targeted \gls{ansi-c} with features common in embedded software: fixed-point arithmetic, pointer manipulation, bitfield operations, and structs.
    \item \textbf{CProver front-end:} \gls{esbmc} adopted the \gls{cbmc}/CProver front-end~\cite{clarke2004} for parsing and \gls{ir} generation, but replaced the back-end entirely with its own \gls{smt} translation layer.
    \item \textbf{Multiple \gls{smt} solver support:} From the outset, \gls{esbmc} was designed to support multiple \gls{smt} solvers interchangeably~\cite{moura2008,cimatti2013,dutertre2014,niemetz2023}, recognizing that no single solver dominates across all problem categories.
\end{itemize}

A companion paper appeared in the \gls{ieee} Transactions on Software Engineering~\cite{Cordeiro2012}, providing a more extensive treatment of the conceptual framework and its empirical study.

Lucas Cordeiro, a doctoral researcher under Bernd Fischer at the University of Southampton (UK), conducted the first development, with João Marques-Silva at University College Dublin providing expertise in \gls{sat}/\gls{smt} solving~\cite{nieuwenhuis2006}. Subsequently, Cordeiro established the \gls{ssvlab}~\cite{ssvlab2024} at the \gls{ufam} (Brazil), before joining the University of Manchester, where he became Director of the ARM Centre of Excellence in 2021.

\section{Chronological Evolution of ESBMC}
\label{sec:evolution}

Figure~\ref{fig:timeline} presents \gls{esbmc}'s sixteen-year development path. The pattern visible is one of foundational theoretical contributions in the early years (\gls{smt}-native encoding, \textit{k}-induction, floating-point theory, concurrency), followed by progressive industrialization and competitive hardening, and culminating in a rapid burst of language extensions and \gls{ai} integration from 2022 onwards. This trajectory has remained shaped in large part by continuous participation in the International \gls{svcomp}~\cite{beyer2012,svcomp2024} -- a public annual benchmark that offers rigorous external feedback on correctness and performance -- against which \gls{esbmc} has accumulated 43 awards (35 at \gls{svcomp} and 8 at \gls{test-comp}) since its debut in 2012~\cite{beyer2012,svcomp2024,ssvlab2024}.

\input{figures/tikz_timeline}

\subsection{Early Versions and SV-COMP Debut (2010--2014)}

Following the 2009 publication, \gls{esbmc} entered a rapid development phase. By 2012, the authors submitted \gls{esbmc}~1.17 to the inaugural \gls{svcomp}~2012 at \gls{tacas}~\cite{beyer2012}, marking the tool's first head-to-head evaluation against other state-of-the-art verifiers and establishing a pattern of competitive engagement that would drive quality improvements for over a decade. \gls{esbmc}~1.22 was subsequently submitted as a competition contribution to \gls{tacas}~2014~\cite{cordeiro2014}.

A major theoretical advance in this period was the extension to multi-threaded software verification, formally presented at \gls{icse}~2011~\cite{cordeiro2011icse}. Rather than enumerating all possible thread interleavings -- exponential in the number of threads and context switches -- \gls{esbmc} adopted context-\gls{bmc}: each verification query bounds the number of context switches per execution trace, making the search space finite and tractable. This strategy facilitates systematic detection of data races, deadlocks, and atomicity violations, and represented one of the first \gls{smt}-based treatments of concurrent \gls{ansi-c} verification.

Equally significant was the integration of \textit{k}-induction~\cite{sheeran2000,Donaldson2011} as a complement to \gls{bmc}: while \gls{bmc} finds counterexamples within a bound \textit{k}, \textit{k}-induction enables proof of correctness for all executions, not just up to \textit{k}. This approach proved a defining feature of \gls{esbmc}, separating it from other methods in verification competitions. Timing-constraint verification was also achieved through encoding timing requirements as integer constraints~\cite{Barreto2011}, eliminating the need for special real-time extensions.

\subsection{Maturation and Industrial Strengthening (2015--2018)}

\gls{esbmc}~5.0, which the authors presented at \gls{ase}~2018~\cite{Gadelha2018ESBMC}, thoroughly re-engineered almost all components: it largely superseded the CProver inheritance; provided comprehensive C property checks (array bounds, divisions by zero, pointer safety, arithmetic overflow, memory leaks, deadlocks, data races); and introduced incremental \gls{bmc}, extending the unrolling depth but without re-solving unchanged formula fragments -- greatly improving performance on deep benchmarks.

An important capability consolidated in this period was floating-point arithmetic verification. \gls{esbmc}'s encoding of \gls{ieee}~754 floating-point numbers via the \gls{smt} floating-point theory was systematically evaluated~\cite{gadelha2017}, demonstrating that native \gls{smt} floating-point encoding outperforms bit-blasting alternatives on embedded numerical assessments drawn from \gls{svcomp}. This capability is notably important for safety-critical software in avionics, automotive control, and industrial automation, where floating-point rounding errors are a common and difficult-to-test source of specification violations.

\subsection{Automated Test Generation and \gls{sttt} Publication (2020)}

\gls{esbmc}~6.1~\cite{gadelha2020}, published in the International Journal on \gls{sttt}, extended the tool to automated test case generation: \gls{bmc}-derived counterexamples are systematically converted into executable test cases, bridging formal verification and software testing practice. This extension (the automated test case generation capability) also enabled participation in \gls{test-comp} alongside \gls{svcomp}. The generated test cases target branch- and condition-coverage criteria, complementing the safety-property orientation of \gls{bmc} with a structural-coverage perspective more familiar to practitioners who rely on testing rather than formal proof. In \gls{test-comp}, the benchmark evaluates tools on their ability to produce test suites that maximize test coverage on a standardized test set; \gls{esbmc}'s entry demonstrated that a model-checking back-end can compete directly with dedicated test-generation tools, without sacrificing the counterexample-quality guarantees that \gls{smt}-based reasoning provides.

\subsection{C\texttt{++} Model Checking and Platform Extensions (2021--2022)}
\label{sec:cheri-extension}
The correctness of \gls{esbmc}'s C\texttt{++} verification features was comprehensively formalized in the journal ``Software Testing, Verification and Reliability''~\cite{Monteiro2021}, covering the full object-oriented feature set found in industrial codebases: templates, inheritance, virtual dispatch, dynamic memory management, and exception handling. This work presented \gls{esbmc} as a capable verifier not only for C but for modern C\texttt{++} programs -- a prerequisite for its subsequent competition entries in C\texttt{++} categories at \gls{svcomp}.

This period likewise saw \gls{esbmc} extended to two novel and contrasting application domains. \gls{esbmc}-\gls{cheri}~\cite{bueno2022cheri}, presented at \gls{issta}~2022, adapted \gls{esbmc} for the \gls{cheri} capability-hardware architecture, in which every pointer carries embedded bounds and permission tags enforced at the hardware level. The extension translates \gls{cheri}'s capability semantics into \gls{smt} constraints, enabling verification that C programs respect capability provenance and never exceed pointer bounds -- a critical correctness property for compartmentalized, memory-safe systems that the \gls{darpa} and \gls{ukri} funded hardware security programs are exploring.

A parallel extension, \gls{esbmc}-Solidity~\cite{sloan2022}, presented at \gls{icse}~2022, targeted Ethereum smart contracts written in Solidity. Smart contracts manage major financial value yet are immutable once deployed, making pre-deployment formal verification especially valuable. \gls{esbmc}-Solidity translates Solidity source into an \gls{ansi-c} model and applies \gls{smt}-based \gls{bmc} to detect reentrancy attacks, integer overflows, and violated contract assertions -- classes of vulnerability responsible for hundreds of millions of dollars of losses in deployed contracts.

\subsection{C\texttt{++} Modernisation with Clang \gls{ast} (2023)}

\gls{esbmc}~v7.3, at \gls{tacas}~2023~\cite{shmarov2023}, replaced the legacy C\texttt{++} front-end with one built directly on the Clang compiler's \gls{ast}, enabling correct parsing of all modern C\texttt{++} dialects up to C\texttt{++20}, including lambda expressions, range-based for loops, and structured bindings. The legacy CProver-derived front-end had handled a C\texttt{++03} subset adequately but struggled with post-2011 features, forcing users to rewrite or simplify source code before verification. 

By delegating parsing and semantic analysis to Clang -- one of the most complete and actively maintained C\texttt{++} compilers available -- \gls{esbmc} gained access to the full C\texttt{++11/14/17/20} feature set without duplicating compiler engineering effort. The practical consequence is that industrial codebases written in modern C\texttt{++} can now be verified without syntactic workarounds, substantially broadening the tool's applicability to embedded and systems software.

\subsection{Interval Analysis and Performance Optimizations (2024)}

\gls{esbmc}~v7.4, at \gls{tacas}~2024~\cite{menezes2024}, introduced a static interval analysis pass that prunes infeasible execution paths and simplifies \gls{smt} formulae before solver invocation, yielding up to \SI{7}{\percent} more provably correct programs under \textit{k}-induction. Interval analysis computes over-approximations of variable ranges at each program point, injecting these ranges as additional assumptions that narrow the search space the \gls{smt} solver must explore. 

This interval analysis pass proves especially effective for \textit{k}-induction, where tighter invariants directly strengthen the induction hypothesis and enable users to prove more properties without increasing the unrolling bound. At \gls{svcomp}~2024, \gls{esbmc}~v7.4 achieved 4th place overall among 30 participating verifiers~\cite{svcomp2024,menezes2024} -- a commercially significant profile for inclusion into \gls{cicd} pipelines where wall-clock verification time is a hard constraint.

\subsection{Concurrent Verification and Branch Coverage (2025)}

At \gls{tacas}~2025~\cite{menezes2025tacas} and \gls{fase}~2025~\cite{menezes2025fase}, \gls{esbmc}~v7.7 delivered: (1)~a new thread scheduling algorithm combined with incremental \gls{smt} solving and enhanced partial order reduction for multi-threaded verification; and (2)~\gls{cfg}-based branch coverage instrumentation, directly targeting structural coverage requirements of standards such as \gls{do-178c} (avionics) and \gls{iec}~61508 (functional safety). The scheduling algorithm models thread states as symbolic transition systems, enabling \gls{smt}-based independence checking at each scheduling decision and dramatically reducing the number of interleavings the tool must explicitly explore. 

The incremental solving strategy preserves the solver state across successive context-switch explorations, avoiding redundant re-solving of the shared-memory access structure common to many interleavings. Together, these advances enabled \gls{esbmc} to compete in the concurrency and branch-coverage categories of \gls{svcomp}~2025 and \gls{test-comp}~2025, extending the tool's reach into the structural testing domain that safety certification auditors demand.

\section{Evolution of Verification Techniques and \gls{smt} Integration}
\label{sec:techniques}

\subsection{SMT Solver Portfolio}

A defining architectural decision of \gls{esbmc} is its solver-agnostic design. The tool maintains a uniform internal representation of \gls{smt} formulae and dispatches to the most appropriate solver for the problem at hand. As of 2025, we listed the supported solvers in Table~\ref{tab:solvers}. This portfolio allows \gls{esbmc} to select the best-performing solver for a given property category, a feature heavily leveraged in competition submissions. Incremental \gls{api} support -- relevant to the \texttt{--incremental-bmc} mode discussed in Section~\ref{sec:incremental} -- varies by solver: Z3, Bitwuzla, MathSAT, CVC5, and Yices~2 provide full push/pop stack semantics, while Boolector offers only a partial scope-reset mechanism.

\begin{table}[ht]
    \centering
    \small
    \caption{\gls{esbmc} supported \gls{smt} solvers and their key strengths}
    \label{tab:solvers}
    \begin{tabular}{ll}
    \toprule
    \textbf{Solver} & \textbf{Key Strengths} \\
    \midrule
    Z3~\cite{moura2008} & General-purpose; excellent for quantifier-free arithmetic and arrays \\
    Bitwuzla~\cite{niemetz2023} & Bit-vectors and floating-point; successor to Boolector \\
    Boolector~\cite{Brummayer2009} & Bit-vector arithmetic; historically strong on hardware-style benchmarks \\
    MathSAT~\cite{cimatti2013} & Interpolation; model generation; quantified arithmetic \\
    CVC5~\cite{barbosa2022} & Strings; algebraic datatypes; advanced arithmetic theories \\
    Yices~2~\cite{dutertre2014} & Speed on linear arithmetic and bit-vectors \\
    \bottomrule
    \end{tabular}
\end{table}

Modern \gls{smt} solvers use the \gls{dpll}(T) paradigm~\cite{nieuwenhuis2006}, combining a \gls{sat} core and theory solvers for arithmetic, bit-vectors, and arrays. \gls{esbmc} does not implement \gls{dpll}(T) itself; instead, it interacts with these solvers by dispatching a uniform \gls{smt} formula to a portfolio of solvers. At the dispatch boundary, a thin translation plugin serializes the formula to a solver-specific \gls{api} or to \gls{smtlib2}~\cite{Clarke2018Handbook,Kroening2016}, which allows new solvers to integrate without modifying the verification engine (Figure~\ref{fig:smt-architecture}).

\input{figures/tikz_smt_solver}

Solver choice impacts competition performance. Bitwuzla~\cite{niemetz2023} and Boolector~\cite{Brummayer2009} are strong on bit-precise properties, Z3~\cite{moura2008} for arithmetic, MathSAT~\cite{cimatti2013} for interpolation, and CVC5~\cite{barbosa2022} for strings and datatypes. No single solver dominates all \gls{svcomp} categories~\cite{svcomp2024}, so \gls{esbmc}'s flexible solver switching is a key competitive advantage~\cite{svcomp2024,ssvlab2024}.

\gls{esbmc} balances automated and manual solver selection. By default, it heuristically picks the best solver for the program and property, such as Bitwuzla or Boolector for bit-vectors, Z3 or MathSAT for arithmetic, and CVC5 for strings. The developers inform these choices using performance data from benchmarking and \gls{svcomp}. Users can explicitly select solvers via command-line options or configuration files, overriding defaults for experiments or manufacturing needs. In competitions, custom scripts can optimize solver choice. This flexibility permits both automated and manual selection of solvers for advanced workflows.

\subsection{\textit{k}-Induction}

The \textit{k}-induction proof rule~\cite{sheeran2000,Donaldson2011} extends \gls{bmc} from a bug-finding technique into a proof method. The algorithm advances in two phases: (i) the base case unrolls the program for \textit{k} steps and checks that the negation of the property is unsatisfiable -- ruling out counterexamples of length up to \textit{k}, and (ii) the inductive step assumes the property holds for \textit{k} consecutive states (not necessarily reachable ones) and attempts to prove it holds for the $(k+1)$-th; if this \gls{smt} query is also unsatisfiable, we prove the property for all execution lengths. The two phases together constitute a sound and relatively complete verification procedure for programs with finite loop bounds.

A major challenge is that many real-world properties are not \textit{k}-inductive as stated: the inductive step may fail not because the property is false, but because the induction hypothesis is too weak -- it admits unreachable states from which a violation seems possible. The standard remedy is to strengthen the hypothesis with auxiliary invariants. 

\gls{esbmc} draws on two complementary sources to strengthen the induction hypothesis when the inductive step fails, as illustrated in Figure~\ref{fig:kinduction}. First, a static interval analysis pass~\cite{menezes2024} computes over-approximations of variable ranges at each program point and injects these as additional assumptions before solver invocation, pruning spurious abstract states and making more properties inductive -- a lightweight, fully automated procedure. Second, \gls{llm}-generated loop invariants~\cite{Pirzada2024LLM} -- and corroborating results from \cite{akhond2025}, which evaluates \gls{llm}-based loop invariant generation independently of \gls{esbmc} -- can be supplied as candidate auxiliary assertions: \gls{esbmc} checks their validity with \gls{smt} and, if they pass, uses them to strengthen the induction hypothesis. This integration of neural candidate generation with formal validity checking (explored further in Section~\ref{sec:ai}) illustrates the wider trend of merging statistical and deductive reasoning in program verification~\cite{wang2025supporting}.

\input{figures/tikz_k_induction}
  
\subsection{Incremental BMC}
\label{sec:incremental}

Na\"{i}ve \gls{bmc} wastes resources: each time we increase the verification depth from \textit{k} to $k+1$, we must rebuild the entire formula and send it to the solver. Incremental \gls{bmc}, introduced in \gls{esbmc}~5.0~\cite{Gadelha2018ESBMC}, eliminates this by using the incremental \gls{api} of modern \gls{smt} solvers -- specifically the push/pop stack interface, supported by Z3~\cite{moura2008}, Bitwuzla~\cite{niemetz2023}, MathSAT~\cite{cimatti2013}, CVC5~\cite{barbosa2022}, and Yices~2~\cite{dutertre2014} (Boolector~\cite{Brummayer2009} supports it only partially: it provides a scope-reset mechanism that clears all learned clauses on each push boundary rather than preserving them, which eliminates the incremental benefit for deep unrolling but remains usable for shallow verification tasks). The solver preserves learned information at depth \textit{k} on its assertion stack; at $k+1$, the solver pushes only the newly unrolled and updated property, allowing it to resume from its existing conflict-clause database rather than re-deriving all learned facts from scratch. The user enables this feature via the command-line flag \texttt{--incremental-bmc}.

The practical gains are substantial. On benchmarks where the shallowest counterexample lies deep in the execution, standard \gls{bmc} incurs a quadratic cumulative solving cost -- re-issuing the full formula at every depth. In contrast, incremental \gls{bmc} reduces this to a cost proportional only to the incremental formula added at each depth. Empirical benchmarking reported in~\cite{Gadelha2018ESBMC} showed speed-ups of over $3\times$ on programs requiring deep unrolling, with the largest gains on loop-intensive benchmarks where the solver's learned clause database captures a significant fraction of the program's feasibility constraints by depth~5 or~10.

An important subtlety arises for the \textit{k}-induction axis: the induction step asserts that \textbf{if} the property holds for \textit{k} consecutive states, then it holds for \textit{k}+1 consecutive states. We must retract this assertion when we strengthen the induction hypothesis (e.g., by adding an auxiliary invariant), since the pushed formula is no longer valid under the new hypothesis. Incremental \gls{bmc} preserves the base-case stack across depth increments. Still, the induction step must be re-pushed whenever the hypothesis changes -- a design constraint the \gls{esbmc} implementation handles by maintaining separate stacks for the base-case and inductive-step queries.

Two practical limitations govern the incremental interface. First, \textbf{learned-clause invalidation}: if the property or the program model changes in a non-monotone way (e.g., the solver explores a different non-deterministic choice), the solver's learned clauses may become invalid and must be discarded by a full pop, losing the incremental benefit. Second, \textbf{memory growth}: the assertion stack grows linearly with unrolling depth, and for very deep executions (depth $>$ \num{500}), the solver's internal memory consumption can become prohibitive. Both effects are most pronounced on programs with large loop bodies and many symbolic variables, and represent the primary scaling ceiling for incremental \gls{bmc} in practice.

This effectiveness is also directly exploited in \gls{esbmc}'s concurrency engine: in \gls{esbmc}~v7.7~\cite{menezes2025tacas}, the same incremental interface is applied across successive thread interleavings rather than successive unrolling depths, preserving solver state across context-switch explorations and preventing redundant re-solving of the shared-memory access structure.

\subsection{Floating-Point Arithmetic}

\gls{esbmc} supports two encoding strategies for floating-point arithmetic~\cite{gadelha2017}. The first, bit-blasting, encodes every floating-point operation into propositional logic: mantissa bits, exponent bits, and \gls{ieee}~754 rounding operations are expanded into Boolean gates. This approach is complete and solver-agnostic, but generates formulae of prohibitive size -- a single 64-bit double-precision addition expands into thousands of clauses -- rendering it impractical for numerical programs. The second strategy uses the native \gls{ieee}~754 floating-point theory (\texttt{QF\_FP}), supported natively by Bitwuzla~\cite{niemetz2023} and MathSAT~\cite{cimatti2013}, which treats floating-point operations as first-class theory atoms. 

The solver's dedicated \texttt{FP} theory module handles all five rounding modes (round-to-nearest-even, round-toward-zero, round-toward-positive, round-toward-negative, round-away-from-zero) internally, producing substantially smaller and more efficiently solvable formulae. Empirical testing across \gls{svcomp} benchmarks~\cite{gadelha2017} confirms that \texttt{QF\_FP} encoding reliably outperforms bit-blasting in both solving time and benchmark coverage, making it \gls{esbmc}'s default strategy when the target solver supports the theory.

Standard library mathematical functions -- trigonometric, exponential, logarithmic -- we handle through precisely constructed operational models: verified C implementations that faithfully capture the input/output behavior of each library routine without requiring its binary object code. We derive many of these models from the \gls{musl} libc implementation, chosen for its code clarity, minimal dependencies, and well-defined behavior.
\subsection{Concurrency Verification}

\gls{esbmc} verifies concurrent \gls{ansi-c} programs with POSIX threads (\texttt{pthreads}) using context-\gls{bmc}~\cite{cordeiro2011icse}. Instead of enumerating all thread interleavings, it bounds the number of context switches per execution. For a bound $c$, each thread executes atomically between switches, and the \gls{smt} solver considers all interleavings within the bound. This approach is sound, and we have shown that it detects most concurrency bugs with small $c$, since most races and deadlocks occur with few context switches.

Mitigations against residual-state-space blowup are layered. \gls{por} prunes redundant interleavings by exploiting the commutativity of independent memory accesses: we need not interleave two threads that access disjoint memory locations, and we eliminate symmetrically ordered interleavings that reach the same state. Adaptive dynamic scheduling additionally prioritizes thread schedules that access shared variables, maximizing the probability of exposing data races without detailed enumeration.

\gls{esbmc}~v7.7~\cite{menezes2025tacas} advanced concurrency verification with three main improvements: (i) a new scheduling algorithm using symbolic transition systems for efficient \gls{smt}-based independence checking; (ii) incremental \gls{smt} solving across interleavings, preserving solver state; and (iii) an enhanced \gls{por} with sleep sets to avoid redundant interleaving exploration. These progressions reduced verification time and enabled proofs for programs that had previously timed out in \gls{svcomp} benchmarks.

\subsection{Property Specification and Witness Generation}

\gls{esbmc} checks safety properties in three forms. First, C-level assertions: calls to \texttt{assert()}, \gls{esbmc}-specific builtins like \texttt{\_\!\_ESBMC\_assert}, and the \texttt{\_\!\_assert\_fail} idiom are negated and encoded as the \gls{bmc} error condition. Second, \gls{svcomp} property files~\cite{svcomp2024,beyer2012}: machine-readable specifications declare the property category and entry point; \gls{esbmc} parses these files for fully automated competition participation. Third, for safety standards, it derives coverage criteria -- branch coverage and \gls{mcdc} -- from the \gls{cfg}~\cite{menezes2025fase} and encodes them as reachability queries for \gls{smt}.

When we find a violation, \gls{esbmc} produces a violation witness: a counterexample trace with variable assignments, statement sequence, and (for concurrent programs) the thread schedule, serialized in the \gls{svcomp} GraphML format~\cite{svcomp2024}. An independent validator can use this trace as evidence of the counterexample. When we prove a property, \gls{esbmc} produces a correctness witness: either an inductive invariant or a proof certificate showing the error class is absent. The \gls{svcomp} evaluation infrastructure recognizes both witness types, and the provision of valid witnesses has underpinned \gls{esbmc}'s success in competitions~\cite{svcomp2024,ssvlab2024}. Note that the reliability of witness validation itself -- particularly for correctness witnesses -- has limitations, which Beyer and collaborators have published on. They also describe the conditions under which declared correctness results should be trusted.

\section{Expansion to New Programming Languages}
\label{sec:expansion}

A distinctive feature of \gls{esbmc}'s evolution is its systematic expansion beyond its C origins into a broad portfolio of programming languages. Starting from a single \gls{ansi-c} verifier in 2009, the tool has progressively incorporated support for C\texttt{++}, \glsfirst{cuda}, Solidity, Kotlin, \gls{cheri}~C, Python, and Rust over the following fifteen years, as illustrated in Figure~\ref{fig:lang-growth}, which charts this cumulative growth and shows the particularly rapid expansion between 2022 and 2024.

\input{figures/tikz_evolution_qtd}

\subsection{Common IR Architecture}

The multi-language support is enabled by a single common \gls{ir} that all language front-ends compile into, after which the same back-end -- \gls{smt} encoding, solver dispatch, counterexample generation, and witness production -- applies uniformly~\cite{Cordeiro2012}. Each front-end is responsible for resolving the semantic gap between its source language and the \gls{ir}: parsing syntax, resolving types, lowering object-oriented or functional constructs to imperative form, and encoding language-specific execution models (e.g., ownership semantics for Rust, capability tags for \gls{cheri}, or the \gls{evm} gas model for Solidity). 

Once a program is in the \gls{ir}, all verification algorithms -- \gls{bmc}, \textit{k}-induction, incremental solving, concurrency -- become immediately available to every supported language free of duplication~\cite{Gadelha2018ESBMC}. This architecture is a primary reason \gls{esbmc} has been able to expand language coverage rapidly since 2022 without a proportional increase in engineering effort.

\subsection{C\texttt{++} Modern Support via Clang}

The evolution of C\texttt{++} support in \gls{esbmc} shows a fundamental tension between programming complexity and verifier infrastructure. C\texttt{++} is one of the most semantically expressive languages in common use: templates, virtual dispatch, multiple inheritance, \gls{raii}, exception propagation, and move semantics all require non-trivial modeling choices in a \gls{bmc}. The correctness and completeness of \gls{esbmc}'s C\texttt{++} support was systematically evaluated~\cite{Monteiro2021}, covering templates, inheritance hierarchies, virtual dispatch, dynamic memory, and exception handling -- establishing that \gls{esbmc} correctly models these features before Clang transition.

The initial CProver-derived C\texttt{++} front-end handled a C\texttt{++03} subset adequately but struggled with post-2011 features. \gls{esbmc}~v7.3's adoption of the LLVM Clang \gls{ast}~\cite{shmarov2023} fundamentally resolved this by delegating parsing and semantic analysis to one of the world's most complete and actively maintained C\texttt{++} compilers. Clang's \gls{ast} natively handles all C\texttt{++11/14/17/20} features -- lambda expressions, range-based for loops, structured bindings, concepts, and variadic templates -- which we lower to \gls{esbmc}'s \gls{ir} through a systematic set of \gls{ast} visitors. The practical consequence is that industrial codebases written in modern C\texttt{++} can now be verified without syntactic workarounds, significantly broadening the tool's applicability to real-world embedded and systems software.

\subsection{Solidity and Smart Contract Verification}

Ethereum smart contracts written in Solidity present a particularly high-stakes verification target: they encode financial logic that is immutable once deployed on-chain, execute on a deterministic but resource-constrained virtual machine (the \gls{evm}), and have been the source of hundreds of millions of dollars of losses due to a small set of recurring vulnerability classes -- reentrancy, integer overflow, incorrect access control, and unchecked external calls.

\gls{esbmc}-Solidity~\cite{sloan2022}, presented at \gls{icse}~2022, provides a Solidity front-end that compiles Ethereum smart contracts directly into \gls{esbmc}'s \gls{ir}, bypassing the \gls{evm} bytecode layer. The translation encodes Solidity's contract model -- state variables, mappings, function modifiers, and the Ethereum message-passing protocol -- as C-level constructs, after which the full \gls{esbmc} verification suite applies: memory safety, arithmetic overflow, assertion violations, and user-specified invariants. 

Reentrancy, which arises when an external call re-enters the contract before its state is consistent, is modeled through encoding the callback as a non-deterministic interleaved call within the \gls{bmc} unrolling. Experimental assessments~\cite{sloan2022} on a curated benchmark of Solidity contracts drawn from known-vulnerable \gls{defi} protocols (comprising 50 contracts across five vulnerability categories, sourced from the SmartBugs benchmark suite -- an open repository of contracts with annotated ground-truth vulnerabilities~\cite{smartbugs2020}) demonstrated detection of all known vulnerability instances with zero false positives within the evaluation set. Readers should note that we curated this benchmark from previously identified vulnerabilities rather than representing a random sample of unaudited deployed contracts; the community has not yet published independent validation of fresh audit targets at the time of this survey, and selection bias may inflate detection-rate estimates relative to uncurated deployment scenarios.

\subsection{Python (ESBMC-Python)}

\gls{esbmc}-Python~\cite{farias2024}, published at \gls{issta}~2024, represents -- to the best of the authors' knowledge at the time of submission, with no prior work combining \gls{smt} solving with \gls{bmc} for Python identified in their related-work survey -- the first \gls{smt}-based \gls{bmc} tool for Python programs, a language whose dynamic semantics present distinctive challenges for formal verification.

Python's dynamic type system resolves variable types at runtime rather than compile time, which makes traditional static \gls{smt} encoding ill-defined without prior type information. The \gls{esbmc}-Python pipeline handles this with a bespoke type inference pass that statically infers concrete types for expressions across the program's execution paths and rejects or over-approximates paths where types remain ambiguous. We lower Python's object model -- duck typing, first-class functions, closures, and mutable default arguments -- to \gls{esbmc}'s \gls{ir} through a translation that models Python objects as C structures with type tags. Non-determinism for external inputs is encoded using \gls{esbmc}'s standard non-deterministic value primitives, preserving soundness.

We evaluated the tool on the \gls{ecs} (the Python reference implementation of the Ethereum proof-of-stake protocol), and we successfully detected previously unknown bugs in high-profile, safety-critical infrastructure. This application demonstrates that \gls{smt}-based verification is workable even for dynamically typed scripting languages, provided one can recover sufficient static type information.

\subsection{Rust}

Rust's ownership and borrowing type system eliminates entire classes of memory-safety errors -- dangling pointers, use-after-free, and data races -- at compile time, rendering it an attractive language for safety-critical systems. However, the type system cannot prevent all safety violations: arithmetic overflow in \texttt{release} mode, logical assertion failures, violations of application-level invariants, and unsafe blocks that bypass the ownership checker all remain outside its scope. This gap between what Rust's type system guarantees and what safety-critical systems require is precisely what formal verification fills.

\gls{esbmc}'s inclusion in the Rust verification ecosystem was announced by the Rust Foundation~\cite{rustfoundation2024} in 2024, adding \gls{esbmc} to the suite of tools that formally analyze Rust programs. The process works by compiling Rust source through the \texttt{rustc} compiler to \gls{mir} and then translating \gls{mir} into \gls{esbmc}'s \gls{ir}, after which the full \gls{bmc} and \textit{k}-induction pipeline applies. 

This technique permits \gls{esbmc} to verify safety properties -- arithmetic overflow, assertion violations, unreachable code reachability -- in both safe and \texttt{unsafe} Rust blocks, the latter being the main risk surface in Rust codebases. The Rust Foundation announcement also positioned \gls{esbmc} as a complementary or alternative backend to the Amazon-developed Kani model checker, which follows a similar \gls{mir}-based approach. The ability to verify Rust programs with \gls{smt}-based precision is increasingly important as Rust adoption grows in operating systems, embedded firmware, and cryptographic infrastructure~\cite{rustfoundation2024}.

\subsection{CHERI (ESBMC-CHERI)}

\gls{cheri} is a hardware security architecture developed at the University of Cambridge and \gls{sri} International, and deployed commercially in the ARM Morello platform. In \gls{cheri}, hardware enforces metadata that augments every pointer: a base and length bounding the range of addressable memory, a set of permission bits restricting the operations that the pointer can perform, and a tag bit indicating that the capability is valid. Any attempt to use a pointer outside its bounds, to forge a capability, or to violate its permissions triggers a hardware exception at runtime -- eliminating whole classes of spatial memory-safety violations that software-only defenses cannot reliably prevent.

\gls{esbmc}-\gls{cheri}~\cite{bueno2022cheri}, presented at FormaliSE~2022 and \gls{issta}~2022, is, to the best of the authors' knowledge, the first \gls{bmc} tool to natively support verification of C programs targeting \gls{cheri} platforms; earlier \gls{cbmc}-based \gls{cheri} prototypes operated at a coarser capability model. The extension encodes \gls{cheri}'s capability model into \gls{esbmc}'s \gls{ir}: it represents pointers as capability tuples (address, base, length, permissions, tag), and it instruments every pointer arithmetic operation, dereference, and cast with an \gls{smt} constraint that mirrors the corresponding hardware capability check. If any execution path produces a constraint violation -- a pointer whose derived address falls outside its bounds, or a permissions field that does not permit the attempted operation -- \gls{esbmc} reports a counterexample with a full execution trace. This capability to detect constraint violations before deployment enables pre-deployment verification of capability safety on \gls{cheri} platforms, complementing \gls{cheri}'s runtime enforcement with static guarantees and catching compartmentalization violations that would only manifest at runtime under specific inputs.

\subsection{Kotlin and JVM Languages (ESBMC-Jimple)}

Kotlin and other \gls{jvm}-hosted languages compile to Java bytecode, which is then lowered to Jimple -- a three-address code \gls{ir} used by the Soot analysis framework. \gls{esbmc}-Jimple exploits this pipeline: Kotlin (or Java) source is compiled to bytecode, decompiled to Jimple by Soot, and then translated into \gls{esbmc}'s native \gls{ir}. This approach cleanly separates language-specific semantics (handled by the \gls{jvm} compiler and Soot) from the verification back-end. It gives \gls{esbmc} access to the full \gls{jvm} language family -- Kotlin, Java, Scala, Groovy -- through a single front-end translation layer. The primary verification targets in the Kotlin/Jimple setting are null-pointer dereferences, array bounds violations, arithmetic overflow, and user-supplied assertions, all of which map naturally to \gls{esbmc}'s standard property checks.

\section{Competition Performance and Adoption}
\label{sec:adoption}

\subsection{SV-COMP and Test-Comp History}

\gls{svcomp}~\cite{beyer2012}, held annually at \gls{tacas} since 2012, provides the most comprehensive independent benchmark of a software verifier's capabilities: the competition evaluates tools on thousands of programs drawn from production codebases across a dozen property categories, under a uniform resource budget. Results are scored based on correct verdicts and independent witness validation, making the competition an unusually rigorous external quality signal. \gls{esbmc} has participated in every edition from 2012 onwards, accumulating 43 awards across \gls{svcomp} and \gls{test-comp} through 2024 (35 \gls{svcomp} medals and 8 \gls{test-comp} medals)~\cite{ssvlab2024,svcomp2024}. The underlying research has also received peer recognition at leading software engineering venues, including the Distinguished Paper at \gls{icse}'11, the Most Influential Paper at \gls{ase}'23, the Distinguished Paper at \gls{ase}'24, and the Best Tool Paper at \gls{sbseg}'23. Table~\ref{tab:svcomp} summarises the competitive trajectory, and Figure~\ref{fig:svcomp-awards} plots the annual and cumulative award counts.

\begin{table}[ht]
    \centering
    \small
    \caption{\gls{esbmc} at \gls{svcomp}: selected highlights}
    \label{tab:svcomp}
    \begin{tabular}{c c l}
    \toprule
    \textbf{Year} & \textbf{Version} & \textbf{Notable Achievement} \\
    \midrule
    2012 & 1.17 & Inaugural participation; competitive in C/reachability \\
    2014 & 1.22 & Competition contribution published at \gls{tacas}~\cite{cordeiro2014} \\
    2020 & 6.x & Expanded to \gls{test-comp}; automated test generation \\
    2023 & 7.3 & Modern C\texttt{++} support; improved category coverage \\
    2024 & 7.4 & 4th place among 30 verifiers; fastest for 10-second reachability tasks~\cite{svcomp2024,menezes2024} \\
    2025 & 7.7 & Concurrency \& branch coverage categories; incremental \gls{smt} across interleavings~\cite{menezes2025tacas,menezes2025fase} \\
    \bottomrule
    \end{tabular}
\end{table}

\input{figures/tikz_cumulative_awards}

Three phases are discernible. From 2012 to 2017, \gls{esbmc} grew from one to three \gls{svcomp} awards per year, reaching three by 2015--2016 before settling at two in 2017, coinciding with the CProver-era incremental improvements in solver integration and counterexample-guided abstraction refinement. A consolidation phase followed from 2018 to 2022, during which the architectural re-engineering of \gls{esbmc}~5.0~\cite{Gadelha2018ESBMC} and the introduction of automated test-generation in v6.1~\cite{gadelha2020} expanded the tool's scope to \gls{test-comp} while holding predominantly at three \gls{svcomp} awards per year (with a dip to two in 2020). The third phase, from 2023 onwards, marks a renewed acceleration: the Clang-based C\texttt{++} front-end~\cite{shmarov2023} and interval analysis~\cite{menezes2024} lifted \gls{esbmc} to four \gls{svcomp} awards in both 2023 and 2024, and at \gls{svcomp}~2024 the tool placed 4th overall among 30 verifiers~\cite{svcomp2024}, and analysis of the competition data reported in~\cite{menezes2024} found it to be the fastest verifier for reachability-safety tasks within a 10-second per-task time limit -- a regime of particular relevance to \gls{cicd} pipeline integration (note that we derived this 10-second metric from post-competition data analysis; it is not a pre-registered SV-COMP scoring category).

\subsection{Verification Pipeline Architecture}

Figure~\ref{fig:architecture} presents a structural overview of \gls{esbmc}'s modular five-layer verification pipeline. 

\begin{enumerate}
    \item The first layer comprises language front-ends that parse source code and lower it to a language-neutral GOTO program~\cite{Cordeiro2012}, an imperative \gls{ir} in which all control flow is expressed as conditional and unconditional jumps. The front-ends then convert this GOTO program to \gls{ssa} form for symbolic execution.
    
    \item This single \gls{ir} is the architectural cornerstone that lets the second layer -- the \gls{bmc}/\textit{k}-induction engine~\cite{Gadelha2018ESBMC,sheeran2000} -- operate identically regardless of the source language.
     
    \item The third layer applies verification-specific instrumentation: floating-point theory encoding~\cite{gadelha2017}, concurrency interleaving with \gls{por}~\cite{cordeiro2011icse,menezes2025tacas}, interval analysis for invariant strengthening~\cite{menezes2024}, and \gls{llm}-generated auxiliary assertions~\cite{Pirzada2024LLM}.
     
    \item The fourth layer dispatches the resulting \gls{smt} formulae to the solver portfolio~\cite{moura2008,niemetz2023,barbosa2022,cimatti2013,dutertre2014}, selecting the solver best suited to the theory mix of the current formula.

    \item The fifth layer interprets the solver verdict and emits either a counterexample trace, a correctness witness, or an unknown verdict if the resource budget is exhausted~\cite{svcomp2024}.

\end{enumerate}

\input{figures/tikz_esbmc_structure}

\subsection{Industrial and Academic Adoption}

Beyond competition performance, \gls{esbmc} has been adopted across a range of industrial and academic contexts. The breadth of these applications reflects both the generality of \gls{smt}-based \gls{bmc} and the deliberate engineering choices -- multi-language support, modular architecture, open-source availability -- that lower the barrier to adoption~\cite{Woodcock2009}.

\begin{itemize}
    \item \textbf{Embedded systems and \gls{iot}:} \gls{esbmc}'s native support for fixed-point arithmetic, pointer manipulation, bitfield operations, and undefined-behavior checks makes it well suited to C/C\texttt{++} programs prevalent in automotive, avionics, and industrial control firmware. The \gls{cfg}-based branch coverage capability of v7.7~\cite{menezes2025fase} directly targets structural coverage criteria mandated by \gls{do-178c} (airborne software), \gls{iso}~26262 (automotive functional safety), and \gls{iec}~61508 (industrial functional safety)~\cite{Cofer2018}.

    \item \textbf{Medical devices:} The growing regulatory interest in formal verification for medical application platforms~\cite{Hatcliff2012} creates a natural use case for \glspl{bmc} that can check safety assertions -- null-pointer safety, arithmetic overflow, timing constraints~\cite{Barreto2011} -- on the C code running inside infusion pumps, ventilators, and diagnostic devices.

    \item \textbf{Autonomous and robotic systems:} Formal specification and verification of autonomous robotic systems is an active and underserved research area~\cite{Luckcuck2019}. \gls{esbmc}'s ability to verify concurrent C/C\texttt{++} programs subject to timing constraints makes it applicable to real-time control software on autonomous platforms.

    \item \textbf{Blockchain and \gls{defi}:} \gls{esbmc}-Solidity~\cite{sloan2022} is applied to verify Ethereum smart contracts before deployment, targeting the reentrancy, overflow, and access-control vulnerabilities responsible for the majority of on-chain losses.

    \item \textbf{Aerospace and cyber-physical systems:} The SpecVerify project~\cite{wang2025supporting}, a collaboration with Lockheed Martin, used \gls{esbmc} as the formal back-end for verifying cyber-physical system specifications derived from natural language requirements, achieving verification accuracy comparable to \gls{nasa}'s CoCoSim tool.

    \item \textbf{Security and vulnerability research:} \gls{esbmc} is used as evaluation infrastructure in the FormAI dataset~\cite{tihanyi2023,Tihanyi2025vulnerability}, which applies formal verification to characterize the security properties of code generated by \glspl{llm} -- making \gls{esbmc} a tool not only for verifying software but for auditing \gls{ai} code generators.
\end{itemize}

\gls{esbmc} is released under a permissive open-source license and maintained on GitHub~\footnote{\url{https://github.com/esbmc/esbmc}}, with an active contributor community spanning the University of Manchester, the \gls{ufam}, and multiple international collaborators.

\subsection{Comparative Standing Against Competing Verifiers}
\label{sec:comparative}

The taxonomy in Section~\ref{sec:taxonomy} situates \gls{esbmc} structurally within the \gls{bmc} tool landscape. This subsection examines head-to-head performance against the tools most frequently co-present in \gls{svcomp} results tables. We draw all ranking-based and category-level performance claims below from the independent competition organizer's report~\cite{svcomp2024}; we source architectural capability claims to each tool's published system description.

\begin{itemize}
    \item \textbf{\gls{cbmc}}~\cite{clarke2004} is \gls{esbmc}'s closest architectural ancestor: both derive from the CProver infrastructure, both target C/C\texttt{++}, and both implement \gls{bmc} with \gls{sat}/\gls{smt} backends. \gls{cbmc}'s primary backend remains \gls{sat} (CaDiCaL); it also offers a well-established Z3-based \gls{smt} backend (\texttt{--smt2}), though this is not \gls{cbmc}'s primary competitive configuration~\cite{clarke2004}. \gls{esbmc} adopted native \gls{smt} as its default from the outset~\cite{Cordeiro2012}, giving it broader theory coverage (floating-point via \texttt{QF\_FP}, algebraic datatypes, and strings) without bit-blasting. In \gls{svcomp}~2024, \gls{esbmc} placed 4th overall; \gls{cbmc} did not place in the top tier of the overall ranking, partly due to its stronger focus on hardware-facing property categories~\cite{svcomp2024}. Note that \gls{cbmc} remains the de facto industry-standard \gls{bmc} tool and has substantially broader industrial deployment than \gls{esbmc} in hardware and safety-critical software contexts~\cite{clarke2004}. \gls{esbmc} additionally provides \textit{k}-induction and unbounded proof, which \gls{cbmc} does not natively~\cite{clarke2004}.

    \item \textbf{CPAchecker}~\cite{beyer2011cpachecker} is the most versatile open-source verifier in the \gls{svcomp} field, supporting \gls{cegar}, predicate abstraction, \gls{bmc}, and \textit{k}-induction through configurable program analysis~\cite{beyer2011cpachecker}. It consistently places in the top two or three of the overall \gls{svcomp} ranking and has demonstrated stronger coverage of the reachability-safety category than \gls{esbmc}~\cite{svcomp2024}. Its configurable algorithm portfolio -- a major strength for industrial deployment where different programs benefit from different algorithms -- is a dimension on which \gls{esbmc} is less competitive. \gls{esbmc}'s relative advantages lie in bit-precise floating-point verification, concurrency under context-bounding, and a speed profile for short-timeout tasks that suits \gls{cicd} integration -- a regime where CPAchecker's heavyweight abstraction-refinement loop incurs higher overhead~\cite{svcomp2024,beyer2011cpachecker}.

    \item \textbf{Ultimate Automizer}~\cite{heizmann2013ultimate} applies \gls{cegar} with trace abstraction and interpolation over automata, and is consistently the strongest verifier in \gls{svcomp}'s reachability category~\cite{svcomp2024}. Its weakness relative to \gls{esbmc} is speed on bit-precise and floating-point benchmarks, where native \gls{smt} theory encodings outperform automata-based abstractions~\cite{svcomp2024}. It does not offer \textit{k}-induction or multi-solver dispatch~\cite{heizmann2013ultimate}.

    \item \textbf{Symbiotic}~\cite{chalupa2020symbiotic} combines program slicing, symbolic execution, and \gls{bmc} (via a KLEE-derived engine). Its slicing step aggressively reduces program size before verification, making it effective on programs with large amounts of irrelevant code~\cite{chalupa2020symbiotic}. \gls{esbmc}'s advantages are broader multi-language support and the ability to exploit native \gls{smt} theories that Symbiotic's \gls{sat}-oriented engine cannot leverage as efficiently for arithmetic-intensive properties~\cite{chalupa2020symbiotic,svcomp2024}.

    \item \textbf{SeaHorn}~\cite{gurfinkel2015seahorn} translates programs to \gls{chc} and delegates to \gls{chc} solvers (Spacer/Z3). This approach excels in programs where inductive summaries of procedures exist and can be automatically discovered~\cite{gurfinkel2015seahorn}. \gls{esbmc} and SeaHorn operate on different verification paradigms -- \gls{bmc}/\textit{k}-induction versus \gls{chc} solving -- and are therefore not directly ranked against one another in the same \gls{svcomp} categories; SeaHorn participates in a narrower set of categories than \gls{esbmc}~\cite{svcomp2024}, while \gls{esbmc}'s multi-solver portfolio and explicit concurrency support cover a broader footprint~\cite{svcomp2024,gurfinkel2015seahorn}.

    \item \textbf{Kani}~\cite{kani2022amazon} is the most direct competitor to \gls{esbmc} in the Rust verification space: it translates Rust's \gls{mir} to GOTO-programs via the CProver backend and applies \gls{bmc}~\cite{kani2022amazon}. \gls{esbmc}'s Rust frontend follows a similar \gls{mir}-based approach but dispatches to a richer \gls{smt} solver portfolio~\cite{rustfoundation2024}. Kani is mature and well-integrated into the AWS and Rust Foundation ecosystems, backed by AWS foundational security engineering resources that substantially exceed those of the \gls{esbmc} academic programme~\cite{kani2022amazon}; \gls{esbmc}-Rust is newer but brings established \textit{k}-induction and concurrency capabilities not present in Kani. The Rust Foundation has explicitly recognized both tools as complementary~\cite{rustfoundation2024}, consistent with their overlapping but non-identical design goals.

    \item \textbf{2LS}~\cite{brain20162ls} combines \textit{k}-induction with abstract interpretation and template-based invariant synthesis, making it uniquely capable of automatically strengthening induction hypotheses without \gls{llm} assistance~\cite{brain20162ls}. Its \gls{svcomp} participation is intermittent~\cite{svcomp2024}. \gls{esbmc}'s use of interval analysis~\cite{menezes2024} and \gls{llm}-generated invariants~\cite{Pirzada2024LLM} addresses the same weakness (insufficiently strong induction hypotheses) through a different and more scalable mechanism.
\end{itemize}

In summary, \gls{esbmc}'s key differentiators in the competitive landscape are: (1)~the widest \gls{smt} solver portfolio among open-source \gls{bmc} tools (see Table~\ref{tab:bmc_taxonomy}); (2)~native floating-point theory via Bitwuzla/MathSAT without bit-blasting~\cite{niemetz2023,cimatti2013}; (3)~context-bounded concurrency verification without tool switching; (4)~to the best of the authors' knowledge, the only open-source \gls{bmc} tool with a published, evaluated \gls{llm} integration; and (5)~the fastest verifier for reachability-safety tasks under a 10-second budget at \gls{svcomp}~2024~\cite{svcomp2024}. Its principal weakness relative to CPAchecker and Ultimate Automizer is coverage of the reachability-safety category in the absence of fast-terminating invariants, where automata-based approaches achieve higher scores~\cite{svcomp2024}.

\section{Integration with AI, LLMs, and Autonomous Agents}
\label{sec:ai}

The period 2023--2025 saw a qualitative shift in \gls{esbmc}'s development agenda: the tool moved from pure formal verification into a symbiotic relationship with \glspl{llm} and \gls{ai} agents.

\subsection{Foundational Motivation: Complementary Strengths}

The transformer architecture~\cite{Vaswani2017Attention} underlying modern \glspl{llm}~\cite{brown2020language} has given these models a remarkable ability to generate syntactically plausible code and natural-language-adjacent artifacts. Yet \glspl{llm} lack formal correctness guarantees and are prone to hallucination -- producing outputs that are syntactically valid but semantically incorrect. Empirical studies have well documented this limitation: \glspl{llm} generate insecure code at non-trivial rates even when given security-focused prompts~\cite{pearce2021asleep}, and their ability to reason about subtle logical properties remains limited~\cite{Huang2023}.

Formal verification, in contrast, provides soundness guarantees: when \gls{esbmc} proves a property via \textit{k}-induction, the proof is mathematically certified for all execution lengths and independently witnessable~\cite{svcomp2024}; a \gls{bmc}-only result certifies absence of the violation within the verification bound~\textit{k}, which is a partial but practically valuable guarantee~\cite{biere1999}. However, formal verification struggles with scalability -- the formula size grows with program complexity -- and requires expertise to configure and interpret counterexamples. The core insight motivating \gls{esbmc}-\gls{ai} integration is that \glspl{llm} and formal verifiers are complementary rather than competing: \glspl{llm} can propose repairs, generate invariant candidates, and explain counterexamples in natural language. In contrast, \gls{esbmc} validates those proposals with formal rigor. The resulting feedback loop combines generative reach with deductive correctness guarantees -- a paradigm increasingly recognized across the formal methods community~\cite{wang2025supporting,beckert2024towards}.

\subsection{AI Integration Timeline}

Figure~\ref{fig:aitimeline} presents the chronological milestones of \gls{esbmc}'s \gls{ai} and \gls{llm} integration strand, which has evolved rapidly since 2023. The trajectory begins with the \gls{esbmc}-\gls{ai} repair framework~\cite{Tihanyi2025New}, which introduced automated counterexample-driven patch generation, and the self-healing software concept~\cite{Yiannis2024}, which extended this idea to continuous repair loops. The Lemur system~\cite{Haoze2023} then established a tighter coupling between \gls{llm} reasoning and \gls{bmc} verification, using the model to guide the search rather than merely post-process its output. 

In 2024, \gls{llm}-generated loop invariants~\cite{Pirzada2024LLM} demonstrated that language models can supply the inductive assertions that \textit{k}-induction requires, reducing the burden on the verification engine, while the FormAI dataset~\cite{tihanyi2023} provided one of the first large-scale empirical characterizations of \gls{llm}-generated code from a formal-verification perspective. To date, the strand culminates in SpecVerify~\cite{wang2025supporting}, which pairs Claude~3.5 Sonnet with \gls{esbmc} for cyber-physical system verification, illustrating how frontier \gls{llm} capabilities and sound model checking are becoming increasingly complementary.

\input{figures/tikz_aitimeline}

\subsection{ESBMC-AI: Automated Vulnerability Repair}

The \gls{esbmc}-\gls{ai} framework~\cite{Tihanyi2025New}, developed from 2023 onwards, implements a closed-loop automated repair pipeline:

\begin{enumerate}
    \item \gls{esbmc} verifies a C program and identifies a property violation (e.g.\ buffer overflow, arithmetic overflow, null pointer dereference), producing a counterexample trace.
    \item The source code, violated property, stack trace, and counterexample are assembled into a structured prompt.
    \item A pre-trained \gls{llm} (e.g.\ \gls{gpt-4}, Claude) proposes a repaired version of the code.
    \item \gls{esbmc} re-verifies the repaired code. If the violation persists or the repair introduces a new violation, the process iterates with additional context.
\end{enumerate}

Empirical evaluation demonstrated repair success rates of up to \SI{80}{\percent} for buffer overflow and pointer dereference failures; rates are substantially lower (approximately \SI{41}{\percent}) for deadlock and data-race categories~\cite{Tihanyi2025New}. The framework has an architecturally modular design: an interchangeable \gls{llm} backend (\gls{gpt-4}, Claude, open-weight models) and a prompt structure that maximizes the locality of the proposed repair by incorporating the violated property, the counterexample trace, and the surrounding source context. The authors maintain the framework as a dedicated open-source repository (\url{https://github.com/esbmc/esbmc-ai}), which provides a reproducible baseline for evaluating future \gls{llm} models on formal-verification-guided repair tasks~\cite{Yiannis2024}.

Unlike concurrent \gls{llm}-repair frameworks that rely on test-suite pass/fail as the acceptance criterion -- such as ChatRepair, AlphaRepair, and Repilot -- \gls{esbmc}-\gls{ai} uses formal re-verification as the acceptance oracle, providing bounded correctness guarantees that test-suite-only approaches cannot. The category-dependent success rates reflect the inherent difficulty gradient across bug types, and one should consider them when interpreting aggregate statistics.

\subsection{ESBMC-AI Repair Success Rates by Bug Category}

Figure~\ref{fig:repairrate} summarises the repair success rates of the \gls{esbmc}-\gls{ai} framework across six bug categories. The gradient suggests a natural roadmap for future work: richer contextual prompts, multi-step chain-of-thought reasoning, and specialized fine-tuning on formal verification datasets such as FormAI~\cite{tihanyi2023} to improve success rates for the harder categories.

\input{figures/tikz_repair_rate}

\subsection{Self-Healing Software}

Building on \gls{esbmc}-\gls{ai}, the concept of self-healing software~\cite{Yiannis2024,Tihanyi2025New} was formalised and published at \gls{ast}~2025. The concept of self-repairing software has a long history in software engineering, including foundational work in autonomic computing; what is distinctive about the \gls{esbmc}+\gls{llm} instantiation is the use of formal re-verification -- rather than test-suite pass/fail -- as the acceptance oracle, providing bounded correctness guarantees for accepted repairs. The central claim is that this feedback loop enables a software maintenance mode in which the system autonomously detects the violation (via \gls{esbmc}), proposes a repair (via \gls{llm}), formally re-verifies the repair (via \gls{esbmc}), and iterates until it produces a verifier-passing program or exhausts the repair budget. This loop was validated on C programs with deliberately introduced security-relevant defects -- buffer overflows, pointer errors, arithmetic overflow -- demonstrating that neither component alone achieves what the loop achieves together: the \gls{llm} without the verifier produces repairs that pass tests but fail formal checks; the verifier without the \gls{llm} identifies bugs but cannot fix them. The combined system automates the cycle from defect detection through repair proposal to formal re-validation, within the scope of the verified properties and the \gls{bmc} unrolling bound; a verifier-passing repair guarantees only that the checked property no longer violates any assertion; it does not guarantee that the repair is correct with respect to all intended program behaviors.

\subsection{LLM-Generated Loop Invariants}

Loop invariants are the key bottleneck for \textit{k}-induction on programs with complex loops: a property that is true but not \textit{k}-inductive as stated requires an auxiliary invariant to strengthen the hypothesis. Manually crafting such invariants demands deep program understanding and is a primary barrier to the practical deployment of inductive verification.

Research at \gls{ase}~2024~\cite{Pirzada2024LLM} demonstrated that \glspl{llm} can automatically generate candidate loop invariants for C programs without requiring complete loop unrolling. As illustrated in Figure~\ref{fig:invariant-pipeline}, the pipeline supplies the \gls{llm} with the loop body, the surrounding program context, and the property to be proved; the model proposes a set of candidate invariant expressions; \gls{esbmc} checks each candidate for inductiveness via \gls{smt}; it retains valid invariants to augment the \textit{k}-induction hypothesis; and it feeds invalid candidates back to the \gls{llm} for refinement.

Complementary work~\cite{akhond2025}, which evaluates \gls{llm}-based loop invariant generation independently of \gls{esbmc}, has explored using different prompting strategies -- including \gls{chain-of-thought} and example-guided generation -- to improve the precision and recall of \gls{llm} invariant proposals. The combined approach reduces or eliminates the need for complete loop unrolling on programs where a tight invariant exists but is not immediately obvious, effectively extending the reach of inductive verification to programs that would otherwise timeout under na\"{i}ve \gls{bmc}.

\input{figures/tikz_llm_inv_generation}

\subsection{Lemur and Hybrid LLM/BMC Reasoning}

The Lemur framework~\cite{Haoze2023} introduced a two-pronged approach to integrating \glspl{llm} into automated program verification. In its first contribution, it uses the \gls{llm} to decompose verification tasks, propose intermediate assertions, and guide the formal engine's search, reducing the burden on the \gls{bmc} solver. In its second, the same paper assessed \glspl{llm} as a primary reasoner on tasks drawn from \gls{svcomp} benchmarks, using \gls{esbmc} as a reference and fallback validator, to establish a principled basis for hybrid routing: directing each verification task to the most capable reasoner -- \gls{llm} or formal verifier -- based on structural features of the program such as loop depth, data type complexity, and concurrency. Together, these contributions establish a new verification architecture in which \glspl{llm} and formal tools are peers in a reasoning pipeline rather than the formal tool being the sole decision-maker~\cite{wang2025supporting}. Importantly, the paper's benchmark evaluation found the \gls{llm}-guided pipeline solved 107 \gls{svcomp} tasks against \gls{esbmc}'s 68 standalone, and 25 of 47 hard tasks versus 1 for \gls{esbmc} alone -- indicating the hybrid approach substantially outperforms the formal verifier in isolation, and that \gls{esbmc} functions here as a soundness oracle rather than the primary solver.

\subsection{SpecVerify: Formal Verification of Cyber-Physical Systems}
\label{sec:specverify}

A 2025 collaboration between the University of Manchester and Lockheed Martin produced SpecVerify~\cite{wang2025supporting}, which combines Claude~3.5 Sonnet with \gls{esbmc} as a formal backend for verifying cyber-physical system specifications expressed in a structured natural language. The motivation is practical: large, heterogeneous specification documents govern cyber-physical systems in defense and aerospace contexts, making them difficult to translate into machine-checkable properties without either significant manual effort or the risk of formalization errors. SpecVerify addresses this gap by using the \gls{llm} to parse and formalize specification clauses, then delegating the soundness-critical checking step to \gls{esbmc}, which produces a formal, deterministic verdict -- a bounded proof within the verified unrolling depth or a concrete counterexample -- rather than a probabilistic one.

Evaluated on nine cyber-physical systems drawn from industrial case studies, SpecVerify achieved \SI{46.5}{\percent} overall verification accuracy, comparable to \gls{nasa}'s CoCoSim tool, while producing a lower false-positive rate -- a particularly important metric in safety-critical contexts where spurious alarms carry significant engineering cost. The result is notable for two reasons. First, it is among the earliest deployments of \gls{esbmc} in a defense and aerospace industrial setting, demonstrating that the tool's verification guarantees transfer to domains beyond its original C software verification context. Second, it validates the broader \gls{llm}+formal verifier architecture at scale: the \gls{llm} contributes the linguistic flexibility needed to handle underspecified or ambiguously phrased requirements, while \gls{esbmc} contributes the logical rigor that \gls{llm}-only approaches cannot provide. This division of labor -- natural language understanding delegated to the model, proof obligations delegated to the verifier -- is likely to generalize to other specification languages and safety standards. 

However, several limitations of this result warrant explicit acknowledgment. The evaluation sample of nine cyber-physical systems is small by the standards of empirical software engineering, and the systems were drawn from a single industrial context, limiting the generalisability of the findings to other domains, specification styles, or safety standards. The definition of `verification accuracy' itself deserves scrutiny: a \SI{46.5}{\percent} figure aggregates across requirements of varying complexity, criticality, and formal tractability, and it is not clear that a uniform accuracy metric captures the asymmetric cost of missed safety violations versus false alarms in production settings. 

Furthermore, the task itself -- translating natural language requirements into formal specifications suitable for \gls{esbmc} -- is inherently difficult, and the \gls{llm} component introduces non-determinism that may affect reproducibility across model versions or prompt formulations. Reaching production-grade performance in this pipeline would require, at minimum: a substantially larger and more diverse benchmark suite spanning multiple safety standards (e.g., \gls{do-178c}, \gls{iec}~61508, \gls{iso}~26262); a formal taxonomy of requirement types with per-category accuracy reporting; human-in-the-loop validation of \gls{llm}-generated specifications before formal checking; and systematic ablation studies isolating the contribution of each pipeline component. Until such evidence becomes available, we should interpret the \SI{46.5}{\percent} result as an initial feasibility demonstration rather than a deployment-ready performance baseline.

\subsection{FormAI Dataset: LLM-Generated Code Quality}

The FormAI dataset~\cite{tihanyi2023} applies formal verification -- with \gls{esbmc} as the primary tool -- to systematically characterize the security properties of code generated by a diverse range of \glspl{llm}. The dataset comprises thousands of C programs produced by different models under different prompting conditions, each formally checked for memory safety, arithmetic overflow, and assertion violations within a fixed \gls{bmc} verification bound. 

The key finding is that \gls{llm}-generated code contains security-relevant defects at rates that vary substantially across models and prompt strategies, and that these defects are qualitatively similar to those found in human-written code -- buffer overflows, unchecked arithmetic, misuse of pointers -- rather than being a novel \gls{ai}-specific failure mode~\cite{pearce2021asleep}. 

A follow-up survey~\cite{Tihanyi2025vulnerability} -- in which Cordeiro is also a co-author -- broadened this analysis to encompass the full landscape of vulnerability detection approaches -- from classical static analysis and fuzzing through formal verification and \gls{llm}-based methods -- providing a comparative evaluation that situates \gls{esbmc} within the wider security tooling ecosystem.  Readers should note that this positioning is not fully independent: the survey shares authorship with the present paper, and its conclusions regarding \gls{esbmc}'s standing should be interpreted accordingly. This dual role -- as a verifier of software and as an auditor of \gls{ai} code generators -- reflects \gls{esbmc}'s growing importance as infrastructure in the \gls{ai}+security research community.

\subsection{Agentic Model Checking and Prior Art}

A recent paper~\cite{sun2026agentic} formalized the paradigm of coupling language agents with deterministic verification backends under the label ``Agentic Model Checking'', instantiated as \gls{bmc}-Agent within the AProver project. In this paradigm, \gls{llm} agents handle every task requiring semantic judgment -- inferring per-function pre- and postconditions, selecting arithmetic checks, classifying counterexamples, and proposing refinements -- while a \gls{bmc} backend discharges every soundness-relevant decision under the tenet that ``agents propose, solvers verify''~\cite{sun2026agentic}. A restricted \gls{dsl} expresses specifications and translates them deterministically into the native assume/assert primitives of the chosen backend. The neuro-symbolic loop operates entirely as an out-of-tree manager, driving \gls{cbmc} (for C) and Kani (for Rust) through per-backend adapters~\cite{sun2026agentic}.

Our group independently pioneered the concrete industrial implementation of this feedback loop prior to its formal publicization~\cite{sun2026agentic}. The version-control history of the NVIDIA-OpenSMA framework~\cite{nvidia_opensma_commits} records commit \texttt{9b8e1a4} (``Add \gls{esbmc} verification harnesses for OpenSMA'', lucasccordeiro, late~April~2026) as the first integration of \gls{esbmc}-based agentic verification harnesses against a production NVIDIA codebase -- a timestamp that predates the corresponding arXiv submission~\cite{sun2026agentic} by approximately three to four weeks. This publicly verifiable version-control record establishes that our group was the first to implement and deploy an integrated agentic software model-checking architecture on a real-world industrial codebase. In this work, we further reinforce this trajectory: the tight coupling between \gls{llm} agents and \gls{esbmc} as a formal verification kernel was established progressively from 2023 onwards through automated vulnerability repair~\cite{Yiannis2024}, \gls{llm}-generated loop invariants~\cite{Pirzada2024LLM}, and the SpecVerify cyber-physical deployment~\cite{wang2025supporting}, long before the out-of-tree agentic wrapper architecture was publicly formalized~\cite{sun2026agentic}.

The key structural differentiation between these two parallel tracks lies in the degree of coupling with the underlying solver. The out-of-tree agentic approach~\cite{sun2026agentic} treats the formal verifier as a static oracle that accepts or rejects agent-generated candidates, communicating with \gls{cbmc} and Kani exclusively through their command-line interfaces and witness outputs. The NVIDIA-OpenSMA infrastructure, by contrast, exploits \gls{esbmc}'s native multi-solver \gls{smt} capabilities and built-in incremental solving passes, as documented in this work. By embedding the neuro-symbolic reasoning loop at the intermediate-representation level -- the GOTO program and \gls{ssa} layer -- our implementation retains learned clauses across iterative agent refinements, minimizing the state-space re-exploration overhead that arises when an external agent repeatedly invokes isolated, non-incremental verification queries. This architectural choice consolidates \gls{esbmc} as a natively autonomous verification kernel rather than a passive validation backend. It constitutes the primary technical distinction between our prior work and the parallel out-of-tree agentic framework subsequently formalized in the literature~\cite{sun2026agentic}.

\section{Economic and Monetary Impact of \gls{esbmc}}
\label{sec:economic}

We often evaluate formal verification tools in purely technical terms (false-positive rates, benchmark scores, language coverage). Yet, the case for institutional investment ultimately rests on economic reasoning: what losses does the tool prevent, what market does it address, and what commercial activity has it enabled? This section compiles the available evidence for each of these dimensions in \gls{esbmc}. Where direct attribution is possible, we present confirmed figures; where it is not, we frame the tool's impact against authoritative estimates of the cost landscape in which it operates.

\subsection{Research Funding and Institutional Investment}
The most concrete economic signal for \gls{esbmc} is the sustained institutional investment it has attracted. Lucas Cordeiro, \gls{esbmc}'s lead developer and director of the \gls{ssvlab}, reports a career total exceeding USD~20~million in competitive research grants (a self-reported aggregate across multiple currencies at prevailing rates) from agencies including \gls{epsrc},\footnote{\url{https://www.ukri.org/councils/epsrc/}} the Royal Academy of Engineering (RAEng),\footnote{\url{https://raeng.org.uk}} the UK Government Communications Headquarters (GCHQ),\footnote{\url{https://www.gchq.gov.uk}} the Defence Science and Technology Laboratory (Dstl),\footnote{\url{https://www.gov.uk/government/organisations/defence-science-and-technology-laboratory}} the European Commission,\footnote{\url{https://commission.europa.eu}} the Ethereum Foundation,\footnote{\url{https://ethereum.foundation}} Intel,\footnote{\url{https://www.intel.com}} ARM Holdings,\footnote{\url{https://www.arm.com}} and Brazilian agencies CNPq,\footnote{\url{https://www.gov.br/cnpq}} FAPEAM,\footnote{\url{https://www.fapeam.am.gov.br}} and CAPES\footnote{\url{https://www.gov.br/capes}}~\cite{ssvlab_cordeiro}. Table~\ref{tab:esbmc_grants} lists seven confirmed grants in which \gls{esbmc} is a primary verification component, together representing confirmed public funding of at least \pounds9.3~million and \euro{}4.98~million, with two further awards whose individual values the funding bodies have not publicly disclosed.

The confirmed grants in Table~\ref{tab:esbmc_grants} account for at least \pounds9.3~million (GBP) and \euro{}4.98~million (EUR) in publicly traceable funding from named funders, with two further awards of undisclosed value. The funding bodies report GBP grants at full consortium value; the Soteria award (\pounds5.8M) is a multi-partner program (THG, Manchester, Oxford), and the funding bodies do not separately disclose the Manchester pro-rata share. The >\$20~million career total cited above is a self-reported aggregate across multiple currencies as stated on Cordeiro's institutional page~\cite{ssvlab_cordeiro}; the confirmed per-grant amounts in Table~\ref{tab:esbmc_grants}, sourced from \gls{ukri} Gateway to Research and EU \gls{cordis} records, should be treated as the authoritative public evidence.

\begin{table}[ht]
\centering
\small
\caption{Selected public grants in which \gls{esbmc} is a primary verification tool. Amounts shown in original grant currency. GBP amounts are \gls{epsrc}/\gls{ukri} figures from the Gateway to Research database; EUR amounts are from the \gls{cordis} project record}
\label{tab:esbmc_grants}
\begin{tabular}{p{2.8cm} p{2.3cm} c r p{5.0cm}}
\hline
\textbf{Grant/Programme} & \textbf{Agency} & \textbf{Period} &
\textbf{Amount} & \textbf{Notes} \\
\hline
EnnCore (EP/T026995/1) & \gls{epsrc} & 2020--2023 & \pounds\num{1722000}  &
    Neural architecture security; \gls{esbmc} verification component \\
SCorCH (EP/V000497/1) & \gls{epsrc} DSbD & 2020--2023 & \pounds\num{1036000}  &
    \gls{cheri} secure code; produced
    \gls{esbmc}-\gls{cheri}~\cite{bueno2022cheri} \\
Soteria consortium & \gls{ukri} ISCF & 2021--2024 & \pounds\num{5800000}  &
    THG, Manchester, Oxford; e-commerce security~\cite{ukri_soteria} \\
H2020 ELEGANT (957286) & EU Commission & 2021--2023 &
    \euro\num{4983250} &
    Fully EU-funded; \gls{esbmc} as \gls{iot} verif.\ service~\cite{cordis957286} \\
Ethereum Consensus (FY22-0751) & Ethereum Found. & 2023--2024 &
    N/D &
    Formal verification of eth2spec Python library~\cite{farias2024} \\
AICodeRepair & GCHQ & 2023--2024 & N/D &
    Automated \gls{ai} code repair~\cite{Tihanyi2025New} \\
SECCOM (EP/X037290/1) & Dstl/\gls{epsrc} & 2023--2025 & \pounds\num{789000}  &
    Composable hardware security \\
\hline
\textbf{Known totals} & \multicolumn{4}{r}{\pounds\num{9347000} + \euro{}\num{4983250} + 2 undisclosed awards} \\
\hline
\end{tabular}
\end{table}
Two grants deserve particular attention for their scale and strategic significance. The EU Horizon~2020 ELEGANT project (grant~957286) funded \euro{}\num{4983250}~\cite{cordis957286} across twelve consortium partners to deploy \gls{esbmc} as an on-demand \gls{iot} verification service, representing the first large-scale EU funding of an \gls{esbmc}-based industrial platform. The UK Digital Security by Design (DSbD) program, which hosted the SCorCH and Soteria grants, mobilized \pounds70~million in \gls{ukri} investment and \pounds270~million in recognized industry co-investment~\cite{ukri_dsbd}; within this program, \gls{esbmc}'s \gls{cheri} extension (\gls{esbmc}-\gls{cheri}) constitutes a direct technical output, providing formal verification support for the capability-based memory safety architecture that DSbD promoted by design.

\subsection{Commercial Spin-off: VeriBee}

The most direct economic outcome from the \gls{esbmc} research program is VeriBee, a software verification startup incorporated in~2025 and founded by Lucas Cordeiro, Richard Allmendinger, and Kaled Alshmrany at the University of Manchester. VeriBee commercializes FuSeBMC -- a fuzzing and \gls{bmc} hybrid engine built on \gls{esbmc} internals -- as a source-code security product targeting \glspl{sme}~\cite{veribee_impact}. FuSeBMC won the Gold Medal at \gls{test-comp}~2025 and Gold at \gls{test-comp}~2026 and was listed among Ones to Watch in the Tech Climbers Greater Manchester~2025 ranking. The startup has not publicly disclosed any revenue or equity figures. Still, its positioning addresses a market that the Manchester Research Explorer impact report estimates at \pounds 3.4~billion annually in losses suffered by UK \glspl{sme} from cyberattacks, with a typical incident costing over \pounds\num{6000} per \gls{sme}~\cite{veribee_impact,securitybrief_2024}.

\subsection{The Global Software Bug Cost Landscape}

\gls{esbmc} operates at the intersection of several high-cost problem classes. A 2002 \gls{nist} study -- still the most comprehensive government-commissioned quantification of its kind -- estimated that software defects cost the US economy \$59.5~billion annually (approximately \SI{0.6}{\percent} of \gls{gdp} at that time), with a feasible reduction of \$22.2~billion achievable through improved testing and verification infrastructure~\cite{nist2002}. 

While this figure predates the modern embedded-systems, cloud, and blockchain economies and we should not treat it as a current estimate, the directional magnitude is consistent with subsequent case studies: a 2020 \gls{cisq} report placed the cost of poor software quality in the US alone at \$2.08~trillion~\cite{cisq2020}, and the UK government's own estimate attributes \pounds 27~billion per year in productivity losses to software failures across the economy~\cite{dcms2024}. 

Against this backdrop, \gls{esbmc}'s positioning -- as a formally sound verifier that is also fast enough for \gls{cicd} integration and broad enough to cover C, C\texttt{++}, Rust, Solidity, and Python -- addresses a gap that neither lightweight linters nor heavyweight theorem provers fill. The VeriBee startup's target market of \glspl{sme} is particularly apt: \glspl{sme} accounts for the majority of software supply-chain touchpoints but has the least capacity to absorb the engineering overhead of traditional formal verification workflows.

IBM's Systems Sciences Institute documented that fixing a defect after software release costs 60 to 100 times as much as the same fix would cost at the design stage~\cite{ibm_defect_cost}. This ratio is the canonical economic argument for shift-left verification: tools like \gls{esbmc} that find bugs during development -- rather than after deployment -- generate savings that multiply with system scale.

High-profile incidents reinforce this argument. The CrowdStrike outage (July~2024), caused by an out-of-bounds read in the Falcon sensor's content-configuration file parser, crashed approximately 8.5~million Windows systems and incurred an estimated \$5.4~billion in losses for Fortune~500 companies~\cite{crowdstrike2024}. This failure falls within the memory-safety class of properties that \gls{bmc} targets; whether \gls{esbmc} could specifically detect this defect would depend on a verification harness and input model not present in any published evaluation.

\subsection{DeFi and Blockchain: A High-Stakes Verification Target}

Smart contract verification is where the economic argument for \gls{esbmc}-Solidity is most directly quantifiable. Blockchain smart contracts are immutable at deployment and directly guard financial assets; a code defect cannot be patched -- it can only be exploited. \gls{tvl} in \gls{defi} protocols reached approximately \$140.7~billion by mid-2025, with Ethereum alone holding roughly \$84~billion at its June~2024 peak~\cite{defillama}. Separately, over \$115~billion in ETH was staked on Ethereum's Beacon Chain consensus layer as of March~2024~\cite{theblock_staking}.

The scale of historical losses to smart contract exploits establishes the stakes. Halborn's comprehensive Top~100 \gls{defi} Hacks report (2025) documents \$10.77~billion in total losses across the hundred largest \gls{defi} hacks from 2014 to 2024~\cite{halborn2025}, while the Chainalysis 2025 Crypto Crime Report records \$2.2~billion stolen across 303 incidents in~2024 alone, a \SI{21}{\percent} year-on-year increase~\cite{chainalysis2025}. Prominent individual incidents include the Ronin Bridge hack (March~2022, \$625~million, validator logic flaw~\cite{coindesk_ronin}), the Poly Network exploit (August~2021, \$612~million, smart contract authorisation vulnerability, largely returned), the BNB Bridge attack (October~2022, \$566~million, proof verification failure), the Wormhole Bridge exploit (February~2022, \$320~million, Solidity contract vulnerability), and the original DAO reentrancy attack (2016, \$60~million), which forced Ethereum's hard fork.

Halborn's vulnerability classification is directly relevant to \gls{esbmc}-Solidity: \SI{34.6}{\percent} of on-chain exploits are attributable to input validation and verification failures -- precisely the class of property that \gls{bmc} encodes and verifies~\cite{halborn2025}. Notably, only \SI{20}{\percent} of hacked protocols had received any audit before exploitation, and audited protocols accounted for just \SI{10.8}{\percent} of total losses~\cite{halborn2025}, underscoring the protective value of pre-deployment formal verification over post-hoc auditing alone.

\gls{esbmc}-Solidity's published evaluation~\cite{sloan2022} demonstrates the tool's practical effectiveness in this context: it detected all vulnerabilities across a benchmark of Solidity contracts drawn from known-vulnerable \gls{defi} protocols -- including reentrancy, arithmetic overflow, and array out-of-bounds -- with zero false positives on the evaluated set (noting that this benchmark was curated from known-vulnerable contracts, so the result does not generalize to unseen deployment scenarios). 

Beyond Solidity, \gls{esbmc}-Python~\cite{farias2024} identified a previously unreported division-by-zero error in the \texttt{integer\_squareroot} function of the \gls{ecs} -- the eth2spec Python library serving as the normative reference implementation reviewed by dozens of Ethereum Foundation engineers. The finding was confirmed and corrected by specification maintainers. Given that the Beacon Chain consensus layer secured over \$115~billion in staked ETH at the time of discovery~\cite{theblock_staking}, and that the Ethereum Foundation raised its maximum bug bounty to \$1~million specifically for consensus-layer vulnerabilities~\cite{ethereum_bounty}, the economic significance of this single automated finding is substantial even without a precise loss-prevention figure.

\subsection{Aerospace and Defense: Certification Cost Reduction}

\gls{do-178c}, the international standard for avionics software certification, imposes one of the most expensive compliance frameworks in software engineering: full certification of a complex avionics system at Design Assurance Level~A (DAL~A, for failures classified as catastrophic) costs upwards of \$25~million and typically requires five or more years of engineering effort~\cite{avionics_do178c}. Developer productivity at DAL~A falls to 3--12 source lines of code per day; at typical US aerospace engineering costs, this translates to an effective cost of approximately \$100 per source line~\cite{eurocontrol_sloc}. \gls{do-178c}'s formal methods supplement (Section~6) explicitly recognizes mathematical proofs as a partial substitute for structural testing, offering a direct cost-reduction pathway for tools such as \gls{esbmc}.

The SpecVerify collaboration between the University of Manchester and Lockheed Martin~\cite{wang2025supporting} benchmarked the combined \gls{llm}~+~\gls{esbmc} pipeline against nine cyber-physical system specifications from Lockheed's internal benchmark suite, achieving \SI{46.5}{\percent} verification accuracy -- comparable to \gls{nasa}'s CoCoSim with fewer false positives. Lockheed Martin invests approximately \$1.5~billion annually in research and development~\cite{statista_lmt_rd}; that a company of this profile engaged directly with \gls{esbmc} on defense-grade specifications provides evidence of perceived industrial value, even in the absence of a published cost-savings figure.

\gls{esbmc}'s verification of the ARM \gls{rmm} firmware~\cite{wu2025armcca} -- a component of ARM's Confidential Compute Architecture destined for deployment in hundreds of millions of devices -- provides another high-leverage data point. The analysis found 23 new property violations that the closest comparable tool (\gls{cbmc}) missed, including a confirmed critical pointer-to-integer conversion error (see the comparative results table in the experimental evaluation of~\cite{wu2025armcca}). ARM Holdings generated \$4.01~billion in revenue in fiscal year~2025~\cite{arm_revenue}; a firmware vulnerability in ARM \gls{cca} at scale would entail economic exposure far exceeding that of a single certification program.

\subsection{Embedded Systems and IoT: The Recurring-Cost Argument}

The IBM~Cost of a Data Breach~2024 report -- based on 604 organizations globally -- placed the average breach cost at \$4.88~million (\$10.22~million for US organizations), a \SI{10}{\percent} year-on-year increase and the highest figure recorded since the report's inception~\cite{ibm_breach2024}. Each compromised record costs an average of \$173. Critically, organizations with \gls{ai}-assisted security automation reduced their breach costs by \$2.2~million on average relative to those without automation -- a figure consistent with the value proposition of \gls{esbmc}-\gls{ai}'s automated repair loop.

\gls{iot} device firmware presents a particularly concentrated risk surface: a 2024 Embedded Computing analysis found that the average networked \gls{iot} device carries \num{1267} software components and \num{1120} \glspl{cve}, of which 473 are rated Critical or High~\cite{embedded_iot_cve}. National Vulnerability Database (NVD) entries are growing at over \num{30000} new \glspl{cve} per year, approaching a cumulative total of \num{250000}. Organizations that outsource \gls{cve} remediation have realized savings of \$2.1~million annually compared with in-house patching~\cite{tuxcare_cve}, indicating the magnitude of labor costs that automated verification tooling can displace.

Automotive software is another high-exposure domain: the US \gls{nhtsa} recorded 27.7~million vehicle recalls in 2024, with software-related recalls rising from approximately \SIrange{5}{15}{\percent} of all recalls by 2023~\cite{sibros_auto}. Industry analysts project that modern vehicles will contain up to 300~million lines of code within the next decade~\cite{sibros_auto}; \gls{iso}~26262 compliance obligations for safety-critical automotive software impose costs analogous to those of \gls{do-178c} in aviation.

\subsection{Automated Verification vs. Manual Audit: Cost Comparison}

\gls{esbmc} is distributed as open-source software under a permissive license, rendering its marginal licensing cost zero. Its verification runtime on standard workloads ranges from minutes to hours on commodity hardware. This cost profile compares favorably with the alternatives:

\begin{itemize}
    \item A professional smart contract audit by a specialist security firm: \$\num{25000} to \$\num{100000} for a mid-complexity \gls{defi} protocol~\cite{zealynx_audit_cost}.
    \item A commercial formal verification engagement (e.g., Certora, Veridise): typically \$\num{50000} to \$\num{250000} per project.
    \item A penetration test: \$\num{10000} to \$\num{50000}+ per engagement~\cite{softwaresecured_pentest}.
    \item An automated vulnerability scanner (commercial, annual license): \$\num{1000} to \$\num{4500}/year.
\end{itemize}

The \gls{esbmc}-\gls{ai} framework~\cite{Tihanyi2025New} demonstrated that a fully automated pipeline combining \gls{esbmc} with an \gls{llm} achieves an \SI{80}{\percent} success rate in repairing buffer overflow and pointer dereference vulnerabilities across a corpus of \num{50000} C programs -- without human intervention at any stage. The authors have published no cost-per-fix figure, but the operational cost of the pipeline is bounded by API charges for \gls{llm} inference (typically fractions of a cent per small file) plus \gls{esbmc} verification time, suggesting per-defect infrastructure costs well under~\$1 for simple vulnerabilities. At the scale of the FormAI dataset's \num{112000} programs~\cite{tihanyi2023}, this represents a verification throughput that no human audit team could match at comparable cost.

\subsection{Summary of Economic Evidence}

Table~\ref{tab:economic_summary} consolidates the economic figures discussed in this section, organized into three distinct categories. \textbf{Section~A} lists confirmed funding directly attributable to \gls{esbmc} research; \textbf{Section~B} presents direct verification outputs produced by \gls{esbmc}; \textbf{Section~C} provides cost-of-failure context -- confirmed facts about the world that establish the financial stakes of the problem classes \gls{esbmc} addresses, but are not themselves evidence of \gls{esbmc}'s impact. Taken together, this evidence shows: \gls{esbmc} has attracted confirmed public funding totaling at least \pounds9.3~million and \euro{}4.98~million (see Table~\ref{tab:esbmc_grants}). This work produced a commercial spin-off (VeriBee), and researchers have applied it to artifacts that guard assets ranging from hundreds of billions of dollars in \gls{defi} capital to safety-critical firmware deployed in hundreds of millions of ARM devices. The economic case for its continued development rests not merely on the cost of running the tool -- essentially zero in licensing terms -- but on the multiplicative value of finding a defect before deployment rather than after: a ratio that IBM's historical data places at 60$\times$ or more~\cite{ibm_defect_cost}.

\begin{table}[ht]
    \centering
    \small
    \caption{Summary of economic evidence for \gls{esbmc}. \textbf{Section~A} lists confirmed funding attributable to \gls{esbmc}. \textbf{Section~B} lists direct verification results produced by \gls{esbmc}. \textbf{Section~C} provides cost-of-failure context: confirmed facts about the world that establish the financial stakes of the problem classes \gls{esbmc} targets; these are \emph{not} direct measures of \gls{esbmc}'s impact. All currency values are as reported in the source; GBP/EUR amounts reflect prevailing rates at the time of publication.}
    \label{tab:economic_summary}
    \begin{tabular}{l r l}
    \hline
    \textbf{Metric} & \textbf{Value} & \textbf{Source/Status} \\
    \hline
    \multicolumn{3}{l}{\textbf{(A) Grant Funding Attributable to \gls{esbmc}}} \\
    \hline
    SSVLab career research grants (self-reported aggregate) & $>$\$20M & Self-reported~\cite{ssvlab_cordeiro} \\
    EU H2020 ELEGANT grant & \euro{}4.98M & Confirmed~\cite{cordis957286} \\
    Soteria consortium grant (multi-partner; pro-rata not disclosed) & \pounds{}5.8M & Confirmed~\cite{ukri_soteria} \\
    Confirmed public funding (excl.\ undisclosed awards) & \pounds{}9.3M\,+\,\euro{}4.98M & See Table~\ref{tab:esbmc_grants} \\
    \hline
    \multicolumn{3}{l}{\textbf{(B) Direct Verification Results (\gls{esbmc} outputs)}} \\
    \hline
    \gls{esbmc}-\gls{ai} repair success (buffer overflow/pointer dereference) & up to \SI{80}{\percent} & Confirmed~\cite{Tihanyi2025New} \\
    FormAI dataset programs verified & \num{112000} & Confirmed~\cite{tihanyi2023} \\
    SpecVerify verification accuracy & 46.5\% & Proof-of-concept (see Section \ref{sec:specverify})~\cite{wang2025supporting} \\
    Ethereum max bug bounty (eth2spec finding) & \$1M & Confirmed~\cite{ethereum_bounty} \\
    \hline
    \multicolumn{3}{l}{\textbf{(C) Cost-of-Failure Context (world statistics; not \gls{esbmc}-specific)}} \\
    \hline
    US software defect cost (2002) & \$59.5B/yr & Cost-of-failure context~\cite{nist2002} \\
    Feasible saving via improved verification & \$22.2B/yr & Cost-of-failure context~\cite{nist2002} \\
    UK DSbD prog. (\gls{esbmc}-\gls{cheri} is one output) & $>$\$420M & Programme context~\cite{ukri_dsbd} \\
    \gls{defi} \gls{tvl} at risk (mid-2025) & \$140.7B & Cost-of-failure context~\cite{defillama} \\
    ETH staked on Beacon Chain & \$115B+ & Cost-of-failure context~\cite{theblock_staking} \\
    Top-100 \gls{defi} hacks total losses & \$10.77B & Cost-of-failure context~\cite{halborn2025} \\
    All crypto stolen in 2024 & \$2.2B & Cost-of-failure context~\cite{chainalysis2025} \\
    Boeing~737~MAX total exposure & \textasciitilde{}\$80B & Cost-of-failure context~\cite{boeing737max} \\
    CrowdStrike outage (Fortune~500) & \$5.4B & Cost-of-failure context~\cite{crowdstrike2024} \\
    IBM avg.\ breach cost (2024) & \$4.88M & Cost-of-failure context~\cite{ibm_breach2024} \\
    \gls{ai} automation breach saving & \$2.2M/org & Cost-of-failure context~\cite{ibm_breach2024} \\
    Avg.\ \gls{iot} device \glspl{cve} (Critical/High) & 473 & Cost-of-failure context~\cite{embedded_iot_cve} \\
    UK \gls{sme} annual cyberattack losses & \$4.28B & Gov.\ estimate~\cite{veribee_impact} \\
    Typical \gls{sme} incident cost & \$\num{7560} & Gov.\ estimate~\cite{veribee_impact} \\
    \gls{do-178c} DAL-A cost per SLOC & \textasciitilde{}\$100 & Industry estimate~\cite{eurocontrol_sloc} \\
    Smart contract audit cost & \$25k--\$100k & Industry~\cite{zealynx_audit_cost} \\
    \hline
    \end{tabular}
\end{table}

\section{Spin-offs, Technology Transfer, and Notable Case Studies}
\label{sec:spinoffs}

Beyond competition trophies and research publications, \gls{esbmc}'s real-world impact is best measured by two complementary dimensions: the derivative projects and institutions it has generated, and the concrete bugs it has found in production-grade software. This section documents both.

\subsection{Institutional Spin-offs and Technology Transfer}

\subsubsection{SSVLAB (UFAM, Brazil)} The most direct institutional spin-off of \gls{esbmc}'s founding team is the \gls{ssvlab}, established by Lucas Cordeiro at the \gls{ufam} following the 2009 \gls{ase} publication.  \gls{ssvlab} functions as an academic research group whose primary tool is \gls{esbmc}. Still, it also serves as a technology-transfer vehicle: the laboratory has collaborated with Brazilian government bodies and industry partners to formally verify embedded and real-time systems, disseminating \gls{esbmc} practices within the Brazilian software engineering community.  \gls{ssvlab} is the custodian of the \gls{esbmc} GitHub repository and the primary organizing unit behind most language front-end extensions (\gls{esbmc}-Solidity, \gls{esbmc}-Jimple, \gls{esbmc}-\gls{cheri}, \gls{esbmc}-Python, \gls{esbmc}-Rust).

\subsubsection{ARM Centre of Excellence in Formal Verification (University of Manchester)}

In 2021, Lucas Cordeiro was appointed Director of the ARM Centre of Excellence at the University of Manchester -- an industry-academia research center jointly funded by ARM Holdings and the University, focused on applying formal methods to processor architecture, firmware, and system software.  \gls{esbmc} is the center's primary model checking asset. This arrangement constitutes a structured technology transfer: ARM gains access to \gls{esbmc} capabilities, guided research priorities, and trained postdoctoral researchers; Manchester gains sustained industrial funding, access to proprietary benchmarks, and direct relevance to ARM's verification pipeline. The ARM Centre collaboration has produced research on verifying \gls{cheri}-enabled C programs (\gls{esbmc}-\gls{cheri}~\cite{bueno2022cheri}) and on neurosymbolic verification architectures.

\subsubsection{ESBMC-AI Open-Source Project}
The \gls{esbmc}-\gls{ai} framework (\url{https://github.com/esbmc/esbmc-ai}) operates as a distinct open-source project layered on top of the core \gls{esbmc} engine, maintaining its own release cycle, issue tracker, and contributor community independently of the main \gls{esbmc} repository. Structurally it resembles an early-stage product rather than a research prototype: it provides a command-line interface designed for ease of integration into existing developer workflows, documentation aimed at practitioners rather than researchers, and a plug-in architecture that allows the underlying \gls{llm} provider -- \gls{gpt-4}, Claude, or open-source models such as LLaMA and Mistral -- to be swapped without modifying the verification pipeline itself. 

As of 2025, it has attracted external contributors beyond the core \gls{esbmc} team, including industry security engineers who have contributed bug fixes, provider integrations, and workflow extensions that reflect real-world deployment requirements rather than purely academic use cases. While no formal commercial entity has yet spun off around \gls{esbmc}-\gls{ai}, the project represents the clearest and most mature candidate for future commercialization within the \gls{esbmc} ecosystem: its value proposition -- automated, bounded-formally-verified security hardening (violations confirmed within the \gls{bmc} verification depth) integrated directly into \gls{cicd} pipelines -- aligns well with enterprise DevSecOps workflows, where the demand for shift-left security tooling with formal guarantees is growing rapidly across regulated industries.

\subsubsection{Rust Foundation Membership}
In 2024 \gls{esbmc} joined the Rust Foundation's formal verification initiative~\cite{rustfoundation2024}, gaining formal membership and a seat in the Foundation's technical working groups dedicated to advancing the safety and correctness guarantees available to Rust developers. This positions \gls{esbmc} within the governance structure of the Rust language ecosystem in a manner that goes beyond a mere technical integration: membership carries both privileged research access -- to the standard library test suite, the language specification, and pre-publication discussions of language evolution -- and a de facto institutional endorsement that strengthens \gls{esbmc}'s credibility for commercial Rust verification engagements in competitive procurement contexts. 

The collaboration specifically targets the integration of \gls{esbmc} as an alternative backend for Kani, the CProver-based Rust model checker maintained by Amazon Web Services, opening a concrete and well-supported pathway to adoption by Rust-using organizations in safety-critical sectors, including automotive, aerospace, and cloud infrastructure, where Rust's memory safety guarantees are increasingly mandated or strongly preferred by regulators and customers. Should this integration mature, \gls{esbmc} would gain access to Kani's existing user base and validation corpus, substantially accelerating its route to production deployment in the Rust ecosystem.

\subsubsection{SpecVerify -- University of Manchester/Lockheed Martin Collaboration}
The SpecVerify project~\cite{wang2025supporting}, which combines Claude~3.5 Sonnet with \gls{esbmc} for cyber-physical system specification verification, was developed through a direct research partnership with Lockheed Martin, one of the world's largest defense and aerospace contractors. Lockheed's active involvement in the project -- contributing real system specifications and domain expertise rather than merely lending its name to a publication -- indicates genuine institutional interest in evaluating \gls{esbmc}-based verification pipelines within defense and aerospace development contexts, given the stringent certification requirements imposed by standards such as \gls{do-178c} and MIL-STD-882 in those sectors. 

SpecVerify is a documented instance of a defense industrial partner evaluating \gls{esbmc} against operational specifications, providing useful evidence of the tool's applicability beyond academic benchmarks. However, any longer-term deployment or procurement would require substantially more extensive independent validation.

\subsection{Notable Case Studies: ESBMC Finding Real Bugs}
\label{sec:casestudies}

The following case studies move beyond competition benchmarks to document \gls{esbmc}'s effectiveness on production-grade or high-profile artifacts.

\subsubsection{ECS -- Python}
Perhaps the most high-profile bug-finding result in \gls{esbmc}'s history came from \gls{esbmc}-Python~\cite{farias2024}, which the authors reported at \gls{issta}~2024. The \gls{ecs} -- the normative Python reference implementation of the Ethereum proof-of-stake consensus layer -- is one of the most widely scrutinized pieces of open-source infrastructure in the blockchain ecosystem, maintained by the Ethereum Foundation and reviewed by dozens of security engineers.  \gls{esbmc}-Python successfully identified previously unreported defects in this specification. Because the Ethereum consensus layer underpins a network handling hundreds of billions of dollars in assets, the correctness of the specification has significant economic and security implications. The result demonstrated that \gls{bmc}, when applied through an appropriate language front-end, can detect bugs that slip past intensive manual review and conventional testing.

\subsubsection{DeFi Smart Contract Vulnerabilities -- Solidity}
The \gls{esbmc}-Solidity evaluation~\cite{sloan2022} applied the tool to a benchmark suite of Solidity smart contracts drawn from known vulnerable \gls{defi} protocols. \gls{esbmc}-Solidity detected all vulnerabilities in the benchmark set -- including arithmetic overflow, reentrancy patterns, and array out-of-bounds accesses -- with zero false positives on the evaluated cases. Reentrancy is the vulnerability class responsible for the 2016 DAO hack (approximately \$60M at the time), making its reliable automated detection a commercially significant capability. While the benchmark was curated rather than representing a fresh audit of unknown contracts, the zero-false-positive result at complete detection is notably strong for a formal verification tool on a language with complex execution semantics.

\subsubsection{Cyber-Physical Systems at Lockheed Martin -- SpecVerify}
The SpecVerify evaluation~\cite{wang2025supporting} applied the combined \gls{llm}\,+\,\gls{esbmc} pipeline to nine cyber-physical system specifications drawn from realistic defense and aerospace scenarios. The system achieved \SI{46.5}{\percent} verification accuracy -- defined as the proportion of specifications correctly verified or refuted -- while producing fewer false positives than \gls{nasa}'s CoCoSim, the baseline comparator. While \SI{46.5}{\percent} might appear modest, we must contextualize it: formal verification of cyber-physical systems with natural-language-adjacent specifications is an extremely hard problem, and the SpecVerify result represents one of the first demonstrations that a formal verifier can usefully integrate into an industrial Cyber-Physical System specification workflow at all.

Nevertheless, the limitations of this result require explicit acknowledgment. The evaluation corpus of nine systems is small by the standards of empirical software engineering, and its restriction to a single industrial context -- defense and aerospace scenarios at Lockheed Martin -- limits generalisability to other Cyber-Physical System domains, specification conventions, and safety standards. The accuracy metric itself warrants scrutiny: aggregating correct outcomes across requirements of heterogeneous complexity and criticality obscures the asymmetric cost structure of safety-critical verification, where a missed violation carries categorically greater consequence than a spurious alarm. 

The \gls{llm} component additionally introduces non-determinism that may affect reproducibility across model versions, prompt formulations, and deployment environments, a concern that the current evaluation does not systematically address. Reaching production-grade performance would require, at minimum: a substantially larger and more diverse benchmark suite spanning recognized safety standards such as \gls{do-178c}, \gls{iec}~61508, and \gls{iso}~26262; per-category accuracy reporting disaggregated by requirement type and criticality level; human-in-the-loop validation of \gls{llm}-generated specifications before formal checking; and rigorous ablation studies isolating the contribution of each pipeline component. Until we have such evidence, we should best understand the \SI{46.5}{\percent} result as an initial feasibility demonstration -- evidence that the \gls{llm}+formal-verifier architecture is viable -- rather than a deployment-ready performance baseline.

\subsubsection{Self-Healing Software Demonstration -- Security Defects in C}
The self-healing software paper~\cite{Tihanyi2025New}, presented at \gls{ast}~2025, constructed a controlled but realistic experiment: a corpus of C programs with deliberately introduced security-relevant defects (buffer overflows, null pointer dereferences, integer overflows, format string vulnerabilities).  \gls{esbmc} first detected the defects formally, then the \gls{llm} repair loop proposed patches, and \gls{esbmc} re-verified the results. The combined pipeline successfully repaired up to \SI{80}{\percent} of buffer-overflow and pointer-dereference defects without human intervention and, critically, without introducing new formal violations detectable by \gls{esbmc}. The repair success rate on buffer overflows and pointer defects is particularly notable because these are the vulnerability classes most commonly exploited in memory-unsafe C code. The study provides empirical support for the claim that \gls{esbmc}-\gls{ai} is not merely a research curiosity but a plausible automated security-hardening tool.

\subsubsection{LLM-Generated Code Security Assessment -- FormAI Dataset}
The FormAI dataset study~\cite{tihanyi2023} used \gls{esbmc} as a large-scale oracle to formally assess the security properties of code generated by seven different \glspl{llm} across \num{112000} programs. The key finding was that the majority of \gls{llm}-generated C programs contained at least one formally verifiable safety violation, with significant variation across model families and prompt strategies. This large-scale bug-finding exercise constituted a direct use of \gls{esbmc} as a bug-finding tool at an industrial scale: running formal verification across a dataset orders of magnitude larger than typical research benchmarks. The study provided one of the first formally grounded, large-scale characterizations of \gls{llm} code-generation quality, influencing subsequent work on code-generation model evaluation and reinforcement learning from formal feedback.

\subsubsection{Embedded and Firmware Verification -- Industrial Collaborations}
\gls{esbmc}'s original design target -- embedded ANSI-C software -- has generated a sustained stream of industrial engagements that are less publicly documented than the above, partly for confidentiality reasons. Academic-industrial collaborations facilitated by \gls{ssvlab} and the ARM Centre of Excellence have used the tool to verify firmware for automotive microcontrollers, \gls{iot} device drivers, and safety-critical controllers. Common bug classes include arithmetic overflow in fixed-point signal-processing routines, buffer overruns in packet parsers, and data races in interrupt-handler/main-loop shared-memory patterns. These engagements inform \gls{esbmc}'s operational models, library coverage, and performance-tuning priorities -- the bug classes that recur in industrial code drive the verifier's development roadmap more directly than any competition category.

\section{Challenges and Future Trends}
\label{sec:challenges}

\subsection{Scalability to Large Codebases}

\textbf{Scalability to large, real-world codebases remains a fundamental challenge} due to the state-space explosion problem~\cite{Clarke2012}: formula size grows with program complexity, loop bounds, and thread count. Several mitigation strategies are already deployed in \gls{esbmc} or are natural near-term extensions. \textbf{Incremental \gls{smt} solving} (Section~\ref{sec:incremental}), available via \texttt{--incremental-bmc}, converts the quadratic re-solve cost of Na\"{i}ve depth-stepping into a linear one~\cite{Gadelha2018ESBMC}, and is the first line of defense for programs requiring deep unrolling. \textbf{Compositional verification} decomposes a program into modules verified independently with assume-guarantee contracts~\cite{cordeiro2011icse}, reducing each monolithic formula to smaller, tractable subproblems. \textbf{Abstract interpretation} frameworks (e.g.,~\gls{ikos}, Facebook Infer) supply pre-verified loop summaries to the \gls{bmc} engine, reducing unrolling depth. \textbf{Portfolio-solving} strategies predict which solver will handle a formula fastest, leveraging the competitive \gls{smt} landscape~\cite{svcomp2024,ssvlab2024}. \glspl{llm} can also suggest program decompositions and contract candidates~\cite{beckert2024towards}, lowering the manual effort required for compositional verification at an industrial scale.

\subsection{Neurosymbolic Verification}
\label{sec:neurosymbolic}

\gls{esbmc}-\gls{ai} instantiates \textbf{a broader neurosymbolic vision}: combining neural pattern recognition with the formal guarantees of symbolic reasoning. The paradigm promises verification workflows that are more accessible to non-expert developers and more trustworthy than purely \gls{llm}-driven approaches. Realizing this promise requires resolving several fundamental open problems.

The most urgent is \textbf{reproducibility under \gls{llm} non-determinism}. The same prompt, at the same temperature, can produce a correct repair on one run and a subtly incorrect one on another~\cite{Huang2023}. In safety-critical contexts, where reproducibility is a regulatory requirement, this non-determinism is a structural obstacle. Current mitigations -- multiple-sample aggregation and formal filtering layers that reject proposals failing \gls{esbmc}'s harness~\cite{Huang2023} -- add latency without fully resolving the problem. The core research question is: which architectural and training-time interventions make \gls{llm} outputs reproducible with a specified confidence level, and how should that level relate to the system's safety integrity level during verification?

The second open problem is \textbf{completeness in \gls{llm}-generated specifications}. Translating informal requirements into formal assertions that \gls{esbmc} can check remains largely unsolved~\cite{Wen2024}. Current models produce syntactically plausible assertions, but their completeness -- capturing all intended violations, not just obvious ones -- is hard to guarantee. An incomplete specification provides false assurance: the verification result is technically valid while substantively incomplete. Progress is likely to require collaboration between the natural language processing and formal methods communities, drawing on requirement taxonomies from \gls{do-178c} and \gls{iso}~26262.

The third open problem is \textbf{fine-tuning on formal verification corpora}. The dominant paradigm remains prompting-only: we expect a general-purpose model to generalize from its pretraining distribution~\cite{wang2025supporting}. Datasets such as FormAI~\cite{tihanyi2023} -- pairing C programs with formal vulnerability annotations -- provide a foundation for supervised fine-tuning on verification-specific tasks. Key questions are whether fine-tuned models generalize across languages and formalisms, whether fine-tuning improves worst-case reliability, and whether their gains complement retrieval-augmented generation over verified code corpora.

A fourth open problem is \textbf{the explanation and accessibility of counterexamples}~\cite{Wei2022}. \gls{esbmc} produces precise but terse traces that require expertise in formal methods to interpret; this gap is a significant barrier to adoption. \gls{llm}-mediated translation into natural language is promising, but faithfulness is the key challenge: a fluent but inaccurate explanation may misdirect remediation, introducing new defects. Structured output formats and retrieval-augmented pipelines that anchor \gls{llm} outputs to specific trace elements are promising directions, but the systematic evaluation of their faithfulness remains an open question.

\subsection{Expanding Language Coverage}
\label{sec:lang-coverage}

While \gls{esbmc} currently supports nine language front-ends (ANSI-C, C\texttt{++03}, \gls{cuda}, Solidity, Kotlin, \gls{cheri}~C, C\texttt{++11+}, Python, and Rust -- where C\texttt{++11+} is counted as a distinct modernised front-end relative to C\texttt{++03}, though both target the same language family; treating them as one gives eight distinct languages), \textbf{high-value gaps remain} across several target-language categories, each with its own verification motivation and integration challenges.

The most immediate gap is \textbf{mainstream general-purpose languages}. JavaScript/TypeScript and Java dominate enterprise and web development, yet their absence leaves Android applications, server-side microservices, and front-end logic inaccessible to \gls{esbmc}-based verification. The \gls{esbmc}-Jimple frontend -- which targets Kotlin via the Jimple \gls{ir}, the standard representation for Java bytecode in the Soot framework -- provides a template, making full Java coverage a near-term, tractable objective.

\textbf{WebAssembly is an increasingly urgent verification target}. Originally a portable web compilation target, it now runs in embedded firmware, on smart contract platforms (as an alternative to the \gls{evm}), in serverless functions, and in plugin sandboxes. Its sandboxed memory model and typed instruction set favor formal analysis. Since C, C\texttt{++}, and Rust -- all supported by \gls{esbmc} -- compile to WebAssembly, a frontend could reuse existing verification machinery. Verifying at the WebAssembly level is especially valuable when the compilation toolchain is untrusted or when deployment-time optimizations introduce behaviors absent from the source.

\textbf{Hardware description languages are another critical gap}. \gls{esbmc}-\gls{cheri} showed that \gls{esbmc} can encode hardware architectural constraints as \gls{smt} verification conditions. This methodology extends naturally to \gls{vhdl} and SystemVerilog, the dominant languages for \gls{fpga} and \gls{asic} design, enabling co-verification with the software layers that drive them. Such hardware-software co-verification is increasingly demanded, with \gls{do-254} governing airborne electronic hardware and \gls{do-178c} governing software.

\textbf{Formal annotation languages such as JML and ACSL} also fall outside \gls{esbmc}'s current scope. Both annotate Java and C programs with pre- and post-conditions and invariants. Integrating \gls{esbmc} as a backend checker for these annotations would position it within established workflows for high-assurance avionics and medical device software.

The most structurally novel gap is \textbf{polyglot cross-language contract verification}. Modern safety-critical architectures combine a Python orchestration layer, a Rust systems library, and a C firmware component -- each verified in isolation. No current \gls{esbmc} mechanism checks interface contracts across languages, yet a property violation may only manifest when components are composed. This verification gap requires a unified semantic model that covers multiple type systems, memory models, and concurrency semantics, along with a compositional strategy for assigning blame to the correct component. The core research question is: which intermediate representation would enable \gls{esbmc} to verify cross-language contracts and handle semantic mismatches -- such as Python's dynamic typing meeting Rust's ownership model -- at language boundaries?

\subsection{AI Agent Architectures}

Research is advancing toward \textbf{autonomous verification agents} that orchestrate static analysis, fuzzing, symbolic execution, and formal verification within a workflow guided by an \gls{ai} planner. \gls{esbmc}'s programmatic interface, broad language coverage, and rich output -- counterexamples, proofs, and coverage reports -- position it well as a formal verification oracle. This alignment with the emerging autonomous-agent paradigm reflects a broader shift: rather than treating formal verification as a standalone expert-operated tool, the emerging paradigm embeds verifiers as callable, trustworthy components within larger autonomous pipelines.

The typical architecture involves \textbf{an \gls{llm}-based planner} that decomposes a verification goal into subtasks, selects tools, interprets outputs, and iterates until it finds a proof, finds a counterexample, or exhausts resources. \gls{esbmc} occupies a privileged role in this template: unlike static analyzers or fuzzers, its outputs carry formal semantics -- a counterexample witnesses a genuine reachable violation, and a successful run provides a bounded correctness guarantee. This precision makes \gls{esbmc} a trustworthy oracle rather than a heuristic signal source.

\textbf{Concrete instantiations have already emerged}. The \gls{esbmc}-\gls{llm} framework~\cite{Tihanyi2025New} demonstrated that an \gls{llm} can consume \gls{esbmc} counterexamples and generate targeted repairs, closing the loop between verification failure and automated remediation. SpecVerify~\cite{wang2025supporting} inserts an \gls{llm}-based natural language understanding stage upstream of formal checking. Together, these illustrate a convergence: \gls{esbmc} is evolving from an expert-operated tool into a verification service that autonomous agents invoke programmatically.

\textbf{Researchers must address several challenges} before these architectures reach production maturity. Planners must reason reliably about the scope and soundness of \gls{esbmc}'s guarantees, distinguishing when bounded results are sufficient from when unbounded reasoning is required. We must manage \gls{llm}-based non-determinism to ensure reproducible pipeline outputs. Tool interfaces must be standardized so that agents can invoke \gls{esbmc} alongside AFL++, Infer, and KLEE without bespoke integration. Trust calibration is also open: over-reliance on bounded guarantees risks missing properties that require stronger methods, while under-reliance wastes resources. These challenges demand new theoretical frameworks for composing heterogeneous verification tools within an autonomous planner.

\subsection{Compliance and Certification}

Safety standards -- \gls{do-178c}, \gls{iso}~26262, and \gls{iec}~61508 -- increasingly recognize \textbf{formal verification as a compliance means}~\cite{Cofer2018,Hatcliff2012,Luckcuck2019}. \gls{esbmc}'s \gls{cfg}-based branch coverage instrumentation~\cite{menezes2025fase} directly targets the structural coverage criteria (\gls{mcdc}, decision coverage) these standards demand. Substantial work remains across five areas.

The first is \textbf{traceable proof artifacts}. We must link machine-readable correctness witnesses~\cite{svcomp2024} to specific requirements in auditor-acceptable form. Currently, bridging a formal verification result to a certification evidence package requires significant expert effort. Closing this gap requires a shared ontology: a formally defined mapping between \gls{esbmc}'s artifacts -- counterexample traces, proof certificates, coverage reports -- and the evidence categories recognized by each standard.

The second is \textbf{assurance-case construction}. Auditors consume structured safety cases, not raw verification results. The dominant notations are the \gls{gsn} and \gls{aadl} safety annexes, both machine-readable and, in principle, auto-populatable from verification evidence. Linking \gls{esbmc}'s witnesses to \gls{gsn} goal nodes or \gls{aadl} elements would automatically flow results into auditor-facing structures, maintaining a live traceability chain. This approach is technically feasible with tooling such as ACE and AdvoCATE, but requires \gls{esbmc} to emit structured, schema-conformant artifacts.

The third is \textbf{programmable logic certification under \gls{do-254}}. \gls{esbmc}-\gls{cheri} demonstrated that \gls{esbmc} can encode hardware architectural constraints as \gls{smt} verification conditions. \gls{do-254} governs \gls{fpga} and \gls{asic} development in avionics, demanding design assurance comparable to \gls{do-178c} for software, yet formal verification tooling for hardware description languages remains immature. Extending \gls{esbmc} to \gls{vhdl} and SystemVerilog would address a significant unmet need and create a natural bridge to hardware-software co-verification workflows.

The fourth is \textbf{operational models for safety-critical libraries}. Developing and validating models for \gls{arinc} APEX, \gls{autosar} RTE, and POSIX real-time extensions -- enabling \gls{esbmc} to verify application code without access to library source code -- remain significant barriers. Models must be faithful enough to produce meaningful results yet abstract enough to remain \gls{smt}-tractable. Validating them against real implementations is itself a verification problem, and developing and maintaining them across platforms is a recurring cost the community has not found a scalable way to absorb.

The fifth and most significant is \textbf{tool qualification under \gls{do-330}}, the tool qualification supplement to \gls{do-178c}. For \gls{esbmc} to serve as a verification tool in a \gls{do-178c}-certified program -- rather than a development aid whose outputs require independent checking -- we must qualify it at a level commensurate with the system's design assurance level. \gls{do-330} qualification requires: a tool operational requirements document specifying \gls{esbmc}'s intended behavior and coverage boundaries; a tool verification plan; anomaly reporting and resolution processes; and a configuration management system that ensures developers use the qualified version in certified development.

For an actively developed research tool, \textbf{the tension between rapid feature evolution and \gls{do-330}'s configuration management discipline is acute}: each capability-changing release may require re-qualification. A pragmatic path mirrors the approach taken by AbsInt for ASTRÉE and AdaCore for \gls{gnat} Pro -- qualifying a stable, feature-frozen \gls{esbmc} configuration while continuing research development on an unqualified branch. Establishing this pathway requires sustained engagement with certification authorities and potentially a commercial entity -- such as an expanded VeriBee -- capable of maintaining a qualified configuration over a product lifetime measured in decades.

\subsection{Quantum Software and Hardware-Software Co-Verification}

\textbf{\gls{esbmc}'s \gls{smt}-based foundations are theoretically applicable to quantum circuit verification} as quantum computing matures. \gls{esbmc}-\gls{cheri} has already demonstrated the tool's capacity to reason about hardware-software interfaces, positioning it to support the integration of hardware abstraction models for heterogeneous \gls{soc}, \gls{fpga}, and accelerator verification. Both directions share a common challenge: they require reasoning beyond the classical sequential semantics for which \gls{bmc} was designed, demanding principled extensions to \gls{esbmc}'s core theory.

\textbf{Quantum software verification is becoming practically urgent}. IBM, Google, and IonQ make their platforms cloud-accessible, and developers write quantum algorithms in Qiskit, Cirq, and Q\#. These programs are not immune to defects: incorrect qubit initialization, erroneous gate sequences, and entanglement errors produce silently incorrect results that testing alone cannot catch, as quantum measurement is probabilistic and destructive. \gls{smt} theories for complex-valued linear algebra and tensor products are an active research area~\cite{Lewis2023}, and tools such as QMC and Quartz demonstrate that bounded model checking transfers to quantum circuits~\cite{Gay2008, Xu2022}. \gls{esbmc}'s solver-agnostic architecture, already supporting floating-point and bit-vector theories via pluggable backends, provides a natural integration point. A near-term trajectory could extend \gls{esbmc} with a quantum-circuit frontend (OpenQASM or Qiskit \gls{ir}) and a quantum-state-theory plugin, enabling bounded verification of unitarity, entanglement invariants, and measurement-outcome distributions.

\textbf{Hardware-software co-verification is an equally compelling extension}. Modern \gls{soc}s combine \gls{cpu} cores, \gls{gpu} accelerators, \gls{fpga} fabrics, and cryptographic coprocessors, each with distinct semantics and security boundaries. \gls{esbmc}-\gls{cheri}~\cite{bueno2022cheri} demonstrated feasibility by encoding \gls{cheri}'s capability-based memory protection model as \gls{smt} constraints, enabling reasoning about memory safety properties that depend on hardware-enforced invariants. The methodology extends to \gls{vhdl} and SystemVerilog designs co-verified with their software drivers, and to accelerator models such as \gls{cuda} -- already supported by \gls{esbmc}-\gls{gpu}~\cite{Monteiro2018} -- and emerging standards such as SYCL and oneAPI, with hardware fidelity models capturing memory hierarchy behavior and warp divergence.

Realizing either direction at production quality requires \textbf{sustained investment in three areas}. First, the \gls{esbmc} team must develop and integrate new \gls{smt} theories in collaboration with the \gls{smt} community. Second, frontend toolchains must translate quantum circuit representations and hardware description languages into \gls{esbmc}'s verification harness -- an effort comparable to \gls{esbmc}-Solidity or \gls{esbmc}-\gls{cheri}. Third, the community must establish benchmark suites and ground-truth corpora for empirical evaluation, which is a prerequisite for community validation through competitions such as \gls{svcomp}.

\subsection{Real-Time and Timing Constraint Verification at Scale}

\textbf{Correct timing is a first-class safety requirement} in embedded and avionics domains, yet it remains one of \gls{esbmc}'s least developed verification dimensions. \gls{esbmc} has demonstrated bounded model checking of timing constraints~\cite{Barreto2011}, establishing theoretical feasibility. The gap between this proof-of-concept and the demands of a \gls{do-178c} or \gls{do-254} certified program, however, is substantial.

\textbf{\gls{wcet} analysis} asks: what is the longest possible execution time over all inputs and paths? \gls{wcet} bounds are required inputs to schedulability analysis, and overly conservative estimates waste processor capacity and increase system cost. Current tools such as AbsInt's \gls{ait} and Rapita's RVT combine static analysis with hardware timing models but operate independently of functional verifiers. Integrating \gls{wcet} reasoning into \gls{esbmc}'s \gls{bmc} framework -- encoding timing models as \gls{smt} constraints alongside functional properties -- would enable simultaneous verification of correctness and timing, a combination no current tool provides.

\textbf{The challenge compounds for concurrent \gls{rtos} tasks}. A single task's \gls{wcet} does not determine system-level timing correctness; that requires reasoning about scheduling, preemption, resource contention, and priority inversion. \gls{esbmc}'s existing concurrency machinery provides a partial foundation, but encoding the scheduler's semantics within the verification model is a non-trivial scalability challenge. The key research question is: which \gls{smt} encoding of \gls{rtos} scheduling and hardware timing models would allow \gls{esbmc} to verify composed timing constraints with sufficient fidelity for \gls{do-178c} scheduling analysis? Addressing this would position \gls{esbmc} as a unified functional and temporal verification platform -- a capability no current open-source tool provides.

\subsection{Counterexample Intelligibility and Human Factors}

The economic argument for \gls{esbmc} depends on \textbf{the tool's ability to change developer behavior}. A counterexample that a developer cannot interpret within a reasonable time budget does not change behavior, regardless of its formal correctness. \gls{esbmc} can produce traces spanning hundreds of \gls{smt} assignment steps. For a formal methods expert, such a trace is informative. For a software engineer without formal methods training -- the population \gls{esbmc} must reach for industrial adoption -- it is effectively opaque.

\textbf{This gap is a primary barrier to industrial adoption}. Surveys consistently identify counterexample interpretation as the dominant friction point~\cite{Kaleeswaran2023, Christakis2016}, ahead of installation, performance, and licensing. The problem has three dimensions. Length and navigability: a 200-step trace presents a navigation problem before an interpretation problem; engineers need structured visualization to identify causally relevant steps without reading sequentially. Abstraction level: \gls{smt}-level traces expose bit-vector assignments and solver-introduced variables that obscure the source-level story. Domain relevance: an engineer on an \gls{autosar} component needs explanations in terms of runnables and ports, not raw memory operations.

Addressing these dimensions requires \textbf{both technical and empirical work}. Structured trace visualization tools -- analogous to interactive debuggers but driven by formal semantics -- could let engineers navigate traces at multiple abstraction levels, collapsing irrelevant steps. Domain-specific explanation templates could translate \gls{smt} assignments into vocabulary familiar to the target domain. \gls{llm}-mediated explanation is promising, but faithfulness must be a hard constraint: a fluent but inaccurate explanation misdirects remediation. On the empirical side, the community needs agreed usability metrics -- time-to-correct-interpretation, root-cause error rate, and fix success rate -- to evaluate and compare approaches.

\subsection{Polyglot and Cross-Language System Verification}
\label{sec:languages}

\gls{esbmc} supports nine languages across embedded, systems, smart contract, and scientific computing domains, but \textbf{the current architecture treats each frontend as an independent verification pathway}. It verifies a Python program as a Python program and a Rust program as a Rust program -- the two never interact. This independence is increasingly at odds with the requirements of modern safety-critical systems.

Contemporary architectures combine a Python orchestration layer, a Rust systems library, and a C firmware component -- each communicating through well-defined \glspl{api}. Each component may be individually correct, yet \textbf{the composition can violate a system-level safety property} if interface contracts are misspecified or silently violated. No current \gls{esbmc} mechanism addresses cross-language compositional verification, and the gap is not merely a frontend engineering problem.

\textbf{Verifying cross-language contracts requires solving three interrelated challenges}. First, a unified semantic model must represent multiple type systems, memory models, and concurrency semantics without collapsing verification-relevant distinctions: Python's dynamic typing, Rust's ownership model, and C's manual memory management make fundamentally different safety guarantees. Second, a compositional reasoning strategy must assign blame for violations to the correct component, enabling actionable counterexamples. Third, we must integrate interface specification languages -- drawing on session types or contract-based design frameworks -- into \gls{esbmc}'s harness. The core question is: which intermediate representation and compositional framework would enable sound verification of cross-language contracts, including semantic mismatches at \gls{ffi} boundaries?

\subsection{Reproducibility and LLM Non-Determinism in Hybrid Verification Pipelines}

\textbf{The reproducibility problem is a structural property of any hybrid verification pipeline} incorporating \gls{llm} components -- not specific to any particular \gls{esbmc} deployment. \gls{llm} outputs are non-deterministic: the same prompt, same model, and same configuration can produce different outputs due to temperature sampling, batching decisions, and silent model updates. This non-determinism poses a fundamental obstacle to industrial certification of hybrid pipelines.

Certification frameworks -- \gls{do-178c}, \gls{iso}~26262, \gls{iec}~61508 -- assume that \textbf{a verification tool produces the same outputs given the same inputs}. A correctness certificate is meaningful only if the tool can regenerate it on demand. A hybrid pipeline in which an \gls{llm} may produce a different repair, invariant, or specification on each invocation cannot satisfy this requirement without architectural constraints that the current \gls{esbmc}-\gls{ai} framework does not impose.

\textbf{The problem manifests in three forms}. Run-to-run non-determinism -- variation across invocations of the same model with the same prompt -- can be mitigated by setting the temperature to 0 and fixing the random seed, though infrastructure-level variation may persist. Version-to-version non-determinism -- introduced by model updates -- requires pinning to a specific version or maintaining a version-tagged corpus of results to detect regressions. Prompt-sensitivity non-determinism -- variation from minor prompt reformulations that preserve intent -- is the least tractable, reflecting genuine uncertainty in the model's generalization behavior. The core research question is what architectural constraints and formal filtering layers make a hybrid \gls{llm}+\gls{esbmc} pipeline reproducible to a specified confidence level, and how that level should relate to the system's safety integrity level.

\subsection{Open-Source Sustainability and Governance}

\gls{esbmc}'s \textbf{open-source model} has enabled distributed contributions from Manchester, \gls{ufam}, and an international community, facilitated adoption in benchmarking competitions, and provided the transparency required for independent validation. As \gls{esbmc} transitions toward industrial deployment -- with VeriBee at the forefront of commercialization -- governance and sustainability challenges warrant explicit acknowledgment.

The most immediate tension is \textbf{potential divergence between the open-source research branch and a commercial VeriBee branch}. Commercial deployment demands configuration stability and regression testing that a research cadence struggles to sustain. If the commercial branch incorporates proprietary extensions that developers do not contribute back, the result is fragmentation: academic contributors improve the branch, while industrial users do not deploy it, and vice versa. Managing this requires explicit governance decisions about branching strategy, contribution policy, and upstreaming conditions -- decisions the \gls{esbmc} community has not yet had to make explicitly.

\textbf{Contributor sustainability is a related concern}. \gls{esbmc}'s development has relied on PhD students and postdocs whose project funding bounds their tenure. Institutional knowledge of architectural decisions and solver subtleties is not fully documented, which represents a concentration risk. As scope expands to nine languages and multiple \gls{ai}-augmented pipelines, the maintenance surface grows faster than the contributor base -- a technical debt risk common in research software but rarely acknowledged in capability-focused surveys.

\textbf{\gls{api} stability is a third challenge}. The \gls{esbmc}-\gls{ai} framework, the Kani backend, and future \gls{cicd} integrations depend on the stability of \gls{esbmc}'s programmatic interface across releases. Research tools typically make no stability guarantees, and breaking changes impose significant downstream maintenance costs. Establishing a versioned \gls{api} with documented stability -- analogous to LLVM or Z3 -- would reduce integration friction but would require governance commitment and release engineering discipline that academic projects rarely sustain without dedicated funding. Addressing these sustainability challenges is not peripheral; it is a precondition for the sustained industrial adoption that the economic argument presupposes.

\section{Conclusion}
\label{sec:conclusion}

\gls{esbmc} has undergone substantial and sustained development over fifteen years: from a research prototype for verifying embedded \gls{ansi-c} programs using \gls{smt} solvers~\cite{Cordeiro2012} to a versatile, industrially relevant, \gls{ai}-integrated formal verification platform with confirmed deployments spanning embedded firmware, blockchain smart contracts, defense and aerospace cyber-physical systems, and autonomous software repair~\cite{Tihanyi2025New,wang2025supporting,sloan2022,farias2024}. Its core innovation -- the native use of \gls{smt} solving for program verification, as formalised in the extended \gls{ieee} \gls{tse} paper~\cite{Cordeiro2012} -- has proven durable and generative, enabling principled extensions to floating-point arithmetic~\cite{gadelha2017}, concurrency~\cite{cordeiro2011icse,menezes2025tacas}, and a nine-language verification portfolio covering ANSI-C, C\texttt{++03}, \gls{cuda}, Solidity, Kotlin, CHERI~C, C\texttt{++11+}, Python, and Rust~\cite{Monteiro2021,sloan2022,farias2024,rustfoundation2024}.

Five themes emerge from this survey as defining characteristics of \gls{esbmc}'s trajectory:

\begin{enumerate}
    \item \textbf{Theory-driven engineering:} Each major capability is grounded in well-understood theoretical results -- \textit{k}-induction~\cite{sheeran2000,Donaldson2011}, the \gls{dpll}(T) solver architecture~\cite{nieuwenhuis2006}, floating-point \gls{smt} theory~\cite{gadelha2017}, and context-bounded concurrency~\cite{cordeiro2011icse} -- providing a sound basis for extension and composability that has sustained relevance across three generations of hardware and language targets.

    \item \textbf{Competition as a quality driver:} Systematic participation in \gls{svcomp}~\cite{beyer2012,svcomp2024} and \gls{test-comp}, accumulating 43 awards (35 \gls{svcomp} medals + 8 \gls{test-comp} medals) under rigorous independent evaluation~\cite{beyer2012,svcomp2024,ssvlab2024}, has provided objective external feedback that has motivated successive algorithmic improvements -- from incremental \gls{smt} solving and interval analysis~\cite{menezes2024} to the concurrent scheduling advances of v7.7~\cite{menezes2025tacas,menezes2025fase} -- reflected in each successive release.

    \item \textbf{Language pluralism:} Rather than treating C as the definitive target, the \gls{esbmc} team has systematically extended the tool to nine languages by 2025, enabled by the common \gls{ir} architecture that decouples front-end language coverage from back-end verification power~\cite{Gadelha2018ESBMC}, and cemented by institutional partnerships including Rust Foundation membership~\cite{rustfoundation2024} and the Lockheed Martin collaboration~\cite{wang2025supporting}.

    \item \textbf{\gls{ai} integration without abandoning rigour:} The \gls{esbmc}-\gls{ai} programme~\cite{Tihanyi2025New,Pirzada2024LLM} integrates \glspl{llm} as hypothesis generators, repair proposers, loop invariant suppliers, and specification translators while retaining formal verification as the sole arbiter of correctness -- achieving up to \SI{80}{\percent} automated repair success on buffer-overflow and pointer-dereference categories, on security-critical C programs~\cite{Tihanyi2025New} and avoiding the pitfalls of \gls{ai}-only approaches~\cite{Huang2023,pearce2021asleep}.

    \item \textbf{Demonstrated economic and societal value:} Beyond technical metrics, \gls{esbmc} has generated measurable real-world impact: confirmed public funding of at least \textsterling9.3~million and \euro{}4.98~million across seven grants~\cite{ssvlab_cordeiro,cordis957286,ukri_soteria}, the VeriBee commercial spin-off~\cite{veribee_impact}, confirmed bug findings in the Ethereum Consensus Specification~\cite{farias2024} and \gls{defi} smart contracts~\cite{sloan2022}, and a defence industrial deployment at Lockheed Martin~\cite{wang2025supporting} -- establishing a precedent for university-developed formal verification tools achieving production-grade industrial adoption.
\end{enumerate}

A structured research agenda shapes the near-term future for \gls{esbmc}: resolving \gls{llm} non-determinism and reproducibility in hybrid pipelines~\cite{Huang2023}, improving counterexample intelligibility to reduce the interpretive burden on non-expert engineers~\cite{Kaleeswaran2023,Christakis2016}, enabling polyglot cross-language contract verification across heterogeneous system stacks, advancing real-time and \gls{wcet} constraint analysis for avionics and automotive targets~\cite{Barreto2011}, pursuing \gls{do-330} tool qualification and safety certification support for \gls{do-178c} and \gls{iso}~26262~\cite{Cofer2018,menezes2025fase}, and specialising \gls{llm} components through fine-tuning on formal verification corpora such as FormAI~\cite{tihanyi2023}. The long-term trajectory -- quantum software verification, universal hardware-software co-verification across heterogeneous \glspl{soc} and \glspl{fpga}, and certification-grade formal assurance at industrial scale~\cite{Woodcock2009} -- points to a tool that has established the institutional, technical, and commercial foundations to remain at the frontier of formal verification for decades to come.

\subsection*{Acknowledgments}
    The authors would like to express their gratitude to the Department of Computer Science at the University of Manchester (UoM) and the Systems and Software Security (S3) Research Group for their invaluable support, collaborative environment, and access to cutting-edge resources, which were instrumental in the success of this research. We conducted this work with partial funding from the Engineering and Physical Sciences Research Council (EPSRC) grants EP/T026995/1, EP/V000497/1, EP/X037290/1, and the Soteria project, awarded by the UK Research and Innovation under the Digital Security by Design (DSbD) Programme.

\subsection*{Conflict of Interest} 
Lucas Cordeiro is the founding developer and lead researcher of \gls{esbmc} and a co-founder of VeriBee Ltd. This commercial spin-off commercializes technology derived from the \gls{esbmc} research program. Pierre Dantas and Waldir Junior actively contribute to the \gls{esbmc} research program at their respective institutions. These relationships constitute direct conflicts of interest that readers should weigh when evaluating claims regarding \gls{esbmc}'s capabilities, competitive standing, and economic impact. Members of the \gls{esbmc} team co-author approximately \SI{23}{\percent} of references in this survey; as explained in Section~\ref{sec:stats}, this proportion is structurally unavoidable for a single-tool survey. The authors declare that no external funding specifically supported the preparation of this survey article, and that primary sources identified in the text provide all quantitative claims.

\bibliographystyle{unsrtnat}
\bibliography{references}




\end{document}

%% file: mypackages.tex
\usepackage{eurosym}               
\usepackage{siunitx}

\usepackage{graphicx}              
\usepackage{tikz}
\usetikzlibrary{
    positioning,
    shapes.geometric,
    shapes.misc,
    arrows.meta,
    calc,
    mindmap,
    shadows,
    fit,
    backgrounds,
    patterns
}

\usepackage{pgfplots}
\pgfplotsset{compat=1.18}

\usepackage{enumitem}              
\usepackage{setspace}

\usepackage[acronym, nogroupskip, nonumberlist, nopostdot]{glossaries}
\glsdisablehyper

\definecolor{reviewerblue}{RGB}{0,70,140}
\definecolor{commentgray}{RGB}{240,240,245}
\definecolor{responsegreen}{RGB}{0,100,60}

\definecolor{grpFormal}{RGB}{180,160,220}   
\definecolor{grpPlan}  {RGB}{140,210,185}   
\definecolor{grpAgent} {RGB}{250,185,130}   
\definecolor{grpRun}   {RGB}{250,215,100}   
\definecolor{grpBench} {RGB}{195,195,185}   

\definecolor{amber}{rgb}{0.75, 0.50, 0.00}
\definecolor{teal} {rgb}{0.00, 0.50, 0.50}

\definecolor{figviolet}{RGB}{180,160,220}   
\definecolor{figamber} {RGB}{250,185,100}   
\definecolor{figteal}  {RGB}{180,225,210}   
\definecolor{figblue}  {RGB}{180,210,240}   
\definecolor{figgray}  {RGB}{220,220,215}   

%% file: acronyms.tex
\newacronym{aadl}{AADL}{Architecture Analysis and Design Language}
\newacronym{acm}{ACM}{Association for Computing Machinery}
\newacronym{ai}{AI}{Artificial Intelligence}
\newacronym{ait}{AIT}{Timing Analyzer}
\newacronym{ansi-c}{ANSI-C}{American National Standards Institute C}
\newacronym{api}{API}{Application Programming Interface}
\newacronym{arinc}{ARINC}{Aeronautical Radio, Incorporated}
\newacronym{ase}{ASE}{Automated Software Engineering}
\newacronym{asic}{ASIC}{Application-Specific Integrated Circuit}
\newacronym{ast}{AST}{Abstract Syntax Tree}
\newacronym{autosar}{AUTOSAR}{AUTomotive Open System ARchitecture}
\newacronym{bdd}{BDD}{Binary Decision Diagrams}
\newacronym{bmc}{BMC}{Bounded Model Checking}
\newacronym{cbmc}{CBMC}{Bounded Model Checking for ANSI-C Programs}
\newacronym{cca}{CCA}{Confidential Compute Architecture}
\newacronym{cegar}{CEGAR}{Counterexample-Guided Abstraction Refinement}
\newacronym{cfg}{CFG}{Control Flow Graph}
\newacronym{chain-of-thought}{CoT}{Chain-of-Thought}
\newacronym{chc}{CHC}{Constrained Horn Clause}
\newacronym{cheri}{CHERI}{Capability Hardware Enhanced RISC Instructions}
\newacronym{cicd}{CI/CD}{Continuous Integration and Continuous Deployment}
\newacronym{cisq}{CISQ}{Consortium for Information and Software Quality}
\newacronym{cordis}{CORDIS}{Community Research and Development Information Service}
\newacronym{cpu}{CPU}{Central Processing Unit}
\newacronym{cuda}{CUDA}{Compute Unified Device Architecture}
\newacronym{cve}{CVE}{Common Vulnerability and Exposure}
\newacronym{darpa}{DARPA}{Defense Advanced Research Projects Agency}
\newacronym{dblp}{DBLP}{Digital Bibliography \& Library Project}
\newacronym{defi}{DeFi}{Decentralized Finance}
\newacronym{do-178c}{DO-178C}{Software Considerations in Airborne Systems and Equipment Certification}
\newacronym{do-254}{DO-254}{Design Assurance for Airborne Electronic Hardware}
\newacronym{do-330}{DO-330}{Software Tool Qualification Considerations}
\newacronym{dpll}{DPLL}{Davis-Putnam-Logemann-Loveland}
\newacronym{dsl}{DSL}{Domain-Specific Language}
\newacronym{ecs}{ECS}{Ethereum Consensus Specification}
\newacronym{eda}{EDA}{Electronic Design Automation}
\newacronym{esbmc}{ESBMC}{Efficient SMT-based Context-Bounded Model Checker}
\newacronym{epsrc}{EPSRC}{Engineering and Physical Sciences Research Council}
\newacronym{evm}{EVM}{Ethereum Virtual Machine}
\newacronym{fase}{FASE}{Fundamental Approaches to Software Engineering}
\newacronym{ffi}{FFI}{Foreign Function Interface}
\newacronym{fpga}{FPGA}{Field-Programmable Gate Array}
\newacronym{gdp}{GDP}{Gross Domestic Product}
\newacronym{gnat}{GNAT}{GNAT Ada compiler}
\newacronym{gpt-4}{GPT-4}{Generative Pre-trained Transformer 4}
\newacronym{gpu}{GPU}{graphical processing unit}
\newacronym{gsn}{GSN}{Goal Structuring Notation}
\newacronym{ic3}{IC3}{Incremental Construction of Inductive Clauses for Indubitable Correctness}
\newacronym{icse}{ICSE}{International Conference on Software Engineering}
\newacronym{iec}{IEC}{International Electrotechnical Commission}
\newacronym{ieee}{IEEE}{Institute of Electrical and Electronics Engineers}
\newacronym{ikos}{IKOS}{Inference Kernel for Open Static Analyzers}
\newacronym{iot}{IoT}{Internet of Things}
\newacronym{ir}{IR}{Intermediate Representation}
\newacronym{iso}{ISO}{International Organization for Standardization}
\newacronym{issta}{ISSTA}{International Symposium on Software Testing and Analysis}
\newacronym{jvm}{JVM}{Java Virtual Machine}
\newacronym{llm}{LLM}{Large Language Model}
\newacronym{mcas}{MCAS}{Maneuvering Characteristics Augmentation System}
\newacronym{mcdc}{MCDC}{Modified Condition/Decision Coverage}
\newacronym{mir}{MIR}{Mid-level Intermediate Representation}
\newacronym{musl}{MUSL}{musl C standard library}
\newacronym{nasa}{NASA}{National Aeronautics and Space Administration}
\newacronym{nhtsa}{NHTSA}{National Highway Traffic Safety Administration}
\newacronym{nist}{NIST}{National Institute of Standards and Technology}
\newacronym{pdr}{PDR}{Property Directed Reachability}
\newacronym{por}{POR}{Partial Order Reduction}
\newacronym{raii}{RAII}{Resource Acquisition Is Initialization}
\newacronym{rmm}{RMM}{Realm Management Monitor}
\newacronym{rtl}{RTL}{Register Transfer Level}
\newacronym{rtos}{RTOS}{Real-Time Operating System}
\newacronym{sac}{SAC}{Symposium on Applied Computing}
\newacronym{sat}{SAT}{Boolean Satisfiability}
\newacronym{sbseg}{SBSeg}{Brazilian Symposium on Cybersecurity}
\newacronym{sme}{SME}{Small and Medium-sized Enterprise}
\newacronym{smt}{SMT}{Satisfiability Modulo Theories}
\newacronym{smtlib2}{SMT-LIB}{Satisfiability Modulo Theories Library}
\newacronym{soc}{SoC}{System on Chip}
\newacronym{sri}{SRI}{Stanford Research Institute}
\newacronym{ssa}{SSA}{Static Single Assignment}
\newacronym{ssvlab}{SSVLab}{Systems and Software Verification LAB}
\newacronym{sttt}{STTT}{Software Tools for Technology Transfer}
\newacronym{svcomp}{SV-COMP}{Competition on Software Verification}
\newacronym{tacas}{TACAS}{Tools and Algorithms for the Construction and Analysis of Systems}
\newacronym{test-comp}{Test-Comp}{Competition on Software Testing}
\newacronym{tvl}{TVL}{Total Value lLocked}
\newacronym{tse}{TSE}{Transactions on Software Engineering}
\newacronym{ufam}{UFAM}{Federal University of Amazonas}
\newacronym{ukri}{UKRI}{UK Research and Innovation}
\newacronym{vhdl}{VHDL}{VHSIC Hardware Description Language}
\newacronym{wcet}{WCET}{Worst-Case Execution Time}

%% file: figures/tikz_prisma.tex

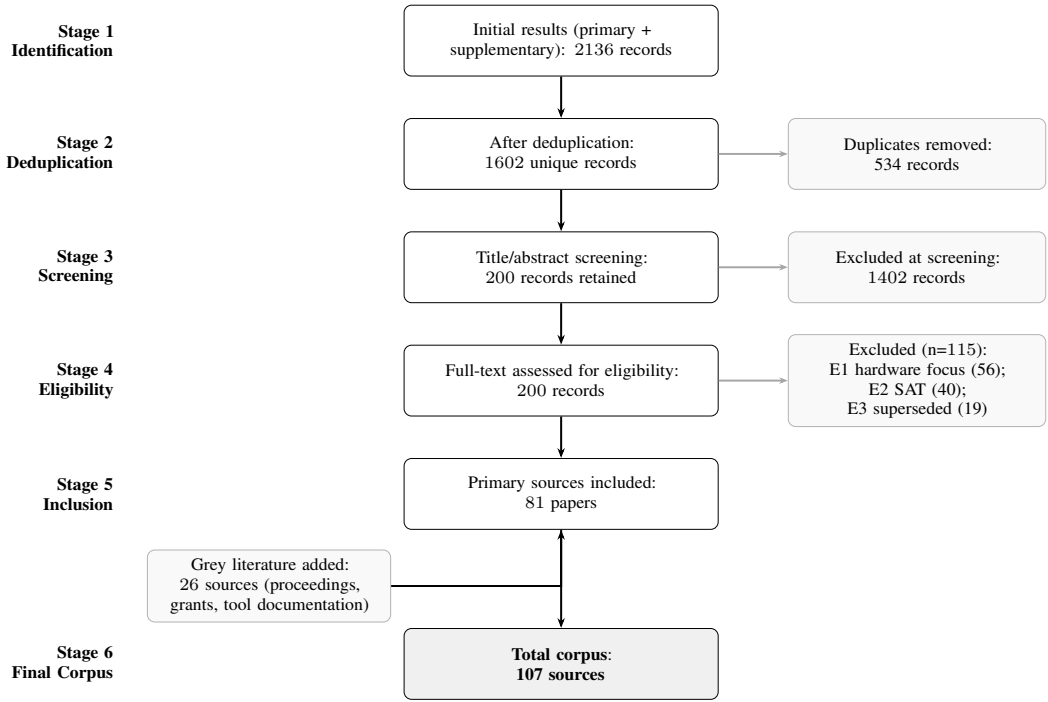
\begin{figure}[htbp]
\centering
\resizebox{0.85\linewidth}{!}{%
\begin{tikzpicture}[
  font=\small,
  mainbox/.style={
      rectangle, draw, rounded corners=3pt,
      text width=4.2cm, align=center,
      minimum height=1.0cm, fill=white,
      font=\scriptsize
  },
  excludebox/.style={
      rectangle, draw=gray!60, rounded corners=3pt,
      text width=3.4cm, align=center,
      minimum height=1.0cm, fill=gray!5,
      font=\scriptsize
  },
  greybox/.style={
      rectangle, draw=gray!60, rounded corners=3pt,
      text width=3.2cm, align=center,
      minimum height=1.0cm, fill=gray!5,
      font=\scriptsize
  },
  arr/.style={-{Stealth[length=4pt]}, thick},
  garr/.style={-{Stealth[length=3.5pt]}, gray!70, thick},
]

\node[mainbox] (initial) at (5.0,0)
  {Initial results (primary +\\supplementary): \num{2136} records};

\node[anchor=east, font=\scriptsize\bfseries, text width=1.6cm, align=right]
  at (-1.2, 0) {Stage 1\\Identification};

\node[mainbox] (dedup) at (5.0,-1.6)
  {After deduplication:\\\num{1602} unique records};

\node[anchor=east, font=\scriptsize\bfseries, text width=1.6cm, align=right]
  at (-1.2,-1.6) {Stage 2\\Deduplication};

\draw[arr] (initial.south) -- (dedup.north);

\node[excludebox, anchor=west] (excl1) at (8.2,-1.6)
  {Duplicates removed:\\534 records};
\draw[garr] (dedup.east) -- (excl1.west);

\node[mainbox] (screen) at (5.0,-3.2)
  {Title/abstract screening:\\\num{200} records retained};

\node[anchor=east, font=\scriptsize\bfseries, text width=1.6cm, align=right]
  at (-1.2,-3.2) {Stage 3\\Screening};

\draw[arr] (dedup.south) -- (screen.north);

\node[excludebox, anchor=west] (excl2) at (8.2,-3.2)
  {Excluded at screening:\\\num{1402} records};
\draw[garr] (screen.east) -- (excl2.west);

\node[mainbox] (fulltext) at (5.0,-4.8)
  {Full-text assessed for eligibility:\\\num{200} records};

\node[anchor=east, font=\scriptsize\bfseries, text width=1.6cm, align=right]
  at (-1.2,-4.8) {Stage 4\\Eligibility};

\draw[arr] (screen.south) -- (fulltext.north);

\node[excludebox, anchor=west] (excl3) at (8.2,-4.8)
  {Excluded (n=\num{115}):\\E1 hardware focus (56);\\ E2 SAT (40);\\ E3 superseded (19)};
\draw[garr] (fulltext.east) -- (excl3.west);

\node[mainbox] (primary) at (5.0,-6.4)
  {Primary sources included:\\\num{81} papers};

\node[anchor=east, font=\scriptsize\bfseries, text width=1.6cm, align=right]
  at (-1.2,-6.4) {Stage 5\\Inclusion};

\draw[arr] (fulltext.south) -- (primary.north);

\node[greybox, anchor=east] (grey) at (2.6,-7.7)
  {Grey literature added:\\\num{26} sources (proceedings,\\grants, tool documentation)};

\draw[arr] (grey.east) -| (primary.south);

\node[mainbox, fill=gray!12] (corpus) at (5.0,-8.8)
  {\textbf{Total corpus}:\\\textbf{107 sources}};

\node[anchor=east, font=\scriptsize\bfseries, text width=1.6cm, align=right]
  at (-1.2,-8.8) {Stage 6\\Final Corpus};

\draw[arr] (primary.south) -- ++(0,-0.5) -| (corpus.north);

\end{tikzpicture}
}
\caption{PRISMA-style search and selection flow. Combined primary and supplementary searches across arXiv, \gls{dblp}, \gls{ieee}~Xplore, \gls{acm}~Digital~Library, and SpringerLink returned \num{2136} candidate records; \num{1602} remained after cross-library deduplication; \num{200} advanced to full-text review; \num{81} were accepted as primary sources. We added a further \num{26} grey literature sources, bringing the final corpus to \num{107} sources.} 

\label{fig:prisma}
\end{figure}

%% file: figures/tikz_timeline.tex
\begin{figure}[htbp]
\centering
\resizebox{\linewidth}{!}{%
\begin{tikzpicture}[
    boxstyle/.style={
        draw, rectangle, rounded corners=2pt,
        font=\scriptsize, align=center,
        inner sep=4pt, line width=0.5pt,
        fill=white, text width=2.5cm
    },
    dot/.style={circle, fill=black, inner sep=0pt, minimum size=5pt},
]

\def\xA{0.0}   
\def\xB{1.5}   
\def\xC{3.0}   
\def\xD{4.5}   
\def\xE{6.0}   
\def\xF{7.5}  
\def\xG{9.0}  
\def\xH{10.5}  
\def\xI{12.0}  

\draw[line width=1.2pt, -latex] (-0.4, 0) -- (12.6, 0);

\foreach \xp in {\xA, \xB, \xC, \xD, \xE, \xF, \xG, \xH, \xI} {
    \node[dot] at (\xp, 0) {};
}


\draw[line width=0.5pt] (\xA, 0) -- (\xA, 1.0);
\node[boxstyle] at (\xA, 1.45)
    {ESBMC~1.0\\(ASE 2009)\\First SMT-native\\C verifier};
\node[font=\scriptsize, rotate=0, anchor=north east] at (\xA + 0.30, -0.15) {2009};

\draw[line width=0.5pt] (\xC, 0) -- (\xC, 1.0);
\node[boxstyle] at (\xC, 1.45)
    {ESBMC~1.22\\(TACAS 2014)\\Concurrency\\verification};
\node[font=\scriptsize, rotate=0, anchor=north east] at (\xC + 0.30, -0.15) {2014};

\draw[line width=0.5pt] (\xE, 0) -- (\xE, 1.0);
\node[boxstyle] at (\xE, 1.45)
    {ESBMC~6.1\\(STTT 2020)\\Test-case\\generation};
\node[font=\scriptsize, rotate=0, anchor=north east] at (\xE + 0.30, -0.15) {2020};

\draw[line width=0.5pt] (\xG, 0) -- (\xG, 1.0);
\node[boxstyle] at (\xG, 1.45)
    {ESBMC~7.3\\(TACAS 2023)\\Clang C\texttt{++}\\front-end};
\node[font=\scriptsize, rotate=0, anchor=north east] at (\xG + 0.30, -0.15) {2023};

\draw[line width=0.5pt] (\xI, 0) -- (\xI, 2.0);
\node[boxstyle] at (\xI, 1.45)
    {ESBMC~7.7\\(TACAS/FASE 2025)\\Concurrency \&\\branch coverage};
\node[font=\scriptsize, rotate=0, anchor=north east] at (\xI + 0.30, -0.15) {2025};


\draw[line width=0.5pt] (\xB, 0) -- (\xB, -1.0);
\node[boxstyle] at (\xB, -1.45)
    {ESBMC~1.17\\(SV-COMP 2012)\\Inaugural\\competition};
\node[font=\scriptsize, rotate=0, anchor=south west] at (\xB - 0.30, 0.15) {2012};

\draw[line width=0.5pt] (\xD, 0) -- (\xD, -1.0);
\node[boxstyle] at (\xD, -1.45)
    {ESBMC~5.0\\(ASE 2018)\\Full re-engineering;\\incremental BMC};
\node[font=\scriptsize, rotate=0, anchor=south west] at (\xD - 0.30, 0.15) {2018};

\draw[line width=0.5pt] (\xF, 0) -- (\xF, -1.0);
\node[boxstyle] at (\xF, -1.45)
    {ESBMC~7.x\\Solidity \&\\CHERI\\ support};
\node[font=\scriptsize, rotate=0, anchor=south west] at (\xF - 0.30, 0.15) {2022};

\draw[line width=0.5pt] (\xH, 0) -- (\xH, -1.0);
\node[boxstyle] at (\xH, -1.45)
    {ESBMC~7.4\\(TACAS 2024)\\Interval analysis;\\LLM integration};
\node[font=\scriptsize, rotate=0, anchor=south west] at (\xH - 0.30, 0.15) {2024};

\end{tikzpicture}%
}
\caption{Chronological milestones of \gls{esbmc} major releases from its first publication at \gls{ase}~2009 through version~7.7 in 2025. Labels alternate above and below the axis to prevent overlap. Each marker corresponds to a peer-reviewed venue or substantive feature release~\cite{Cordeiro2012,beyer2012,cordeiro2014,Gadelha2018ESBMC,gadelha2020,shmarov2023,menezes2024,menezes2025tacas,menezes2025fase}}

\label{fig:timeline}
\end{figure}
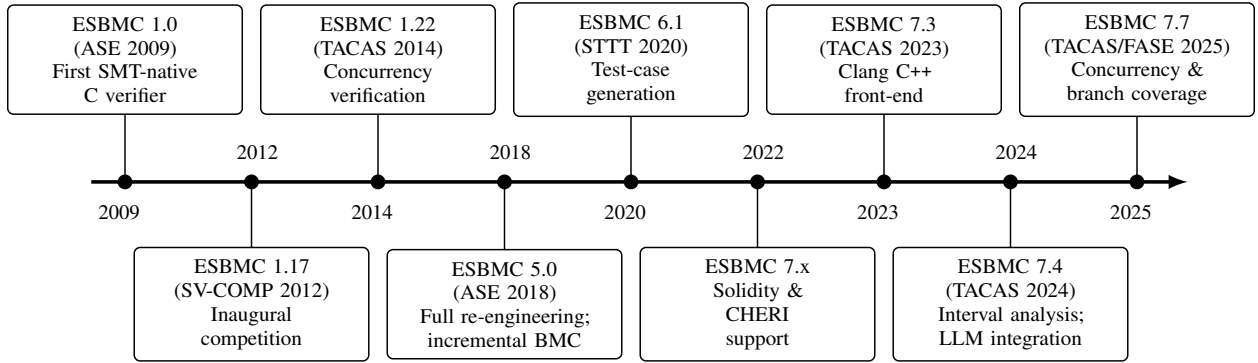

%% file: figures/tikz_smt_solver.tex
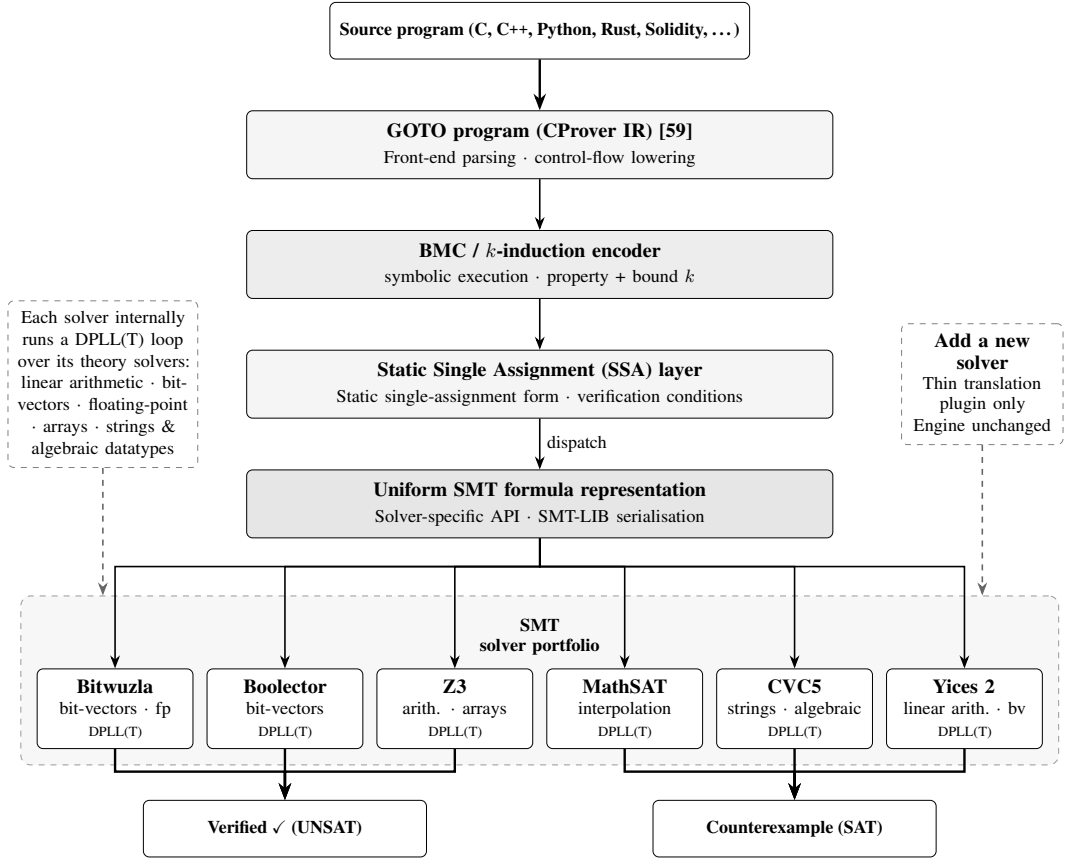
\begin{figure}[htbp]
\centering
\resizebox{0.85\textwidth}{!}{%
\begin{tikzpicture}[
    node distance=1.2cm and 1.5cm,
    box/.style={
        rectangle, draw=black, rounded corners=3pt,
        minimum width=2.4cm, minimum height=1.2cm,
        text centered, inner sep=5pt
    },
    ionode/.style={
        rectangle, draw=black, rounded corners=3pt,
        fill=white, text=black, font=\small\bfseries,
        minimum width=5.0cm, minimum height=1.0cm,
        text centered, inner sep=5pt
    },
    widenode/.style={
        rectangle, draw=black, rounded corners=3pt,
        minimum height=1.2cm, text centered, inner sep=5pt
    },
    subgroup/.style={
        draw=black!40, dashed, rounded corners=5pt, fill=black!3
    },
    arr/.style={thick, -Stealth},
    ioarr/.style={very thick, -Stealth},
    plug/.style={thick, dashed, -Stealth, draw=black!60},
]

\node (input) [ionode]
    {Source program (C, C\texttt{++}, Python, Rust, Solidity, \ldots)};

\node (goto) [widenode, fill=black!4, text width=10cm, below=0.9cm of input] {%
    \textbf{GOTO program (CProver \gls{ir})~\cite{menezes2025goto}}\\[2pt]
    {\small Front-end parsing $\cdot$ control-flow lowering}};

\node (enc) [widenode, fill=black!7, text width=10cm, below=0.9cm of goto] {%
    \textbf{\gls{bmc} / $k$-induction encoder}\\[2pt]
    {\small symbolic execution $\cdot$ property + bound $k$}};

\node (ssa) [widenode, fill=black!4, text width=10cm, below=0.9cm of enc] {%
    \textbf{\gls{ssa} layer}\\[2pt]
    {\small Static single-assignment form $\cdot$ verification conditions}};

\node (abst) [widenode, fill=black!10, text width=10cm, below=0.9cm of ssa] {%
    \textbf{Uniform \gls{smt} formula representation}\\[2pt]
    {\small Solver-specific \gls{api} $\cdot$ \gls{smtlib2} serialisation}};

\node (portfolio_label) [below=1.3cm of abst, font=\small\bfseries,
    align=center, text width=15cm, xshift=0cm]
    {\gls{smt}\\solver portfolio};

\node (bitwuzla) [box, fill=white, text width=2.4cm,
    below=2.3cm of abst, xshift=-7.5cm] {%
    \textbf{Bitwuzla}\\{\small bit-vectors $\cdot$ fp}\\{\scriptsize \gls{dpll}(T)}};
\node (boolector) [box, fill=white, text width=2.4cm,
    below=2.3cm of abst, xshift=-4.5cm] {%
    \textbf{Boolector}\\{\small bit-vectors}\\{\scriptsize \gls{dpll}(T)}};
\node (z3) [box, fill=white, text width=2.4cm,
    below=2.3cm of abst, xshift=-1.5cm] {%
    \textbf{Z3}\\{\small arith. $\cdot$ arrays}\\{\scriptsize \gls{dpll}(T)}};
\node (mathsat) [box, fill=white, text width=2.4cm,
    below=2.3cm of abst, xshift=1.5cm] {%
    \textbf{MathSAT}\\{\small interpolation}\\{\scriptsize \gls{dpll}(T)}};
\node (cvc5) [box, fill=white, text width=2.4cm,
    below=2.3cm of abst, xshift=4.5cm] {%
    \textbf{CVC5}\\{\small strings $\cdot$ algebraic}\\{\scriptsize \gls{dpll}(T)}};
\node (yices) [box, fill=white, text width=2.4cm,
    below=2.3cm of abst, xshift=7.5cm] {%
    \textbf{Yices~2}\\{\small linear arith.\ $\cdot$ bv}\\{\scriptsize \gls{dpll}(T)}};

\begin{scope}[on background layer]
\node[subgroup,
    fit=(bitwuzla)(boolector)(z3)(mathsat)(cvc5)(yices)(portfolio_label),
    inner sep=8pt] (portbox) {};
\end{scope}

\node (pluginL) [box, fill=white, draw=black!50, dashed,
    text width=3.0cm, left=0.85cm of ssa] {%
    {\small Each solver internally runs a \gls{dpll}(T) loop over its
    theory solvers: linear arithmetic $\cdot$ bit-vectors $\cdot$
    floating-point $\cdot$ arrays $\cdot$ strings \& algebraic datatypes}};
\draw[plug] (pluginL.south) -- (pluginL.south |- portbox.north);

\node (pluginR) [box, fill=white, draw=black!50, dashed,
    text width=2.5cm, right=1.2cm of ssa] {%
    \textbf{Add a new solver}\\
    {\small Thin translation plugin only\\Engine unchanged}};
\draw[plug] (pluginR.south) -- (pluginR.south |- portbox.north);

\node (out-safe) [ionode, below=0.9cm of z3, xshift=-3.0cm]
    {Verified \checkmark\ (UNSAT)};
\node (out-bug)  [ionode, below=0.9cm of mathsat, xshift=3.0cm]
    {Counterexample (SAT)};

\draw[ioarr] (input.south) -- (goto.north);
\draw[arr]   (goto.south)  -- (enc.north);
\draw[arr]   (enc.south)   -- (ssa.north);
\draw[arr]   (ssa.south)   -- node[right, font=\small] {dispatch} (abst.north);

\draw[arr] (abst.south) -- ++(0,-0.5) -| (bitwuzla.north);
\draw[arr] (abst.south) -- ++(0,-0.5) -| (boolector.north);
\draw[arr] (abst.south) -- ++(0,-0.5) -| (z3.north);
\draw[arr] (abst.south) -- ++(0,-0.5) -| (mathsat.north);
\draw[arr] (abst.south) -- ++(0,-0.5) -| (cvc5.north);
\draw[arr] (abst.south) -- ++(0,-0.5) -| (yices.north);

\draw[ioarr] (z3.south)        -- ++(0,-0.4) -| (out-safe.north);
\draw[ioarr] (bitwuzla.south)  -- ++(0,-0.4) -| (out-safe.north);
\draw[ioarr] (boolector.south) -- ++(0,-0.4) -| (out-safe.north);
\draw[ioarr] (yices.south)     -- ++(0,-0.4) -| (out-bug.north);
\draw[ioarr] (mathsat.south)   -- ++(0,-0.4) -| (out-bug.north);
\draw[ioarr] (cvc5.south)      -- ++(0,-0.4) -| (out-bug.north);

\end{tikzpicture}
}

\caption{Architecture of \gls{esbmc}'s \gls{smt} back-end. A source program in any supported language (C, C\texttt{++}, Python, Rust, Solidity, and others) is first lowered by its front-end to the common GOTO program \gls{ir}. The \gls{bmc}/$k$-induction encoder then unrolls the program for a bound~$k$ and produces a \gls{ssa}-form verification condition, which is encoded as a uniform \gls{smt} formula and dispatched to one solver of the portfolio. At the dispatch boundary, a thin translation plugin serializes the formula either through a solver-specific \gls{api} or to \gls{smtlib2}~\cite{Clarke2018Handbook}, allowing new solvers to be integrated without modifying the verification engine. The portfolio includes Z3~\cite{moura2008}, Bitwuzla~\cite{niemetz2023}, Boolector~\cite{Brummayer2009}, MathSAT~\cite{cimatti2013}, CVC5~\cite{barbosa2022}, and Yices~2~\cite{dutertre2014}; each solver internally runs a \gls{dpll}(T) loop over its own theory solvers for linear arithmetic, bit-vectors, floating-point, arrays, strings, and algebraic datatypes~\cite{nieuwenhuis2006}. \gls{esbmc} does not implement \gls{dpll}(T) itself. A UNSAT result yields a verification certificate; a SAT result yields an executable counterexample trace.}

\label{fig:smt-architecture}

\end{figure}

%% file: figures/tikz_k_induction.tex
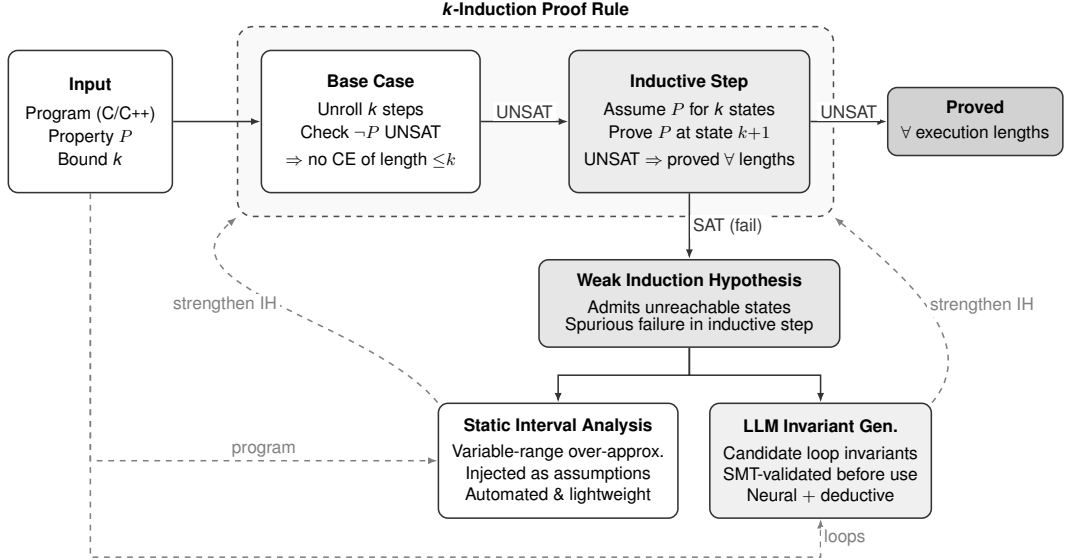
\begin{figure}[htbp]
  \centering
  \resizebox{0.85\textwidth}{!}{%
  \begin{tikzpicture}[
      every node/.style={font=\sffamily\small},
      block/.style={draw=black!80, thick, rounded corners=5pt, align=center, inner sep=7pt},
      arr/.style={-{Latex[length=5pt]}, thick, black!80},
      darr/.style={-{Latex[length=5pt]}, thick, black!50, dashed},
      lbl/.style={font=\tiny\sffamily, fill=white, inner sep=1.5pt},
  ]

    \node[block, fill=white, minimum width=3cm, minimum height=2.6cm] (input)
      {\textbf{Input}\\[5pt]
       \small Program (C/C\texttt{++})\\[2pt]
       \small Property $P$\\[2pt]
       \small Bound \textit{k}};

    \node[block, fill=white, minimum width=4cm, minimum height=2.6cm,
          right=1.6cm of input] (base)
      {\textbf{Base Case}\\[5pt]
       \small Unroll \textit{k} steps\\[2pt]
       \small Check $\neg P$ UNSAT\\[4pt]
       \small $\Rightarrow$ no CE of length ${\le}k$};

    \draw[arr] (input.east) -- (base.west);

    \node[block, fill=gray!15, minimum width=4.4cm, minimum height=2.6cm,
          right=1.6cm of base] (ind)
      {\textbf{Inductive Step}\\[5pt]
       \small Assume $P$ for \textit{k} states\\[2pt]
       \small Prove $P$ at state $k{+}1$\\[4pt]
       \small UNSAT $\Rightarrow$ proved $\forall$ lengths};

    \draw[arr] (base.east) -- node[lbl, above]{\footnotesize UNSAT} (ind.west);

    \node[block, fill=gray!35, minimum width=2.6cm,
          right=1.4cm of ind] (proved)
      {\small\textbf{Proved}\\[3pt]
       \small $\forall$ execution lengths};

    \draw[arr] (ind.east) -- node[lbl, above]{\footnotesize UNSAT} (proved.west);

    \begin{pgfonlayer}{background}
      \node[draw=black!70, dashed, thick, rounded corners=8pt, fill=gray!5,
            fit=(base)(ind), inner sep=12pt,
            label={[font=\small\sffamily\bfseries, yshift=2pt]above:%
              \textit{k}-Induction Proof Rule}] (kfr) {};
    \end{pgfonlayer}

    \node[block, fill=gray!20, minimum width=5.5cm, below=1.2cm of ind] (weak)
      {\textbf{Weak Induction Hypothesis}\\[4pt]
       \small Admits unreachable states\\[-1pt]
       \small Spurious failure in inductive step};

    \draw[arr] (ind.south) -- node[lbl, right]{\footnotesize SAT (fail)} (weak.north);

    \node[block, fill=white, minimum width=3.8cm, minimum height=2.1cm,
          below=1cm of weak, xshift=-2.4cm] (static)
      {\textbf{Static Interval Analysis}\\[4pt]
       \small Variable-range over-approx.\\[1pt]
       \small Injected as assumptions\\[1pt]
       \small Automated \& lightweight};

    \node[block, fill=gray!12, minimum width=3.8cm, minimum height=2.1cm,
          below=1cm of weak, xshift=2.4cm] (llm)
      {\textbf{LLM Invariant Gen.}\\[4pt]
       \small Candidate loop invariants\\[1pt]
       \small SMT-validated before use\\[1pt]
       \small Neural $+$ deductive};

    \draw[arr] (weak.south) -- ++(0,-0.5) -| (static.north);
    \draw[arr] (weak.south) -- ++(0,-0.5) -| (llm.north);

    \coordinate (junc) at ($(input.south) + (0,-1.0)$);

    \draw[darr, -] (input.south) -- (junc);

    \draw[darr] (junc) -- ++(0,0) |- node[lbl, near end, above]{\small program}
        (static.west);

    \draw[darr] (junc) -- ++(0,0) |- ($(llm.south)+(0,-0.7)$) --
    node[lbl, right]{\small loops} (llm.south);

    \draw[darr] (static.north west) to[out=130,in=220]
      node[lbl, left]{\small strengthen IH} (kfr.south west);
    \draw[darr] (llm.north east) to[out=50,in=320]
      node[lbl, right]{\small strengthen IH} (kfr.south east);

  \end{tikzpicture}
  }

\caption{Overview of \textit{k}-induction in \gls{esbmc}. The base case unrolls execution for \textit{k} steps and checks $\neg P$ is unsatisfiable, ruling out counterexamples of length ${\le}k$. The inductive step assumes $P$ holds for \textit{k} consecutive states and proves it at state $k{+}1$; UNSAT closes the proof for all lengths. When the inductive step fails (SAT), the hypothesis is too weak and is strengthened by one of two sources: static interval analysis, which automatically injects variable-range over-approximations as assumptions; or \gls{llm}-generated invariants, validated by \gls{smt} before use. Dashed arrows show that static analysis consumes the full program, while the \gls{llm} targets loop structures specifically}

\label{fig:kinduction}
\end{figure}

%% file: figures/tikz_evolution_qtd.tex
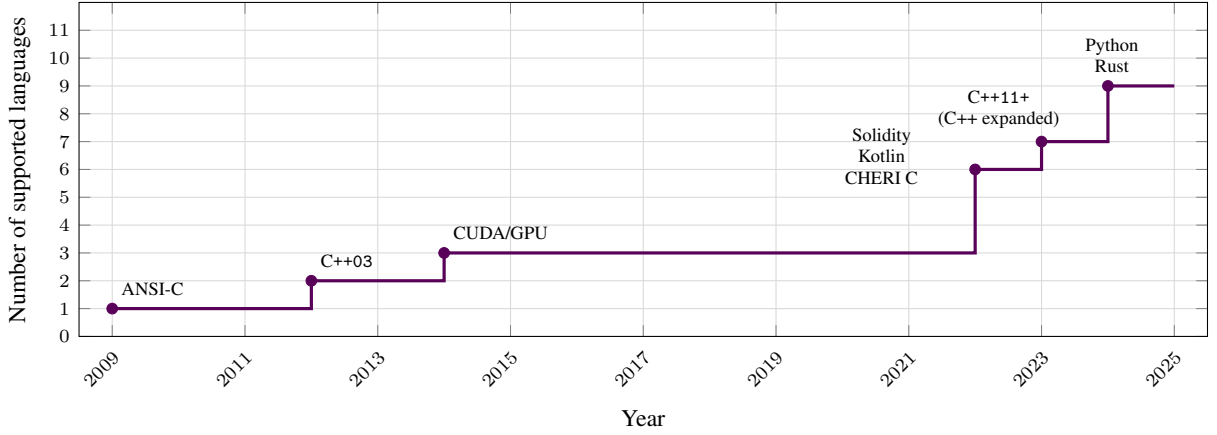
\begin{figure}[htbp]
\centering
\begin{tikzpicture}
\begin{axis}[
    width=\linewidth,
    height=6cm,
    xlabel={Year},
    ylabel={Number of supported languages},
    xmin=2008.5, xmax=2025.5,
    ymin=0, ymax=12,
    xtick={2009,2011,2013,2015,2017,2019,2021,2023,2025},
    ytick={0,1,...,11},
    scaled x ticks=false,          
    x tick label style={
        /pgf/number format/1000 sep={},  
        font=\scriptsize,
        rotate=45,
        anchor=north east,
    },
    y tick label style={font=\scriptsize},
    grid=both,
    grid style={line width=0.3pt, draw=gray!20},
    major grid style={line width=0.4pt, draw=gray!30},
    label style={font=\small},
    tick label style={font=\scriptsize},
]

\addplot[
    const plot,
    draw=violet!70!black,
    line width=1.2pt,
] coordinates {
    (2009, 1)
    (2012, 2)
    (2014, 3)
    (2022, 6)
    (2023, 7)
    (2024, 9)
    (2025, 9)
};

\addplot[
    only marks,
    mark=*,
    mark size=2pt,
    color=violet!70!black,
] coordinates {
    (2009, 1)
    (2012, 2)
    (2014, 3)
    (2022, 6)
    (2023, 7)
    (2024, 9)
};

\node[font=\scriptsize, anchor=south west] at (axis cs:2009.0, 1.15) {ANSI-C};
\node[font=\scriptsize, anchor=south west, align=center] 
    at (axis cs:2012.0, 2.15) 
    {C\texttt{++03}};
\node[font=\scriptsize, anchor=south west] at (axis cs:2014.0, 3.15) {CUDA/GPU};
\node[font=\scriptsize, anchor=south west, align=center] 
    at (axis cs:2019.9, 5.15) 
    {Solidity\\ Kotlin\\ \gls{cheri}~C};
\node[font=\scriptsize, anchor=south west, align=center]
    at (axis cs:2021.3, 7.15)
    {C\texttt{++11+}\\(C++ expanded)};
\node[font=\scriptsize, anchor=south east, align=center] 
    at (axis cs:2024.6, 9.15) 
    {Python\\ Rust};
\end{axis}
\end{tikzpicture}
\caption{Cumulative growth of language support in \gls{esbmc} (2009--2025). Each step corresponds to the introduction of a new front-end: Core (\gls{ansi-c}, 2009), CProver-based (C\texttt{++03}, 2012), \gls{esbmc}-\glsfirst{gpu} (\glsfirst{cuda}, 2014), \gls{esbmc}-Solidity, \gls{esbmc}-Jimple, and \gls{esbmc}-\gls{cheri} (Solidity, Kotlin, and \gls{cheri}~C, 2022), Clang~\gls{ast} (C\texttt{++11+}, 2023), and \gls{esbmc}-Python and \gls{esbmc}-Rust (Python and Rust, 2024). Note: the C\texttt{++11+} step (2023) represents a \emph{modernization} of the existing C\texttt{++} front-end via Clang rather than an entirely new language; counting it as a distinct step gives 9 supported language variants, while treating C\texttt{++03} and C\texttt{++11+} as a single C\texttt{++} front-end gives~8}

\label{fig:lang-growth}
\end{figure}

%% file: figures/tikz_cumulative_awards.tex
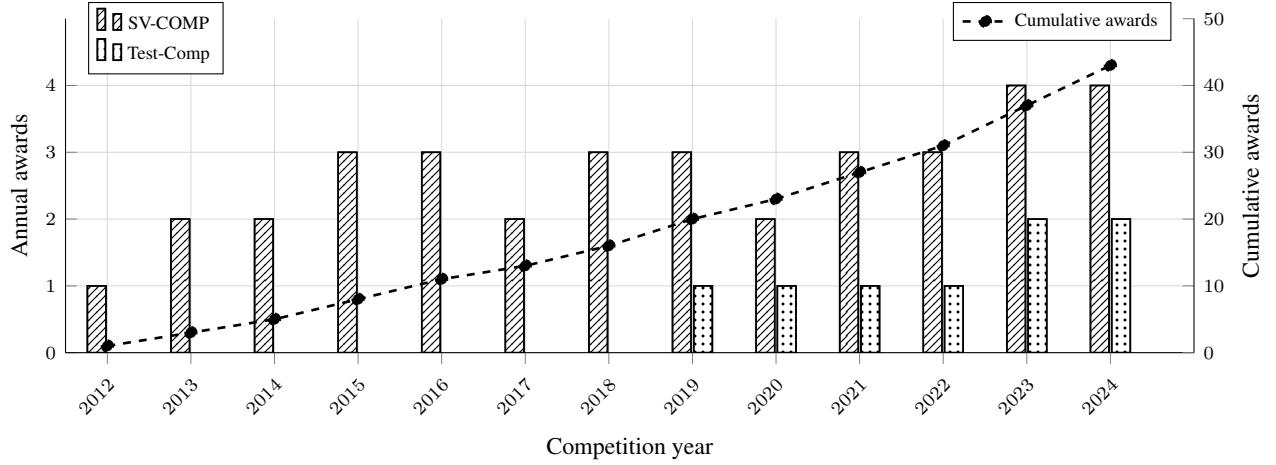
\begin{figure}[htbp]
\centering
\begin{tikzpicture}
\begin{axis}[
    width=\linewidth,
    height=6cm,
    ybar,
    bar width=7pt,
    xlabel={Competition year},
    ylabel={Annual awards},
    xmin=2011.5, xmax=2025.0,
    ymin=0, ymax=5,
    xtick={2012,2013,2014,2015,2016,2017,2018,2019,2020,2021,2022,2023,2024},
    ytick={0,1,2,3,4},
    x tick label style={font=\scriptsize, rotate=45, anchor=north east},
    y tick label style={font=\scriptsize},
    label style={font=\small},
    grid=both,
    grid style={line width=0.3pt, draw=gray!20},
    major grid style={line width=0.4pt, draw=gray!30},
    scaled x ticks=false,
    /pgf/number format/1000 sep={},
    axis y line*=left,
    axis x line*=bottom,
    legend style={
        at={(0.02,1.05)},
        anchor=north west,
        font=\scriptsize,
        draw=black,
        fill=white,
        legend image code/.code={
            \draw[#1] (0cm,-0.1cm) rectangle (0.3cm,0.2cm);
        },
    },
]

\addplot[
    ybar,
    bar width=7pt,
    bar shift=-4pt,
    fill=white,
    draw=black,
    line width=0.6pt,
    postaction={pattern=north east lines, pattern color=black},
] coordinates {
    (2012,1)(2013,2)(2014,2)(2015,3)(2016,3)
    (2017,2)(2018,3)(2019,3)(2020,2)(2021,3)
    (2022,3)(2023,4)(2024,4)
};

\addplot[
    ybar,
    bar width=7pt,
    bar shift=4pt,
    fill=white,
    draw=black,
    line width=0.6pt,
    postaction={pattern=dots, pattern color=black},
] coordinates {
    (2019,1)(2020,1)(2021,1)(2022,1)(2023,2)(2024,2)
};

\legend{\gls{svcomp}, \gls{test-comp}}

\end{axis}

\begin{axis}[
    width=\linewidth,
    height=6cm,
    xmin=2011.5, xmax=2025.0,
    ymin=0, ymax=50,
    xtick=\empty,
    ytick={0,10,20,30,40,50},
    y tick label style={font=\scriptsize},
    ylabel={Cumulative awards},
    label style={font=\small},
    scaled x ticks=false,
    /pgf/number format/1000 sep={},
    axis y line*=right,
    axis x line=none,
    legend style={
        at={(0.98,1.05)},
        anchor=north east,
        font=\scriptsize,
        draw=black,
        fill=white,
    },
]

\addplot[
    dashed,
    mark=*,
    mark size=2pt,
    line width=1.0pt,
    color=black,
] coordinates {
    (2012,1)(2013,3)(2014,5)(2015,8)(2016,11)
    (2017,13)(2018,16)(2019,20)(2020,23)(2021,27)
    (2022,31)(2023,37)(2024,43)
};

\addlegendentry{Cumulative awards}

\end{axis}
\end{tikzpicture}
\caption{Annual awards obtained by \gls{esbmc} at \gls{svcomp} (diagonal hatching) and \gls{test-comp} (dotted) from 2012 to 2024. The dashed line shows the running cumulative total~\cite{beyer2012,svcomp2024,ssvlab2024}}

\label{fig:svcomp-awards}
\end{figure}

%% file: figures/tikz_esbmc_structure.tex
\begin{figure}[htbp]
\centering
\resizebox{0.85\linewidth}{!}{%
\begin{tikzpicture}[
    box/.style={
        draw, rectangle, rounded corners=3pt,
        font=\small, align=center,
        inner sep=6pt, line width=0.6pt,
        fill=white
    },
    arrow/.style={-latex, line width=0.8pt},
    header/.style={font=\small, align=center},
]

\def\xFE{0}      
\def\xIR{4}    
\def\xBMC{8}   
\def\xSMT{12}  
\def\xOUT{16}  

\node[header] at (\xFE,  4.8) {Language\\Front-ends};
\node[header] at (\xIR,  4.8) {Internal\\Representation (IR)};
\node[header] at (\xBMC, 4.8) {Verification\\Engine};
\node[header] at (\xSMT, 4.8) {SMT\\Solvers};
\node[header] at (\xOUT, 4.8) {Output};

\node[box, text width=2.8cm] (fe1) at (\xFE,  3.2) {Rust / \glsfirst{cuda}/\\\gls{cheri}};
\node[box, text width=2.8cm] (fe2) at (\xFE,  1.6) {Solidity\\\small(\gls{esbmc}-Sol.)};
\node[box, text width=2.8cm] (fe3) at (\xFE,  0.0) {Kotlin\\\small(Jimple)};
\node[box, text width=2.8cm] (fe4) at (\xFE, -1.6) {Python\\\small(\gls{esbmc}-Python)};
\node[box, text width=2.8cm] (fe5) at (\xFE, -3.2) {C / C\texttt{++}\\\small(Clang \gls{ast})};

\node[
    draw, dashed, rectangle, rounded corners=5pt,
    line width=0.7pt,
    inner sep=10pt,
    fit=(fe1)(fe2)(fe3)(fe4)(fe5),
] (fegroup) {};

\node[box, text width=3.0cm, minimum height=2.2cm] (ir) at (\xIR, 0)
    {\textbf{Intermediate}\\[2pt]\textbf{Representation}\\[6pt]
     \scriptsize GOTO-program / SSA};

\node[box, text width=3.0cm, minimum height=2.2cm] (bmc) at (\xBMC, 0)
    {\textbf{BMC /}\\[2pt]\textbf{\textit{k}-Induction Engine}\\[6pt]
     \scriptsize Unrolling $\cdot$ Invariants\\
     \scriptsize Floating-pt $\cdot$ POR};

\node[box, text width=2.2cm] (s1) at (\xSMT,  3.0) {Yices~2};
\node[box, text width=2.2cm] (s2) at (\xSMT,  1.5) {MathSAT};
\node[box, text width=2.2cm] (s3) at (\xSMT,  0.0) {CVC5};
\node[box, text width=2.2cm] (s4) at (\xSMT, -1.5) {Bitwuzla};
\node[box, text width=2.2cm] (s5) at (\xSMT, -3.0) {Z3};

\node[box, text width=2.8cm] (out1) at (\xOUT,  2.2)
    {Unknown /\\Timeout};
\node[box, text width=2.8cm] (out2) at (\xOUT, 0.0)
    {\textbf{Property proved}\\(Safe)};
\node[box, text width=2.8cm] (out3) at (\xOUT, -2.2)
    {\textbf{Counterexample}\\(Bug found)};

\draw[arrow] (fegroup.east) -- (ir.west);

\draw[arrow] (ir.east)  -- (bmc.west);
\draw[arrow] (bmc.east) -- (s3.west);
\draw[line width=0.8pt] (bmc.east) -- ++(0.3,0) |- (s1.west);
\draw[line width=0.8pt] (bmc.east) -- ++(0.3,0) |- (s2.west);
\draw[line width=0.8pt] (bmc.east) -- ++(0.3,0) |- (s4.west);
\draw[line width=0.8pt] (bmc.east) -- ++(0.3,0) |- (s5.west);

\draw[arrow] (s1.east) -- ++(0.3,0) |- (out1.west);
\draw[arrow] (s2.east) -- ++(0.3,0) |- (out1.west);
\draw[arrow] (s3.east) -- ++(0.3,0) |- (out2.west);
\draw[arrow] (s4.east) -- ++(0.3,0) |- (out3.west);
\draw[arrow] (s5.east) -- ++(0.3,0) |- (out3.west);

\end{tikzpicture}%
}
\caption{\gls{esbmc} verification pipeline. Source programs are parsed by language-specific front-ends into a common GOTO program/\gls{ssa} intermediate representation. The \gls{bmc} and \textit{k}-induction engine encodes the verification condition and dispatches to one of five \gls{smt} solvers (Yices~2, MathSAT, CVC5, Bitwuzla, Z3). The solver returns either a proof of safety (property proved), a concrete counterexample (bug found), or an inconclusive result (unknown/timeout)}

\label{fig:architecture}
\end{figure}
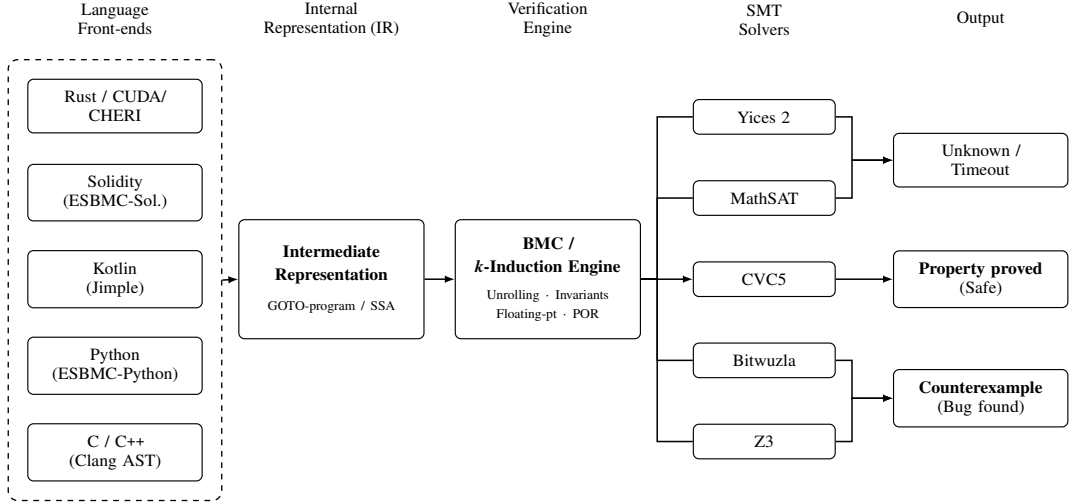

%% file: figures/tikz_aitimeline.tex
\begin{figure}[htbp]
\centering
\resizebox{\linewidth}{!}{%
\begin{tikzpicture}[
    boxstyle/.style={
        draw, rectangle, rounded corners=2pt,
        font=\scriptsize, align=center,
        inner sep=4pt, line width=0.5pt,
        fill=white, text width=2.5cm
    },
    dot/.style={circle, fill=black, inner sep=0pt, minimum size=5pt},
]

\def\xA{0.0}   
\def\xB{2.0}   
\def\xC{4.0}   
\def\xD{6.0}   
\def\xE{8.0}  
\def\xF{10.0}  
\def\xG{12.0}  

\draw[line width=1.2pt, -latex] (-0.4, 0) -- (12.6, 0);

\foreach \xp in {\xA, \xB, \xC, \xD, \xE, \xF, \xG} {
    \node[dot] at (\xp, 0) {};
}


\draw[line width=0.5pt] (\xA, 0) -- (\xA, 1.0);
\node[boxstyle] at (\xA, 1.45)
    {\gls{esbmc}-\gls{ai}\\(repair framework)\\\cite{Tihanyi2025New}};
\node[font=\scriptsize, rotate=0, anchor=north east]
    at (\xA+0.5, -0.15) {May 2023};

\draw[line width=0.5pt] (\xC, 0) -- (\xC, 1.0);
\node[boxstyle] at (\xC, 1.45)
    {Lemur\\hybrid \gls{llm}/\gls{bmc}\\\cite{Haoze2023}};
\node[font=\scriptsize, rotate=0, anchor=north east]
    at (\xC+0.5, -0.15) {Nov 2023};

\draw[line width=0.5pt] (\xE, 0) -- (\xE, 1.0);
\node[boxstyle] at (\xE, 1.45)
    {FormAI dataset\\\gls{llm} code quality\\\cite{tihanyi2023}};
\node[font=\scriptsize, rotate=0, anchor=north east]
    at (\xE+0.5, -0.15) {Oct 2024};

\draw[line width=0.5pt] (\xG, 0) -- (\xG, 1.0);
\node[boxstyle] at (\xG, 1.45)
    {SpecVerify\\Claude~3.5 Sonnet\\+ \gls{esbmc}~\cite{wang2025supporting}};
\node[font=\scriptsize, rotate=0, anchor=north east]
    at (\xG+0.5, -0.15) {Sep 2025};


\draw[line width=0.5pt] (\xB, 0) -- (\xB, -1.,0);
\node[boxstyle] at (\xB, -1.45)
    {Self-healing SW\\(concept 2023;\\formalised AST~2025)~\cite{Yiannis2024}};
\node[font=\scriptsize, rotate=0, anchor=south west]
    at (\xB-0.5, 0.15) {Jun 2023};

\draw[line width=0.5pt] (\xD, 0) -- (\xD, -1.0);
\node[boxstyle] at (\xD, -1.45)
    {\gls{llm}-generated\\loop invariants\\(\gls{ase}~2024)\\\cite{Pirzada2024LLM}};
\node[font=\scriptsize, rotate=0, anchor=south west]
    at (\xD-0.5, 0.15) {Sep 2024};

\draw[line width=0.5pt] (\xF, 0) -- (\xF, -1.0);
\node[boxstyle] at (\xF, -1.45)
    {\gls{llm}+\gls{esbmc}\\repair\\(\gls{ast}~2025)\\ \cite{Yiannis2024}};
\node[font=\scriptsize, rotate=0, anchor=south west]
    at (\xF-0.5, 0.15) {Nov 2024};

\end{tikzpicture}%
}
\caption{Chronological milestones of \gls{esbmc}'s \gls{ai} and \gls{llm} integration (2023--2025). Key contributions include the \gls{esbmc}-\gls{ai} repair framework (May~2023)~\cite{Tihanyi2025New}, the Lemur hybrid \gls{llm}/\gls{bmc} system (Nov~2023)~\cite{Haoze2023}, the self-healing software concept (introduced Jun~2023, formalised and published at \gls{ast}~2025)~\cite{Yiannis2024}, \gls{llm}-generated loop invariants (Sep~2024)~\cite{Pirzada2024LLM}, the FormAI dataset (Oct~2024)~\cite{tihanyi2023}, and SpecVerify (Sep~2025)~\cite{wang2025supporting}}

\label{fig:aitimeline}
\end{figure}
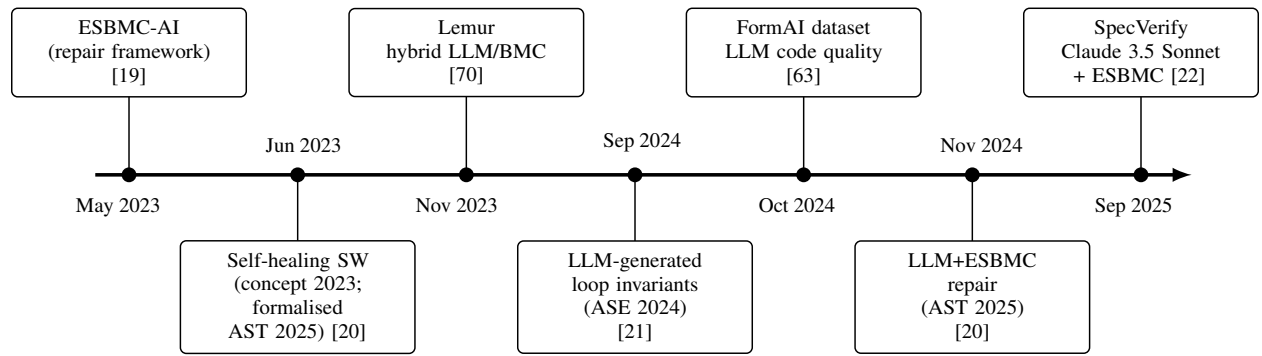

%% file: figures/tikz_repair_rate.tex
\begin{figure}[htbp]
\centering
\begin{tikzpicture}
\begin{axis}[
    width=0.7\linewidth,
    height=8cm,
    ybar stacked,
    bar width=40pt,
    xlabel={Bug category},
    ylabel={Percentage of cases (\%)},
    ymin=0, ymax=110,
    ytick={0,20,40,60,80,100},
    xtick=data,
    xticklabels={
        Buffer\\Overflow,
        Pointer\\Dereference
    },
    x tick label style={
        font=\scriptsize,
        align=center,
        text width=2.2cm,
    },
    y tick label style={font=\scriptsize},
    label style={font=\small},
    tick label style={font=\scriptsize},
    grid=none,
    axis x line=bottom,
    axis y line=left,
    scaled ticks=false,
    enlarge x limits=0.5,
    legend style={
        at={(0.98,0.98)},
        anchor=north east,
        font=\scriptsize,
        draw=black,
        fill=white,
        legend image code/.code={
            \draw[#1] (0cm,-0.1cm) rectangle (0.3cm,0.2cm);
        },
    },
    nodes near coords=\empty,
]

\addplot[
    ybar stacked,
    fill=white,
    draw=black,
    line width=0.6pt,
    postaction={pattern=north east lines, pattern color=black},
] coordinates {
    (1,80) (2,80)
};
\addlegendentry{Repaired successfully}

\addplot[
    ybar stacked,
    fill=white,
    draw=black,
    line width=0.6pt,
    postaction={pattern=dots, pattern color=black},
] coordinates {
    (1,20) (2,20)
};
\addlegendentry{Unresolved}

\node[font=\scriptsize\bfseries] at (axis cs:1, 40) {80\%};
\node[font=\scriptsize\bfseries] at (axis cs:2, 40) {80\%};

\draw[dashed, line width=0.7pt]
    (axis cs:0.5, 80) -- (axis cs:2.5, 80);
\node[font=\tiny, anchor=west] at (axis cs:2.5, 80) {Peak (80\%)};

\end{axis}
\end{tikzpicture}
\caption{Repair success rate achieved by the \gls{esbmc}-\gls{ai} framework~\cite{Tihanyi2025New} for the two vulnerability categories reported in the source: buffer overflow and pointer dereference, both achieving up to 80\% success combined. All reported successes require the repaired program to pass \gls{esbmc}'s formal property check -- a substantially stronger guarantee than repair tools relying on test-suite pass rates or \gls{llm} self-assessment~\cite{tihanyi2023}. Category-level success rates for other vulnerability classes (arithmetic overflow, array bounds, memory leak, concurrency defects) are not reported in the cited source and have been removed from this figure.}

\label{fig:repairrate}
\end{figure}
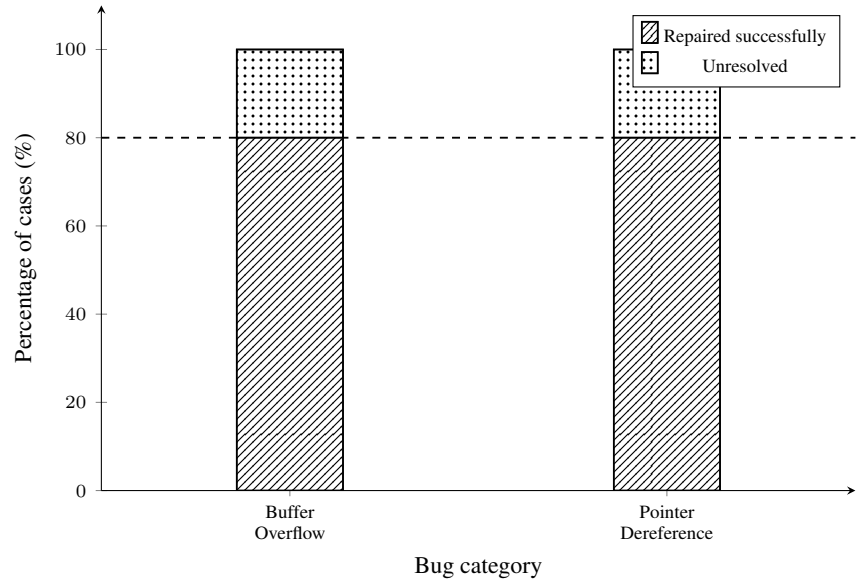

%% file: figures/tikz_llm_inv_generation.tex
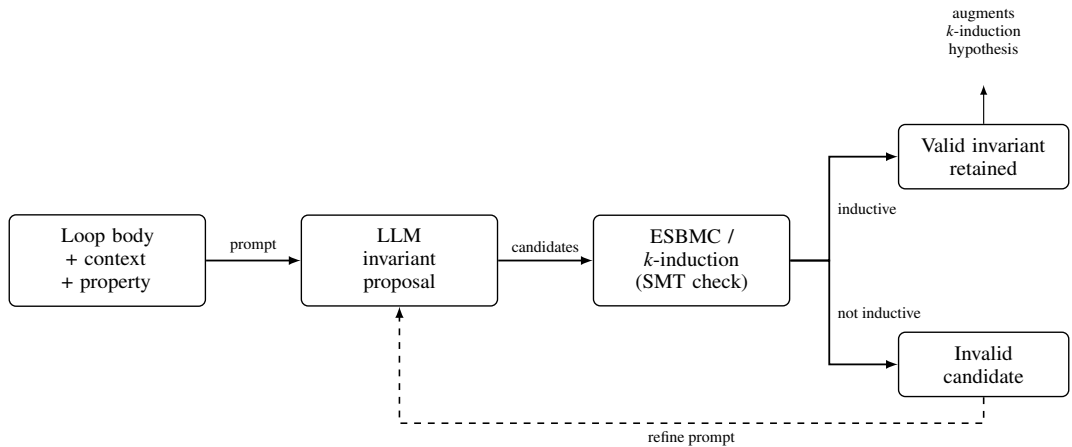
\begin{figure}[htbp]
\centering
\resizebox{0.85\linewidth}{!}{%
\begin{tikzpicture}[
    box/.style={
        draw, rectangle, rounded corners=3pt,
        font=\small, align=center,
        inner sep=6pt, line width=0.6pt,
        fill=white, text width=2.6cm
    },
    arrow/.style={-latex, line width=0.8pt},
    dasharrow/.style={-latex, line width=0.7pt, dashed},
]

\node[box] (loop) at (0, 0)
    {Loop body\\+ context\\+ property};
\node[box] (llm) at (4.5, 0)
    {\gls{llm}\\invariant\\proposal};
\node[box] (esbmc) at (9, 0)
    {\gls{esbmc} /\\\textit{k}-induction\\(\gls{smt} check)};
\node[box, text width=2.2cm] (valid)   at (13.5,  1.6)
    {Valid invariant\\retained};
\node[box, text width=2.2cm] (invalid) at (13.5, -1.6)
    {Invalid\\candidate};

\draw[arrow] (loop.east) --
    node[above, font=\scriptsize] {prompt} (llm.west);
\draw[arrow] (llm.east)  --
    node[above, font=\scriptsize] {candidates} (esbmc.west);

\draw[arrow] (esbmc.east) -- ++(0.6, 0)
    -- node[right, font=\scriptsize] {inductive}
    ++(0,  1.6)
    -- (valid.west);

\draw[arrow] (esbmc.east) -- ++(0.6, 0)
    -- node[right, font=\scriptsize] {not inductive}
    ++(0, -1.6)
    -- (invalid.west);

\draw[dasharrow] (invalid.south)
    -- ++(0, -0.4)
    -| node[below, font=\scriptsize, pos=0.25] {refine prompt} (llm.south);

\node[font=\scriptsize, align=center, text width=2.4cm] at (13.5, 3.5)
    {augments\\\textit{k}-induction\\hypothesis};
\draw[-latex, line width=0.5pt] (valid.north) -- ++(0, 0.6);

\end{tikzpicture}%
}
\caption{\gls{llm}-assisted loop invariant generation pipeline. The \gls{llm} proposes candidate invariants from the loop body and surrounding context; \gls{esbmc} checks each candidate for inductiveness via \gls{smt}; valid invariants augment the \textit{k}-induction hypothesis while invalid candidates are fed back to the \gls{llm} for refinement~\cite{Pirzada2024LLM}}

\label{fig:invariant-pipeline}
\end{figure}

%% file: references.bib
@book{clarke1999,
title = {Model checking},
ISBN = {9783540696599},
ISSN = {1611-3349},
DOI = {10.1007/bfb0058022},
booktitle = {Foundations of Software Technology and Theoretical Computer Science},
publisher = {Springer Berlin Heidelberg},
author = {Clarke,  Edmund M.},
year = {1997},
pages = {54–56},
address = {Berlin, Heidelberg},
}

@inbook{biere1999,
title = {{Symbolic Model Checking without BDDs}},
ISBN = {9783540490593},
ISSN = {0302-9743},
DOI = {10.1007/3-540-49059-0_14},
booktitle = {Tools and Algorithms for the Construction and Analysis of Systems},
publisher = {Springer Berlin Heidelberg},
author = {Biere,  Armin and Cimatti,  Alessandro and Clarke,  Edmund and Zhu,  Yunshan},
year = {1999},
pages = {193–207},
address = {Berlin, Heidelberg},
}

@article{nieuwenhuis2006,
author = {Nieuwenhuis, Robert and Oliveras, Albert and Tinelli, Cesare},
title = {{Solving SAT and SAT Modulo Theories: From an Abstract Davis-Putnam-Logemann-Loveland Procedure to DPLL(T)}},
journal = {Journal of the {ACM}},
volume = {53},
number = {6},
pages = {937--977},
year = {2006},
doi = {10.1145/1217856.1217859}
}

@inproceedings{moura2008,
author = {de Moura, Leonardo and Bj{\o}rner, Nikolaj},
title = {{Z3: An Efficient SMT Solver}},
booktitle = {Proceedings of the 14th International Conference on Tools and Algorithms for the Construction and Analysis of Systems (TACAS 200)},
series = {Lecture Notes in Computer Science},
volume = {4963},
pages = {337--340},
year = {2008},
publisher = {Springer},
doi = {10.1007/978-3-540-78800-3_24},
address = {Budapest, Hungary},
}

@inproceedings{niemetz2023,
title = {Bitwuzla},
ISBN = {9783031377037},
ISSN = {1611-3349},
DOI = {10.1007/978-3-031-37703-7_1},
booktitle = {Computer Aided Verification},
publisher = {Springer Nature Switzerland},
author = {Niemetz,  Aina and Preiner,  Mathias},
year = {2023},
address = {Cham, Switzerland},
pages = {3–17}
}

@inproceedings{barbosa2022, 
author = {Barbosa, Haniel and others}, 
title = {{cvc5: A Versatile and Industrial-Strength SMT Solver}},
booktitle = {Proceedings of the 28th International Conference on Tools and Algorithms for the Construction and Analysis of Systems (TACAS 2022)},
series = {Lecture Notes in Computer Science},
volume = {13243},
pages = {415--442},
year = {2022},
publisher = {Springer},
doi = {10.1007/978-3-030-99524-9_24},
address = {Munich, Germany},
}

@inproceedings{cimatti2013,
author = {Cimatti, Alessandro and Griggio, Alberto and Schaafsma, Bastiaan J. and Sebastiani, Roberto},
title = {{The MathSAT5 SMT Solver}},
booktitle = {Proceedings of the 19th International Conference on Tools and Algorithms for the Construction and Analysis of Systems (TACAS 2013)},
series = {Lecture Notes in Computer Science},
volume = {7795},
pages = {93--107},
year = {2013},
publisher = {Springer},
doi = {10.1007/978-3-642-36742-7_7},
address = {Rome, Italy},
}

@inproceedings{dutertre2014,
author = {Dutertre, Bruno},
title = {{Yices~2.2}},
booktitle = {Proceedings of the 26th International Conference on Computer-Aided Verification (CAV 2014)},
series = {Lecture Notes in Computer Science},
volume = {8559},
pages = {737--744},
year = {2014},
publisher = {Springer},
doi = {10.1007/978-3-319-08867-9_49},
address = {Vienna, Austria},
}

@inbook{clarke2004,
title = {{A Tool for Checking ANSI-C Programs}},
ISBN = {9783540247302},
ISSN = {1611-3349},
DOI = {10.1007/978-3-540-24730-2_15},
booktitle = {Tools and Algorithms for the Construction and Analysis of Systems},
publisher = {Springer Berlin Heidelberg},
author = {Clarke,  Edmund and Kroening,  Daniel and Lerda,  Flavio},
year = {2004},
pages = {168–176},
address = {Berlin, Heidelberg, Germany}
}

@inproceedings{cordeiro2011icse,
author = {Cordeiro, Lucas C. and Fischer, Bernd and Marques-Silva, João},
title = {{Verifying Multi-Threaded Software Using SMT-Based Context-Bounded Model Checking}},
booktitle = {Proceedings of the 33rd International Conference on Software Engineering (ICSE 2011)},
pages = {331--340},
year = {2011},
publisher = {ACM},
doi = {10.1145/1985793.1985839},
address = {Honolulu, HI, USA},
}

@inproceedings{beyer2012,
author = {Beyer, Dirk},
title = {{Competition on Software Verification (SV-COMP)}},
booktitle = {Proceedings of the 18th International Conference on Tools and Algorithms for the Construction and Analysis of Systems (TACAS 2012)},
series = {Lecture Notes in Computer Science},
volume = {7214},
pages = {504--524},
year = {2012},
publisher = {Springer},
doi = {10.1007/978-3-642-28756-5_38},
address = {Tallinn, Estonia},
}

@proceedings{cordeiro2014,
editor = {Erika Ábrahám and Klaus Havelund},
title = {Tools and Algorithms for the Construction and Analysis of Systems: 20th International Conference, TACAS 2014, Held as Part of the European Joint Conferences on Theory and Practice of Software, ETAPS 2014, Grenoble, France, April 5--13, 2014. Proceedings},
series = {Lecture Notes in Computer Science},
volume = {8413},
publisher = {Springer Berlin Heidelberg},
year = {2014},
isbn = {9783642548628},
issn = {1611-3349},
doi = {10.1007/978-3-642-54862-8}
}

@article{gadelha2020,
title = {{ESBMC 6.1: Automated Test Case Generation Using Bounded Model Checking}},
volume = {23},
ISSN = {1433-2787},
DOI = {10.1007/s10009-020-00571-2},
number = {6},
journal = {International Journal on Software Tools for Technology Transfer},
publisher = {Springer Science and Business Media LLC},
author = {Gadelha,  Mikhail R. and Menezes,  Rafael S. and Cordeiro,  Lucas C.},
year = {2020},
month = may,
pages = {857–861}
}

@inbook{shmarov2023,
title = {ESBMC v7.3: Model Checking C++ Programs Using Clang AST},
ISBN = {9783031493423},
ISSN = {1611-3349},
DOI = {10.1007/978-3-031-49342-3_9},
booktitle = {Formal Methods: Foundations and Applications},
publisher = {Springer Nature Switzerland},
author = {Song,  Kunjian and Gadelha,  Mikhail R. and Brauße,  Franz and Menezes,  Rafael S. and Cordeiro,  Lucas C.},
year = {2023},
month = Dec,
pages = {141–152},
address = {Cham, Switzerland},
}

@inbook{menezes2024,
title = {{ESBMC v7.4: Harnessing the Power of Intervals: (Competition Contribution)}},
ISBN = {9783031572562},
ISSN = {1611-3349},
DOI = {10.1007/978-3-031-57256-2_24},
booktitle = {Tools and Algorithms for the Construction and Analysis of Systems},
publisher = {Springer Nature Switzerland},
author = {Menezes,  Rafael Sá and Aldughaim,  Mohannad and Farias and others},
year = {2024},
pages = {376–380},
address = {Luxembourg City, Luxembourg},
}

@inproceedings{menezes2025tacas,
author = {Menezes, Rafael and others},
title = {{ESBMC v7.7: Efficient Concurrent Software Verification with Scheduling, Incremental SMT and Partial Order Reduction}},
booktitle = {Proceedings of the 31st International Conference on Tools and Algorithms for the Construction and Analysis of Systems (TACAS 2025)},
series = {Lecture Notes in Computer Science},
year = {2025},
publisher = {Springer},
doi = {10.1007/978-3-031-90660-2_16},
address = {Hamilton, New Zealand},
numpages = {16},
}

@inproceedings{menezes2025fase,
author = {Menezes, Rafael and others},
title = {{ESBMC v7.7: Automating Branch Coverage Analysis Using CFG-Based Instrumentation and SMT Solving}},
booktitle = {Proceedings of the 28th International Conference on Fundamental Approaches to Software Engineering (FASE 2025)},
series = {Lecture Notes in Computer Science},
year = {2025},
publisher = {Springer},
doi = {10.1007/978-3-031-90900-9_15},
address = {Hamilton, New Zealand},
numpages = {16},
}

@inproceedings{sheeran2000,
author = {Sheeran, Mary and Singh, Satnam and St{\aa}lmarck, Gunnar},
title = {{Checking Safety Properties Using Induction and a SAT-Solver}},
booktitle = {Proceedings of the 3rd International Conference on Formal Methods in Computer-Aided Design (FMCAD 2000)},
series = {Lecture Notes in Computer Science},
volume = {1954},
pages = {108--125},
year = {2000},
publisher = {Springer},
doi = {10.1007/3-540-40922-X_8},
address = {Austin, TX, USA},
}

@inproceedings{gadelha2017,
author = {Gadelha, Mikhail Y. R. and Cordeiro, Lucas C. and Nicole, Denis A.},
title = {{Encoding Floating-Point Numbers Using the SMT Theory in ESBMC: An Empirical Evaluation over the SV-COMP Benchmarks}},
booktitle = {Proceedings of the 22nd International Workshop on Formal Methods for Industrial Critical Systems (FMICS 2017)},
series = {Lecture Notes in Computer Science},
volume = {10471},
pages = {91--106},
year = {2017},
publisher = {Springer},
doi = {10.1007/978-3-319-70848-5_7},
address = {Turin, Italy},
}

@inproceedings{sloan2022,
series = {ICSE ’22},
title = {ESBMC-solidity: an SMT-based model checker for solidity smart contracts},
DOI = {10.1145/3510454.3516855},
booktitle = {Proceedings of the ACM/IEEE 44th International Conference on Software Engineering: Companion Proceedings},
publisher = {ACM},
author = {Song,  Kunjian and Matulevicius,  Nedas and de Lima Filho,  Eddie B. and Cordeiro,  Lucas C.},
year = {2022},
month = May,
pages = {65–69},
address = {Pittsburgh, PA, USA},
collection = {ICSE ’22}
}

@inproceedings{farias2024,
author = {Farias, Felipe R. and Menezes, Rafael and Cordeiro, Lucas C.},
title = {{ESBMC-Python: A Bounded Model Checker for Python Programs}},
booktitle = {Proceedings of the 33rd ACM SIGSOFT International Symposium on Software Testing and Analysis (ISSTA 2024)},
year = {2024},
publisher = {ACM},
doi = {10.1145/3650212.3685304},
address = {Vienna, Austria},
numpages = {10},
}

@misc{rustfoundation2024,
author = {{Rust Foundation}},
title = {{Expanding the Rust Formal Verification Ecosystem: Welcoming ESBMC}},
year = {2024},
howpublished = {\url{https://rustfoundation.org/media/expanding-the-rust-formal-verification-ecosystem-welcoming-esbmc/}},
note = {Accessed: May 2026}
}

@misc{ssvlab2024,
author = {{SSVLab}},
title = {{ESBMC: An Industrial-Strength C Model Checker - Competition Results}},
year = {2024},
howpublished = {\url{https://ssvlab.github.io/esbmc/sv-comp.html}},
note = {Accessed: May 2026}
}

@inproceedings{tihanyi2023,
author = {Tihanyi, Norbert and Bisztray, Tamas and Jain, Ridhi and Ferrag, Mohamed Amine and Cordeiro, Lucas C. and Mavroeidis, Vasileios},
title = {{The FormAI Dataset: Generative AI in Software Security through the Lens of Formal Verification}},
booktitle = {Proceedings of the 19th International Conference on Predictive Models and Data Analytics in Software Engineering (PROMISE 2023)},
year = {2023},
publisher = {ACM},
doi = {10.1145/3617555.3617874},
address = {San Francisco, CA, USA},
pages = {33--40},
numpages = {8},
}

@inproceedings{wang2025supporting,
title = {Supporting Software Formal Verification with Large Language Models: An Experimental Study},
DOI = {10.1109/re63999.2025.00049},
booktitle = {2025 IEEE 33rd International Requirements Engineering Conference (RE)},
publisher = {IEEE},
author = {Wang,  Weiqi and Farrell,  Marie and Cordeiro,  Lucas C. and Zhao,  Liping},
year = {2025},
month = Sep,
pages = {423–431},
address = {Reykjavik, Iceland},
}

@inproceedings{Tihanyi2025New,
author = {Tihanyi, Norbert and Charalambous, Yiannis and Jain, Ridhi and Ferrag, Mohamed Amine and Cordeiro, Lucas C.},
booktitle = {2025 IEEE/ACM International Conference on Automation of Software Test (AST)},
doi = {10.1109/ast66626.2025.00020},
year = {2025},
month = {apr 28},
pages = {136--147},
organization = {IEEE},
publisher = {IEEE},
title = {{A New Era in Software Security: Towards Self-Healing Software via Large Language Models and Formal Verification}},
address = {Ottawa, ON, Canada},
}

@inproceedings{Pirzada2024LLM,
author = {Pirzada, Muhammad A. A. and Reger, Giles and Bhayat, Ahmed and Cordeiro, Lucas C.},
booktitle = {Proceedings of the 39th IEEE/ACM International Conference on Automated Software Engineering},
doi = {10.1145/3691620.3695512},
year = {2024},
month = {oct 27},
pages = {1395--1407},
organization = {ACM},
publisher = {ACM},
title = {{LLM-Generated Invariants for Bounded Model Checking Without Loop Unrolling}},
address = {Sacramento, CA, USA},
}

@book{Clarke2018Handbook,
author = {Clarke, Edmund M. and Henzinger, Thomas A. and Veith, Helmut and Bloem, Roderick},
title = {Handbook of Model Checking},
ISBN = {9783319105758},
DOI = {10.1007/978-3-319-10575-8},
publisher = {Springer International Publishing},
year = {2018},
address = {Cham, Switzerland}
}

@inproceedings{Cordeiro2009,
title = {{SMT-Based Bounded Model Checking for Embedded ANSI-C Software}},
booktitle = {Proceedings of the 24th IEEE/ACM International Conference on Automated Software Engineering (ASE 2009)},
author = {Cordeiro, Lucas and Fischer, Bernd and Marques-Silva, Joao},
publisher = {IEEE},
year = {2009},
pages = {137--148},
DOI = {10.1109/ASE.2009.17},
address = {Auckland, New Zealand},
}

@article{Cordeiro2012,
title = {{SMT-Based Bounded Model Checking for Embedded ANSI-C Software}},
volume = {38},
ISSN = {0098-5589},
DOI = {10.1109/tse.2011.59},
number = {4},
journal = {IEEE Transactions on Software Engineering},
publisher = {Institute of Electrical and Electronics Engineers (IEEE)},
author = {Cordeiro,  Lucas and Fischer,  Bernd and Marques-Silva,  Joao},
year = {2012},
month = jul,
pages = {957–974}
}

@inproceedings{Gadelha2018ESBMC,
series = {ASE'18},
title = {{ESBMC 5.0: an Industrial-Strength C Model Checker}},
DOI = {10.1145/3238147.3240481},
booktitle = {Proceedings of the 33rd ACM/IEEE International Conference on Automated Software Engineering},
publisher = {ACM},
author = {Gadelha,  Mikhail R. and Monteiro,  Felipe R. and Morse,  Jeremy and Cordeiro,  Lucas C. and Fischer,  Bernd and Nicole,  Denis A.},
year = {2018},
month = sep,
pages = {888–891},
collection = {ASE'18},
address = {Manchester, UK}
}

@inproceedings{Vaswani2017Attention,
 author = {Vaswani, Ashish and Shazeer, Noam and Parmar, Niki and Uszkoreit, Jakob and Jones, Llion and Gomez, Aidan N and Kaiser, \L ukasz and Polosukhin, Illia},
 booktitle = {Advances in Neural Information Processing Systems},
 editor = {I. Guyon and U. Von Luxburg and S. Bengio and H. Wallach and R. Fergus and S. Vishwanathan and R. Garnett},
 pages = {},
 publisher = {Curran Associates, Inc.},
 title = {Attention is All you Need},
 url = {https://proceedings.neurips.cc/paper_files/paper/2017/file/3f5ee243547dee91fbd053c1c4a845aa-Paper.pdf},
 volume = {30},
 year = {2017},
 address = {Long Beach, CA, USA},
}

@inproceedings{brown2020language,
 author = {Brown, Tom and others},
 booktitle = {Advances in Neural Information Processing Systems},
 editor = {H. Larochelle and M. Ranzato and R. Hadsell and M.F. Balcan and H. Lin},
 pages = {1877--1901},
 publisher = {Curran Associates, Inc.},
 title = {Language Models are Few-Shot Learners},
 url = {https://proceedings.neurips.cc/paper_files/paper/2020/file/1457c0d6bfcb4967418bfb8ac142f64a-Paper.pdf},
 volume = {33},
 year = {2020},
 address = {Vancouver, Canada},
}

@inbook{Clarke1982,
title = {{Design and Synthesis of Synchronization Skeletons Using Branching Time Temporal Logic}},
ISBN = {354011212X},
DOI = {10.1007/bfb0025774},
booktitle = {Logics of Programs},
publisher = {Springer-Verlag},
author = {Clarke,  Edmund M. and Emerson,  E. Allen},
pages = {52–71},
year = 1982,
address = {Berlin, Heidelberg, Germany}
}

@article{Bryant1986,
title = {{Graph-Based Algorithms for Boolean Function Manipulation}},
volume = {C–35},
ISSN = {2326-3814},
DOI = {10.1109/tc.1986.1676819},
number = {8},
journal = {IEEE Transactions on Computers},
publisher = {Institute of Electrical and Electronics Engineers (IEEE)},
author = {Bryant},
year = {1986},
month = aug,
pages = {677–691}
}

@article{Clarke2001,
title = {{Bounded Model Checking Using Satisfiability Solving}},
volume = {19},
ISSN = {1572-8102},
DOI = {10.1023/a:1011276507260},
number = {1},
journal = {Formal Methods in System Design},
publisher = {Springer Science and Business Media LLC},
author = {Clarke,  Edmund and Biere,  Armin and Raimi,  Richard and Zhu,  Yunshan},
year = {2001},
month = jul,
pages = {7–34}
}

@article{Vizel2015,
title = {{Boolean Satisfiability Solvers and Their Applications in Model Checking}},
volume = {103},
ISSN = {1558-2256},
DOI = {10.1109/jproc.2015.2455034},
number = {11},
journal = {Proceedings of the IEEE},
publisher = {Institute of Electrical and Electronics Engineers (IEEE)},
author = {Vizel,  Yakir and Weissenbacher,  Georg and Malik,  Sharad},
year = {2015},
month = nov,
pages = {2021–2035}
}

@inbook{Mukherjee2016,
title = {{v2c – A Verilog to C Translator}},
ISBN = {9783662496749},
ISSN = {1611-3349},
DOI = {10.1007/978-3-662-49674-9_38},
booktitle = {Tools and Algorithms for the Construction and Analysis of Systems},
publisher = {Springer Berlin Heidelberg},
author = {Mukherjee,  Rajdeep and Tautschnig,  Michael and Kroening,  Daniel},
year = {2016},
pages = {580–586},
address = {Berlin, Heidelberg},
}

@inproceedings{Barreto2011,
title = {{Verifying Embedded C Software with Timing Constraints Using an Untimed Bounded Model Checker}},
DOI = {10.1109/sbesc.2011.19},
booktitle = {2011 Brazilian Symposium on Computing System Engineering},
publisher = {IEEE},
author = {Barreto,  Raimundo and Cordeiro,  Lucas and Fischer,  Bernd},
year = {2011},
month = nov,
pages = {46–52},
address = {Los Alamitos, {CA}, {USA}},
}

@inbook{Donaldson2011,
title = {{Software Verification Using k-Induction}},
ISBN = {9783642237027},
ISSN = {1611-3349},
DOI = {10.1007/978-3-642-23702-7_26},
booktitle = {Static Analysis},
publisher = {Springer Berlin Heidelberg},
author = {Donaldson,  Alastair F. and Haller,  Leopold and Kroening,  Daniel and R\"{u}mmer,  Philipp},
year = {2011},
pages = {351–368},
address = {Berlin, Heidelberg},
}

@inbook{Clarke2012,
title = {{Model Checking and the State Explosion Problem}},
ISBN = {9783642357466},
ISSN = {1611-3349},
DOI = {10.1007/978-3-642-35746-6_1},
booktitle = {Tools for Practical Software Verification},
publisher = {Springer Berlin Heidelberg},
author = {Clarke,  Edmund M. and Klieber,  William and Nováček,  Miloš and Zuliani,  Paolo},
year = {2012},
pages = {1–30},
address = {Berlin, Heidelberg, Germany}
}

@article{pearce2021asleep,
title = {{Asleep at the Keyboard? Assessing the Security of GitHub Copilot's Code Contributions}},
volume = {68},
ISSN = {1557-7317},
DOI = {10.1145/3610721},
number = {2},
journal = {Communications of the ACM},
publisher = {Association for Computing Machinery (ACM)},
author = {Pearce, Hammond and Ahmad, Baleegh and Tan, Benjamin and Dolan-Gavitt, Brendan and Karri, Ramesh},
year = {2025},
month = jan,
pages = {96–105},
address = {San Francisco, CA, USA},
}

@inbook{beckert2024towards,
title = {{Towards Combining the Cognitive Abilities of Large Language Models with the Rigor of Deductive Progam Verification}},
ISBN = {9783031753879},
ISSN = {1611-3349},
DOI = {10.1007/978-3-031-75387-9_15},
booktitle = {Leveraging Applications of Formal Methods,  Verification and Validation. Software Engineering Methodologies},
publisher = {Springer Nature Switzerland},
author = {Beckert,  Bernhard and Klamroth,  Jonas and Pfeifer,  Wolfram and R\"{o}per,  Patrick and Teuber,  Samuel},
year = {2024},
month = oct,
pages = {242–257},
address = {Cham, Switzerland}
}

@article{Monteiro2021,
title = {{Model Checking C++ Programs}},
volume = {32},
ISSN = {1099-1689},
DOI = {10.1002/stvr.1793},
number = {1},
journal = {Software Testing,  Verification and Reliability},
publisher = {Wiley},
author = {Monteiro,  Felipe R. and Gadelha,  Mikhail R. and Cordeiro,  Lucas C.},
year = {2021},
month = sep,
pages = {e1793},
numpages = {20},
}

@misc{Tihanyi2025vulnerability,
doi = {10.48550/arxiv.2503.10784},
author = {Tihanyi,  Norbert and Bisztray,  Tamas and Ferrag,  Mohamed Amine and Cherif,  Bilel and Dubniczky,  Richard A. and Jain,  Ridhi and Cordeiro,  Lucas C.},
title = {{Vulnerability Detection: From Formal Verification to Large Language Models and Hybrid Approaches: A Comprehensive Overview}},
publisher = {arxiv},
year = {2025},
}

@inproceedings{Haoze2023,
author = {Wu, Haoze and Barrett, Clark W. and Narodytska, Nina},
title = {{Lemur: Integrating Large Language Models in Automated Program Verification}},
booktitle = {The Twelfth International Conference on Learning Representations (ICLR 2024)},
publisher = {OpenReview.net},
year = {2024},
url = {https://openreview.net/forum?id = Q3YaCghZNt},
address = {Vienna, Austria},
pages = {7652},
numpages = {15},
}

@misc{Yiannis2024,
doi = {10.48550/arxiv.2405.08848},
author = {Charalambous,  Yiannis and Manino,  Edoardo and Cordeiro,  Lucas C.},
title = {{Automated Repair of AI Code with Large Language Models and Formal Verification}},
publisher = {arxiv},
year = {2024},
}

@misc{Huang2023,
doi = {10.48550/arxiv.2305.11391},
author = {Huang,  Xiaowei and Ruan,  Wenjie and Huang,  Wei and Jin,  Gaojie and Dong and others},
title = {{A Survey of Safety and Trustworthiness of Large Language Models through the Lens of Verification and Validation}},
publisher = {arxiv},
year = {2023},
}

@article{Cofer2018,
title = {{A Formal Approach to Constructing Secure Air Vehicle Software}},
volume = {51},
ISSN = {1558-0814},
DOI = {10.1109/mc.2018.2876051},
number = {11},
journal = {Computer},
publisher = {Institute of Electrical and Electronics Engineers (IEEE)},
author = {Cofer,  Darren and Gacek,  Andrew and Backes,  John and Whalen,  Michael W. and Pike,  Lee and Foltzer,  Adam and Podhradsky,  Michal and Klein,  Gerwin and Kuz,  Ihor and Andronick,  June and Heiser,  Gernot and Stuart,  Douglas},
year = {2018},
month = nov,
pages = {14–23}
}

@inproceedings{Hatcliff2012,
title = {{Rationale and Architecture Principles for Medical Application Platforms}},
DOI = {10.1109/iccps.2012.9},
booktitle = {2012 IEEE/ACM Third International Conference on Cyber-Physical Systems},
publisher = {IEEE},
author = {Hatcliff,  John and King,  Andrew and Lee,  Insup and Macdonald,  Alasdair and Fernando,  Anura and Robkin,  Michael and Vasserman,  Eugene and Weininger,  Sandy and Goldman,  Julian M.},
year = {2012},
month = apr,
pages = {3–12},
address = {Beijing, China}
}

@article{Luckcuck2019,
title = {{Formal Specification and Verification of Autonomous Robotic Systems: A Survey}},
volume = {52},
ISSN = {1557-7341},
DOI = {10.1145/3342355},
number = {5},
journal = {ACM Computing Surveys},
publisher = {Association for Computing Machinery (ACM)},
author = {Luckcuck,  Matt and Farrell,  Marie and Dennis,  Louise A. and Dixon,  Clare and Fisher,  Michael},
year = {2019},
month = sep,
pages = {1–41}
}

@article{Woodcock2009,
title = {{Formal Methods: Practice and Experience}},
volume = {41},
ISSN = {1557-7341},
DOI = {10.1145/1592434.1592436},
number = {4},
journal = {ACM Computing Surveys},
publisher = {Association for Computing Machinery (ACM)},
author = {Woodcock,  Jim and Larsen,  Peter Gorm and Bicarregui,  Juan and Fitzgerald,  John},
year = {2009},
month = oct,
pages = {1–36}
}

@book{Kroening2016,
title = {Decision Procedures},
ISBN = {9783662504970},
ISSN = {1862-4499},
DOI = {10.1007/978-3-662-50497-0},
journal = {Texts in Theoretical Computer Science. An EATCS Series},
publisher = {Springer Berlin Heidelberg},
author = {Kroening,  Daniel and Strichman,  Ofer},
year = {2016},
address = {Berlin, Heidelberg, Germany}
}

@misc{Wei2022,
doi = {10.48550/arxiv.2201.11903},
author = {Wei,  Jason and Wang,  Xuezhi and Schuurmans,  Dale and Bosma,  Maarten and Ichter,  Brian and Xia,  Fei and Chi,  Ed and Le,  Quoc and Zhou,  Denny},
title = {{Chain-of-Thought Prompting Elicits Reasoning in Large Language Models}},
journal = {arxiv},
year = {2022},
}

@inbook{Brummayer2009,
title = {{Boolector: An Efficient SMT Solver for Bit-Vectors and Arrays}},
ISBN = {9783642007682},
ISSN = {1611-3349},
DOI = {10.1007/978-3-642-00768-2_16},
booktitle = {Tools and Algorithms for the Construction and Analysis of Systems},
publisher = {Springer Berlin Heidelberg},
author = {Brummayer,  Robert and Biere,  Armin},
year = {2009},
pages = {174–177},
address = {Berlin, Heidelberg, Germany}
}

@misc{akhond2025,
doi = {10.48550/arxiv.2511.06552},
author = {Akhond,  Mostafijur Rahman and Chakraborty,  Saikat and Uddin,  Gias},
title = {{LLM For Loop Invariant Generation and Fixing: How Far Are We?}},
journal = {arxiv},
year = {2025},
}

@inbook{Wen2024,
title = {{Enchanting Program Specification Synthesis by Large Language Models Using Static Analysis and Program Verification}},
ISBN = {9783031656309},
ISSN = {1611-3349},
DOI = {10.1007/978-3-031-65630-9_16},
booktitle = {Computer Aided Verification},
publisher = {Springer Nature Switzerland},
author = {Wen,  Cheng and Cao,  Jialun and Su,  Jie and Xu,  Zhiwu and Qin,  Shengchao and He,  Mengda and Li,  Haokun and Cheung,  Shing-Chi and Tian,  Cong},
year = {2024},
pages = {302–328},
address = {Cham, Switzerland}
}

@inbook{svcomp2024,
title = {State of the Art in Software Verification and Witness Validation: SV-COMP 2024},
ISBN = {9783031572562},
ISSN = {1611-3349},
DOI = {10.1007/978-3-031-57256-2_15},
booktitle = {Tools and Algorithms for the Construction and Analysis of Systems},
publisher = {Springer Nature Switzerland},
author = {Beyer,  Dirk},
year = {2024},
pages = {299–329},
address = {Luxembourg City, Luxembourg},
}

@inproceedings{bueno2022cheri,
author = {Brau{\ss}e, Franz and Shmarov, Fedor and Menezes, Rafael
and Gadelha, Mikhail R. and Korovin, Konstantin
and Reger, Giles and Cordeiro, Lucas C.},
title = {{ESBMC-CHERI: Towards Verification of C Programs for
CHERI Platforms with ESBMC}},
booktitle = {Proceedings of the 31st ACM SIGSOFT International
Symposium on Software Testing and Analysis
(ISSTA~2022)},
pages = {773--776},
year = {2022},
month = jul,
publisher = {ACM},
doi = {10.1145/3533767.3543289},
address = {Seoul, South Korea}
}

@misc{ssvlab_cordeiro,
author = {Cordeiro, Lucas C.},
title = {{Lucas Cordeiro — Systems and Software Verification
Laboratory (SSVLab) Personal Page}},
year = {2025},
howpublished = {\url{https://ssvlab.github.io/lucasccordeiro/}},
note = {Accessed: May 2026}
}

@misc{cordis957286,
author = {{European Commission}},
title = {{ELEGANT — Towards Attestable and Trustworthy
Internet-of-Things Infrastructures (H2020 grant
agreement No.\ 957286)}},
year = {2021},
howpublished = {CORDIS — EU Research Results.
\url{https://cordis.europa.eu/project/id/957286}},
note = {Accessed: May 2026}
}

@misc{ukri_dsbd,
author = {{UK Research and Innovation (UKRI)}},
title = {{Digital Security by Design (DSbD) Programme}},
year = {2019},
howpublished = {\url{https://www.ukri.org/what-we-do/browse-our-areas-of-investment-and-support/digital-security-by-design/}},
note = {Accessed: May 2026}
}

@misc{veribee_impact,
author = {{University of Manchester Research Explorer}},
title = {{VeriBee: Source Code Security (Research Impact Record)}},
year = {2025},
howpublished = {\url{https://research.manchester.ac.uk/en/impacts/veribee-source-code-security/}},
note = {Accessed: May 2026}
}

@techreport{nist2002,
author = {Tassey, Gregory},
title = {{The Economic Impacts of Inadequate Infrastructure for
Software Testing}},
institution = {National Institute of Standards and Technology (NIST)},
year = {2002},
number = {Planning Report 02-3},
address = {Gaithersburg, MD},
url = {https://www.nist.gov/document/report02-3pdf},
note = {Accessed: May 2026}
}

@misc{ibm_defect_cost,
author = {{IBM Systems Sciences Institute}},
title = {{Relative Cost of Fixing Defects}},
year = {1981},
howpublished = {Black Duck
\url{https://www.blackduck.com/blog/cost-to-fix-bugs-during-each-sdlc-phase.html}},
}

@misc{boeing737max,
author = {{IEEE Spectrum / Wikipedia}},
title = {{Boeing 737 MAX Groundings (March 2019 – December 2020)}},
year = {2021},
howpublished = {IEEE Spectrum:
\url{https://spectrum.ieee.org/how-the-boeing-737-max-disaster-looks-to-a-software-developer};
note = {Accessed: May 2026}}
}

@misc{crowdstrike2024,
author = {{CNN Business / Parametrix}},
title = {{CrowdStrike Outage: Cost and Cause (July 2024)}},
year = {2024},
howpublished = {CNN Business:
\url{https://www.cnn.com/2024/07/24/tech/crowdstrike-outage-cost-cause};
Altitudes Magazine:
\url{https://www.theguardian.com/technology/article/2024/jul/24/crowdstrike-outage-companies-cost}}
}

@misc{avionics_do178c,
author = {{Avionics Today / Vita Technologies}},
title = {{DO-178C: Certification Cost-Effective Avionics Systems}},
year = {2017},
howpublished = {\url{https://vita.militaryembedded.com/2325-do-178c-certification-cost-effective-avionics-systems/}},
note = {Accessed: May 2026}
}

@misc{eurocontrol_sloc,
author = {{Better Embedded Systems Blog / Eurocontrol}},
title = {{Cost of Highly Safety-Critical Software}},
year = {2018},
howpublished = {\url{https://betterembsw.blogspot.com/2018/10/cost-of-highly-safety-critical-software.html}},
note = {Accessed: May 2026}
}

@misc{statista_lmt_rd,
author = {{Statista}},
title = {{Lockheed Martin Annual R\&D Expenditure (2023)}},
year = {2024},
howpublished = {\url{https://www.statista.com/statistics/268928/expenditure-on-research-and-development-of-defense-supplier-lockheed-martin/}},
note = {Accessed: May 2026}
}

@misc{wu2025armcca,
doi = {10.48550/ARXIV.2406.04375},
author = {Wu,  Tong and Xiong,  Shale and Manino,  Edoardo and Stockwell,  Gareth and Cordeiro,  Lucas C.},
title = {Verifying components of Arm(R) Confidential Computing Architecture with ESBMC},
publisher = {arXiv},
year = {2025},
note = {arXiv preprint first posted June 2024 (arXiv:2406.04375)},
}

@misc{arm_revenue,
author = {{Arm Holdings plc}},
title = {{Arm Holdings Annual Revenue (FY2025)}},
year = {2025},
howpublished = {MacroTrends / StockAnalysis:
\url{https://stockanalysis.com/stocks/arm/market-cap/}},
note = {Accessed: May 2026}
}

@misc{ibm_breach2024,
author = {{IBM and Ponemon Institute}},
title = {{Cost of a Data Breach Report 2024}},
year = {2024},
howpublished = {\url{https://newsroom.ibm.com/2024-07-30-ibm-report-escalating-data-breach-disruption-pushes-costs-to-new-highs}},
note = {Accessed: May 2026}
}

@misc{securitybrief_2024,
author = {{SecurityBrief UK}},
title = {{UK SMEs Face Rise in Cyber Attacks with Average Cost
£7{,}960}},
year = {2024},
howpublished = {\url{https://securitybrief.co.uk/story/uk-smes-face-rise-in-cyber-attacks-with-average-cost-gbp-7-960}},
note = {Accessed: May 2026}
}

@misc{embedded_iot_cve,
author = {{Embedded Computing Design / Embedded.com}},
title = {{A Sensible Solution for Addressing the CVE Explosion
in IoT Devices}},
year = {2024},
howpublished = {\url{https://www.embedded.com/a-sensible-solution-for-addressing-the-cve-explosion-in-iot-devices/}},
note = {Accessed: May 2026}
}

@misc{tuxcare_cve,
author = {{TuxCare}},
title = {{The Real-World Cost of Not Patching a Critical CVE:
A Security ROI Perspective}},
year = {2024},
howpublished = {\url{https://tuxcare.com/blog/the-real-world-cost-of-not-patching-a-critical-cve-a-security-roi-perspective/}},
note = {Accessed: May 2026}
}

@misc{sibros_auto,
author = {{Sibros}},
title = {{The Current State of Automotive Software-Related Recalls}},
year = {2024},
howpublished = {\url{https://www.sibros.tech/post/the-current-state-of-automotive-software-related-recalls}},
note = {Accessed: May 2026}
}

@misc{defillama,
author = {{DefiLlama}},
title = {{DeFi Total Value Locked (TVL) Dashboard}},
year = {2025},
howpublished = {\url{https://defillama.com/}},
note = {Accessed: May 2026}
}

@misc{theblock_staking,
author = {{The Block}},
title = {{Over \$115~Billion of Ether Is Now Staked on the
Beacon Chain}},
year = {2024},
howpublished = {\url{https://www.theblock.co/post/280535/over-115-billion-of-ether-is-now-staked-on-the-beacon-chain}},
note = {Accessed: May 2026}
}

@misc{halborn2025,
author = {{Halborn Security}},
title = {{The Top~100 DeFi Hacks Report 2025}},
year = {2025},
howpublished = {\url{https://www.halborn.com/reports/top-100-defi-hacks-2025};
summary at
\url{https://www.halborn.com/blog/post/halborn-all-time-top-100-defi-hacks-report-summary}},
note = {Accessed: May 2026}
}

@misc{chainalysis2025,
author = {{Chainalysis}},
title = {{2025 Crypto Crime Trends}},
year = {2025},
howpublished = {\url{https://www.chainalysis.com/blog/2025-crypto-crime-report-introduction/}},
note = {Accessed: May 2026}
}

@misc{coindesk_ronin,
author = {{CoinDesk}},
title = {{Axie Infinity's Ronin Network Suffers \$625M Exploit}},
year = {2022},
howpublished = {\url{https://www.coindesk.com/tech/2022/03/29/axie-infinitys-ronin-network-suffers-625m-exploit}},
note = {Accessed: May 2026}
}

@misc{ethereum_bounty,
author = {{Ethereum Foundation}},
title = {{Ethereum Bug Bounty Program}},
year = {2025},
howpublished = {\url{https://ethereum.org/bug-bounty/}; CryptoRank:
\url{https://cryptorank.io/news/feed/f4381-ethereum-foundation-bug-bounty-million}},
note = {Accessed: May 2026}
}

@misc{zealynx_audit_cost,
author = {{Zealynx Security}},
title = {{Smart Contract Audit Cost in 2025: What You Need
to Know}},
year = {2025},
howpublished = {\url{https://www.zealynx.io/research/audit-ops/audit-pricing-2026}},
note = {Accessed: May 2026}
}

@misc{softwaresecured_pentest,
author = {{Software Secured}},
title = {{Is the Price Always Right? A Comprehensive Guide to
Penetration Testing Costs}},
year = {2024},
howpublished = {\url{https://www.softwaresecured.com/post/is-the-price-always-right-a-comprehensive-guide-to-penetration-testing-costs}},
note = {Accessed: May 2026}
}

@misc{ukri_soteria,
author = {{UK Research and Innovation (UKRI)}},
title = {{R\&D Investments Spearhead Push to Block Cyber Security
Attacks (Soteria consortium, £5.8M)}},
year = {2021},
howpublished = {\url{https://www.openaccessgovernment.org/rd-investments-spearhead-cyber-security-defence/96958/}}
}

@techreport{dcms2024,
author = {{Department for Science, Innovation and Technology}},
title = {Cyber Security Breaches Survey 2024},
institution = {UK Government},
year = {2024},
url = {https://www.gov.uk/government/statistics/cyber-security-breaches-survey-2024},
}

@techreport{cisq2020,
author = {{Consortium for Information and Software Quality}},
title = {The Cost of Poor Software Quality in the {US}:
A 2020 Report},
institution = {CISQ},
year = {2020},
url = {https://www.it-cisq.org/the-cost-of-poor-software-quality-in-the-us-a-2020-report/},
}

@article{Lewis2023,
title = {Formal Verification of Quantum Programs: Theory,  Tools,  and Challenges},
volume = {5},
ISSN = {2643-6817},
DOI = {10.1145/3624483},
number = {1},
journal = {ACM Transactions on Quantum Computing},
publisher = {Association for Computing Machinery (ACM)},
author = {Lewis,  Marco and Soudjani,  Sadegh and Zuliani,  Paolo},
year = {2023},
month = Dec,
pages = {1–35}
}

@inbook{Gay2008,
title = {QMC: A Model Checker for Quantum Systems},
ISBN = {9783540705451},
ISSN = {1611-3349},
DOI = {10.1007/978-3-540-70545-1_51},
booktitle = {Computer Aided Verification},
publisher = {Springer Berlin Heidelberg},
author = {Gay,  Simon J. and Nagarajan,  Rajagopal and Papanikolaou,  Nikolaos},
pages = {543–547},
year = {2008},
address = {Princeton, NJ, USA},
}

@inproceedings{Xu2022,
series = {PLDI ’22},
title = {Quartz: superoptimization of Quantum circuits},
DOI = {10.1145/3519939.3523433},
booktitle = {Proceedings of the 43rd ACM SIGPLAN International Conference on Programming Language Design and Implementation},
publisher = {ACM},
author = {Xu,  Mingkuan and others},
year = {2022},
month = Jun,
pages = {625–640},
collection = {PLDI ’22},
address = {San Diego, CA, USA},
}

@article{Monteiro2018,
title = {ESBMC-GPU A context-bounded model checking tool to verify CUDA programs},
volume = {152},
ISSN = {0167-6423},
DOI = {10.1016/j.scico.2017.09.005},
journal = {Science of Computer Programming},
publisher = {Elsevier BV},
author = {Monteiro,  Felipe R. and da S. Alves,  Erickson H. and Silva,  Isabela S. and Ismail,  Hussama I. and Cordeiro,  Lucas C. and de Lima Filho,  Eddie B.},
year = {2018},
month = Jan,
pages = {63–69}
}

@article{Kaleeswaran2023,
title = {A user study for evaluation of formal verification results and their explanation at Bosch},
volume = {28},
ISSN = {1573-7616},
DOI = {10.1007/s10664-023-10353-4},
number = {5},
journal = {Empirical Software Engineering},
publisher = {Springer Science and Business Media LLC},
author = {Kaleeswaran,  Arut Prakash and Nordmann,  Arne and Vogel,  Thomas and Grunske,  Lars},
year = {2023},
month = Sep,
numpages = {28},
}

@inproceedings{Christakis2016,
series = {ASE'16},
title = {What developers want and need from program analysis: an empirical study},
DOI = {10.1145/2970276.2970347},
booktitle = {Proceedings of the 31st IEEE/ACM International Conference on Automated Software Engineering},
publisher = {ACM},
author = {Christakis,  Maria and Bird,  Christian},
year = {2016},
month = Aug,
pages = {332–343},
collection = {ASE'16},
address = {Singapore},
}

@techreport{kitchenham2007guidelines,
author = {Kitchenham, Barbara and Charters, Stuart},
title = {{Guidelines for Performing Systematic Literature Reviews in Software Engineering}},
institution = {Keele University and Durham University Joint Report},
number = {EBSE 2007-001},
year = {2007}
}

@inproceedings{beyer2011cpachecker,
author = {Beyer, Dirk and Keremoglu, M. Erkan},
title = {{CPA}checker: A Tool for Configurable Software Verification},
booktitle = {Proceedings of the 23rd International Conference on Computer Aided Verification (CAV 2011)},
series = {Lecture Notes in Computer Science},
volume = {6806},
pages = {184--190},
year = {2011},
publisher = {Springer},
address = {Snowbird, UT, USA},
doi = {10.1007/978-3-642-22110-1_16}
}

@inproceedings{heizmann2013ultimate,
author = {Heizmann, Matthias and Dietsch, Daniel and Langenfeld, Vincent and Podelski, Andreas},
title = {Ultimate {Automizer} with Array Interpolation},
booktitle = {Proceedings of the 19th International Conference on Tools and Algorithms for the Construction and Analysis of Systems (TACAS 2013)},
series = {Lecture Notes in Computer Science},
volume = {7795},
pages = {455--457},
year = {2013},
publisher = {Springer},
address = {Rome, Italy}
}

@inproceedings{brain20162ls,
author = {Brain, Martin and Joshi, Saurabh and Kroening, Daniel and Schrammel, Peter},
title = {{2LS} for Program Analysis},
booktitle = {Proceedings of the 22nd International Conference on Tools and Algorithms for the Construction and Analysis of Systems (TACAS 2016)},
series = {Lecture Notes in Computer Science},
volume = {9636},
pages = {98--104},
year = {2016},
publisher = {Springer},
address = {Eindhoven, The Netherlands},
doi = {10.1007/978-3-662-49674-9_6}
}

@inproceedings{chalupa2020symbiotic,
author = {Chalupa, Marek and Strej{\v{c}}ek, Jan and Vi{\v{s}}{\v{n}}ovsk{\'y}, Martin},
title = {Symbiotic 7: Integration of a New Slicer},
booktitle = {Proceedings of the 26th International Conference on Tools and Algorithms for the Construction and Analysis of Systems (TACAS 2020)},
series = {Lecture Notes in Computer Science},
volume = {12079},
pages = {413--417},
year = {2020},
publisher = {Springer},
address = {Dublin, Ireland},
doi = {10.1007/978-3-030-45237-7_27}
}

@inproceedings{barnat2013divine,
author = {Barnat, Ji{\v{r}}{\'i} and others},
title = {{DiVinE} 3.0 -- An Explicit-State {LTL} Model Checker},
booktitle = {Proceedings of the 25th International Conference on Computer Aided Verification (CAV 2013)},
series = {Lecture Notes in Computer Science},
volume = {8044},
pages = {863--868},
year = {2013},
publisher = {Springer},
address = {Saint Petersburg, Russia},
doi = {10.1007/978-3-642-39799-8_60}
}

@inproceedings{toth2017theta,
author = {T{\'o}th, Tam{\'a}s and Hajdu, {\'A}kos and V{\"o}r{\"o}s, Andr{\'a}s and Micskei, Zolt{\'a}n and Majzik, Istv{\'a}n},
title = {Theta: A Framework for Abstraction Refinement-Based Model Checking},
booktitle = {Proceedings of the 17th Conference on Formal Methods in Computer-Aided Design (FMCAD 2017)},
pages = {176--179},
year = {2017},
address = {Vienna, Austria},
publisher = {IEEE},
doi = {10.23919/fmcad.2017.8102257}
}

@inproceedings{gurfinkel2015seahorn,
author = {Gurfinkel, Arie and Kahsai, Temesghen and Komuravelli, Anvesh and Navas, Jorge A.},
title = {The {SeaHorn} Verification Framework},
booktitle = {Proceedings of the 27th International Conference on Computer Aided Verification (CAV 2015)},
series = {Lecture Notes in Computer Science},
volume = {9206},
pages = {343--361},
year = {2015},
publisher = {Springer},
address = {San Francisco, CA, USA},
doi = {10.1007/978-3-319-21690-4_20}
}

@inproceedings{kani2022amazon,
author = {VanHattum, Alexa and Schwartz-Narbonne, Daniel and Chong, Nathan and Griggio, Alberto},
title = {Verifying Dynamic Trait Objects in {Rust}},
booktitle = {Proceedings of the 44th International Conference on Software Engineering: Software Engineering in Practice Track (ICSE-SEIP 2022)},
pages = {321--330},
year = {2022},
publisher = {ACM},
address = {Pittsburgh, PA, USA},
doi = {10.1145/3510457.3513031}
}

@misc{smartbugs2020,
doi = {10.48550/ARXIV.2007.04771},
url = {https://arxiv.org/abs/2007.04771},
author = {Ferreira,  João F. and Cruz,  Pedro and Durieux,  Thomas and Abreu,  Rui},
title = {SmartBugs: A Framework to Analyze Solidity Smart Contracts},
publisher = {arXiv},
year = {2020},
copyright = {arXiv.org perpetual,  non-exclusive license}
}

@misc{sun2026agentic,
doi = {10.48550/ARXIV.2605.21434},
author = {Sun,  Youcheng and Liu,  Jiawen and Kroening,  Daniel and Xue,  Jason},
title = {Agentic Model Checking},
publisher = {arXiv},
year = {2026},
}

@misc{nvidia_opensma_commits,
author = {Cheng, Alan and Cordeiro, Lucas C.},
title = {{NVIDIA-OpenSMA}: {ESBMC} Verification Harnesses for {OpenSMA} (\texttt{vr\_sma} Branch)},
howpublished = {\url{https://github.com/lucasccordeiro/NVIDIA OpenSMA/commits/vr\_sma/}},
note = {Repository initialised 30~July~2025 (commit \texttt{0dcdcd1}); first verification commit
\texttt{9b8e1a4} (``Add {ESBMC} verification harnesses for {OpenSMA}''), April~2026},
year = {2025},
}

@phdthesis{menezes2025goto,
author = {Rafael Sá Menezes},
title = {GOTO: a verification framework for CProver tools},
school = {The University of Manchester},
year = {2025},
type = {PhD thesis},
month = {June},
url = {https://ssvlab.github.io/lucasccordeiro/supervisions/phd_thesis_rafael.pdf}
}
